\documentclass[fleqn,usenatbib]{mnras}
\usepackage{newtxtext,newtxmath}
\usepackage[]{xcolor}
\usepackage{threeparttable}
\usepackage{soul}
\usepackage{multicol}
\usepackage[T1]{fontenc}
\DeclareRobustCommand{\VAN}[3]{#2}
\let\VANthebibliography\thebibliography
\def\thebibliography{\DeclareRobustCommand{\VAN}[3]{##3}\VANthebibliography}

\usepackage{graphicx}	
\usepackage{amsmath}	

\usepackage{amssymb}	

\newcommand{\XMM}{\textit{XMM}}
\newcommand{\xmm}{{{\it XMM-Newton}}}
\newcommand{\Swift}{\textit{Swift}}
\newcommand{\wise}{\emph{WISE}}

\title[XMMSL2 variable sources]{Populations of highly variable X-ray sources in the {\xmm} slew survey}

\author[Dongyue Li et al.]{
Dongyue Li$^{1,2,3}$\thanks{E-mail: dyli@nao.cas.cn},
R.L.C. Starling$^{3}$, 
R.D. Saxton$^{4}$, 
Hai-Wu Pan$^{1,2}$
and Weimin Yuan$^{1,2}$\thanks{E-mail: wmy@nao.cas.cn}
\\
$^{1}$ 
National Astronomical Observatories, Chinese Academy of Sciences, 20A Datun Road, Chaoyang District, Beijing, 100101, China\\
$^{2}$School of Astronomy and Space Science, University of Chinese Academy of Sciences, 19A Yuquan Road, Beijing, 100049, China\\
$^{3}$School of Physics and Astronomy, University of Leicester, University Road, Leicester LE1 7RH, UK\\
$^{4}$Telespazio UK for ESA, Operations Department, European Space Astronomy Centre,  E-28691 Villanueva de la Ca$\tilde{n}$ada, Spain\\
}

\date{Accepted 2022 March 9. Received 2022 March 8; in original form 2021 September 23}

\pubyear{2022}

\begin{document}
\label{firstpage}
\pagerange{\pageref{firstpage}--\pageref{lastpage}}
\maketitle

\begin{abstract}
We present the identifications of a flux-limited sample of highly variable X-ray sources on long time-scales from the second catalogue of the {\xmm} SLew survey (XMMSL2). The carefully constructed sample, comprising 265 sources (2.5 per cent) selected from the XMMSL2 clean catalogue, displayed X-ray variability of a factor of more than 10 in 0.2--2\,keV compared to the ROSAT All Sky Survey. Of the sample sources, 94.3 per cent are identified. The identification procedure follows a series of cross-matches with astronomical data bases and multiwavelength catalogues to refine the source position and identify counterparts to the X-ray sources. Assignment of source type utilizes a combination of indicators including counterparts offset, parallax measurement, spectral colours, X-ray luminosity, and light-curve behaviour. We identified 40 per cent of the variables with stars, 10 per cent with accreting binaries and at least 30.4 per cent with active galactic nuclei. The rest of the variables are identified as galaxies.
It is found that the mean effective temperatures of the highly variable stars are lower than those of less variable stars. Our sample of highly variable AGN tend to have lower black hole masses, redshifts, and marginally lower soft X-ray luminosities compared to the less variable ones, while no difference was found in the Eddington ratio distributions.
Five flaring events are tidal disruption events published previously. This study has significantly increased the number of variable sources in XMMSL2 with identifications and provides greater insight on the nature of many of the sources, enabling further studies of highly variable X-ray sources.
\end{abstract}

\begin{keywords}
Surveys -- X-rays: general -- galaxies: active -- stars: flare --X-rays: binaries
\end{keywords}


\section{Introduction}
Variability is known to be one of the key characteristics of the X-ray sky. 
Since the early days of X-ray astronomy, the study of the temporal properties of X-ray radiation, alongside spectral and imaging studies, has been a powerful tool with which to explore the radiation processes and nature of X-ray sources, ranging from stellar coronal activity, to accreting stellar compact objects, and to active galactic nuclei (AGN) (see \citealt{seward2010} for a review). Strong X-ray radiation is produced in emitting regions characterized by high temperatures (above $10^6$\,K) and/or high-energy electron populations. Such regions are usually associated with extreme physical conditions such as strong gravity and/or strong magnetic fields, and are involving heating or particle acceleration mechanisms such as shocks. These regions and processes, often subject to certain instabilities, are non-stationary and give rise to variations in the X-ray radiation observed (e.g. \citealp{melia2009}). Therefore, the variability amplitudes and time-scales carry information on the transfer of energy from a certain form (e.g. gravity, magnetic field) to the radiating plasma, or information on the properties and physical processes of the source (e.g. instabilities, dynamics). The study of variability has greatly aided the development of the physical models of various types of X-ray sources known so far. For example, the amplitude and timescale of observed X-ray flares from main-sequence stars provide a measure of the energy released in those events, which fits well with models involving magnetic-reconnection in the stellar coronae.

Some sources were observed to show high-amplitude variations or even outbursts, in which an exceptionally large amount of energy is released. Such events are of particular interest since they are likely produced in a physical condition or environment more extreme than usual (such as extremely strong magnetic fields, or being very close to a compact object), or in a violent or even catastrophic process. For instance, the transition of the accretion flow from a low to a high state can result in an increase in the X-ray luminosity of a black hole X-ray binary by several orders of magnitude \citep{remillard2006}. Another example is a tidal disruption event (TDE), in which a star is tidally disrupted by a massive black hole, producing an X-ray flare at the centre of a non-active galaxy as luminous as an AGN (\citealp{rees1988, KomossaBade1999,komossa2017}, refer to \citealp{saxton2021} for a recent review). 

The variability of X-ray sources has been extensively studied for decades. 
The majority of previous and current X-ray instruments have a small field of view (FoV) and have observed pre-selected individual sources, 
while wide-field X-ray monitor missions, such as {\it Beppo-SAX} WFC \citep{boella1997}, {\it Swift} BAT \citep{barthelmy2005} and {\it MAXI} \citep{matsuoka2009}, have provided a more unbiased census of the variability of X-ray sources. Time-domain survey data are valuable for understanding the statistical properties of the variability, which can help build the overall paradigm for the nature and activity of any type of source. 
They also provide an important discovery mechanism for transients and new types of variables. Current wide-field surveys are limited, however, by a relatively low detection sensitivity and can monitor only bright X-ray sources, mostly Galactic, leaving the ensemble variability of the bulk of the X-ray source populations largely unexplored. On the other hand, some sensitive narrow-field instruments have accumulated huge data sets during their pointed observations or slews, in which many X-ray sources have been detected serendipitously. These valuable archival data can be used to study the long-term variability properties of X-ray sources in a statistical manner by comparing with other surveys/observations performed in similar passbands. There have been many such previous studies. For example, \citet{fuhrmeister2003} studied the variability within the {\it ROSAT} All-Sky Survey (RASS; \citealp{voges1999}) observations and obtained statistical results regarding the variable X-ray populations. By comparing X-ray sources from the {\xmm} serendipitous survey in its first-year observations with those previously measured with {\it ROSAT}, \citet{yuan2006} found a small sample of highly variable sources on time-scales of years. \citet{traulsen2020} studied the long-term variable content in the {\xmm} serendipitous survey, finding that about 4 per cent of the sources show very long-term variability. Recently, \citet{boller2021} published a variable source catalogue obtained from the eROSITA Final Equatorial-Depth Survey (eFEDS; \citealp{brunner2021}), which contains mostly flaring stars.

The {\xmm} SLew survey (XMMSL; \citealp{saxton2008}), with its large and nearly homogeneous sky coverage and comparable soft X-ray sensitivity to that of the RASS, is well suited for the study of long-term variability of relatively faint X-ray sources, and for the discovery of X-ray flaring and transient events. It contains data taken with the EPIC-pn camera during slews from one pointed observation to another, which have been processed separately in three energy bands, 0.2--12 (total band), 2--12 (hard band), and 0.2--2\,keV (soft band).  
XMMSL2\footnote{\url{https://www.cosmos.esa.int/web/xmm-newton/xmmsl2-ug}} was generated from 2114 slews, executed between 2001 August 26 and 2014 December 31, revolutions 314--2758 (see Data Availability). In XMMSL2, 84 per cent of the sky has been covered with a flux limit of 5.7$\times$ 10$^{-13}$\,erg\,s$^{-1}$\,cm$^{-2}$ and a mean exposure time of 6\,s in the 0.2--2\,keV energy band. 
The XMMSL2 clean catalogue contains 22 337 sources in the 0.2--2\,keV soft band with detection likelihoods greater than 10.5 in general and greater than 15.5 for sources found in images with higher than usual background. About 4 per cent\footnote{This number was derived for an earlier version of the slew catalogue, XMMSL1, but is equally valid for XMMSL2.} of these sources are expected to be spurious from statistical considerations \citep{saxton2008}. For each entry, the public XMMSL2 catalogue reports the source type returned by the cross-correlation with different catalogues, including SIMBAD (Set of Identifications, Measurements, and Bibliography for Astronomical Data; \citealp{wenger2000}), NED (NASA/IPAC Extragalactic Database), and other catalogues and data bases. For those sources with multiple counterparts the identification was selected as the closest match. This process led to the identification of about half of the sources in the XMMSL2 clean source catalogue.

A comparison of source fluxes in the XMMSL data with  previous measurements has allowed the detection of individual variable sources and transients, such as novae \citep[e.g.,][]{read2008, read2009}, tidal disruption events \citep[e.g.,][]{esquej2008,saxton2012,saxton2014,li2020} and AGN activity \citep[e.g.,][]{miniutti2013, strotjohann2016}. \citet{kanner2013} performed a search for low-redshift extragalactic soft X-ray transients in XMMSL2 and gave the first estimate of the event rate within the LIGO/Virgo horizon. 
Nevertheless, a systematic census of the overall population of variable sources in XMMSL is still lacking.

In this paper, we first search for  long-term variable soft X-ray sources from the XMMSL2 clean catalogue by comparing the source fluxes with those measured in the earlier RASS, and then perform multiwavelength identification for those highly variable sources using various archival data bases. The RASS catalogue used for the comparison is the second ROSAT All Sky Survey point source catalogue (RASS2RXS; \citealp{voges1999}, \citealp{boller2016}), which contains point-like sources from the RASS observations performed with the Position-Sensitive Proportional Counter (PSPC) between 1990 June and 1991 August. There are about 135 000 X-ray detections in the 0.1--2.4\,keV energy band down to a likelihood threshold of 6.5, with a flux limit of few $\times$\,10$^{-13}$ erg\,s$^{-1}$\,cm$^{-2}$ \citep{boller2016}. 
Given the relatively large positional uncertainties of the XMMSL2 objects and the surface density of the relevant astrophysical source populations, the identification cannot be determined solely by matches at the closest distance.
For the selected sources we refine the identification process performed in the original XMMSL2 catalogue in order to verify the existing identifications and to provide a classification for nearly all the X-ray variable sources considered in this work.

This paper is organized as follows. We describe the XMMSL data and the search of variable sources in Section~\ref{sec:selection}, the identification process in Section~\ref{sec:identification}, and the results of identification in Section~\ref{sec:results}. Section~\ref{sec:discussion} discusses the identification methods and results and comparisons with other surveys, followed by a summary 
in Section~\ref{sec:summary}. Appendix~\ref{app:spurious} provides detailed justification for the matching radii adopted for cross-correlating the catalogues and estimates the potential rate of chance matches. More information for sources with more than one entry in SIMBAD or NED is provided in Appendix~\ref{sec:app_mul}.
All uncertainties in this paper are at the 1$\sigma$ significance level and one-sided limits at the 2$\sigma$ level unless stated otherwise. A flat cosmology with $H_{\rm 0}=70\,$km s$^{-1}$ Mpc$^{-1}$, $\Omega_m=0.3$, and $\Omega_A=0.7$ is assumed.  

\section{sample selection}\label{sec:selection}

We selected an initial sample from the XMMSL2 clean catalogue in the 0.2--2\,keV band using the following criteria:
\begin{itemize}
    \item  {\bf(a)} having detection likelihood$>$10.5 and background rate$<$3.0 (counts\,s$^{-1}$), or having detection likelihood$>$15.5; 
    \item {\bf (b)} detected with more than 5 source counts;
    \item {\bf (c)} free from optical loading\footnote{Optical loading is an effect where the optical photons from an extremely bright sources can excite a significant number of electrons in the X-ray CCDs, and can be falsely recognized as an X-ray source. Smith.M 2008,  PN  optical  loading,  XMM-SOC-CAL-TN-0051. \url{https://xmmweb.esac.esa.int/docs/documents/CAL-TN-0051-1-2.pdf}};  
    \item {\bf (d)} consistent with a point-like source (the fitted source extent $<$5 pixels). 
\end{itemize}
This resulted in 10 463 unique sources. 
For sources observed more than once (2296), the positions given in the observations with the highest detection likelihood were adopted and the highest fluxes were used in the following comparison.

To select long-term variable sources, we compared the XMMSL2 soft band fluxes of the initial sample with those measured in RASS2RXS. The mean positional accuracy for XMMSL2 and RASS2RXS is 15 and 8\,arcsec, respectively. Thus we adopted a 2\,arcmin radius around each of the XMMSL2 sources to search for its RASS2RXS counterpart, which  implies a potential chance match fraction of 15/7848 (see Appendix~\ref{app:spurious} for the justification of the matching radius and estimation of chance match, and Table~\ref{tab:spurious_summary} for a summary).
 
The Upper Limit Server tool, HILIGT \footnote{\url{http://xmmuls.esac.esa.int/hiligt/}} \citep{koenig2022,saxton2022}, was used to obtain the RASS count rates or upper limits at a given XMMSL2 position.
Count rate to flux conversion factors based on the assumption of a power-law spectrum with an index of $\Gamma = 1.7$ and a hydrogen column density of $N_{\rm H}=3 \times 10^{20} \rm cm^{-2}$ were adopted. This is also the spectral shape used in the fourth {\xmm} serendipitous source catalogue (4XMM-DR9; \citealp{webb2020}) and the XMMSL catalogue. 
We calculated ratios between the XMMSL2 and RASS fluxes (or upper limits) and selected sources with a flux ratio greater than 10, which resulted in 295 sources. We note that a relative flux of 10 is not very sensitive to the choice of spectral slope within $1.7 \le \Gamma \le 2.0$. We find a relative flux of 9.5 --11.5 (9.3--11.8) if the column density is varied from 10$^{20}$-10$^{22}$ cm$^{-2}$ with $\Gamma=1.7$ (2.0).

\begin{figure*}
	\includegraphics[width=1.8\columnwidth]{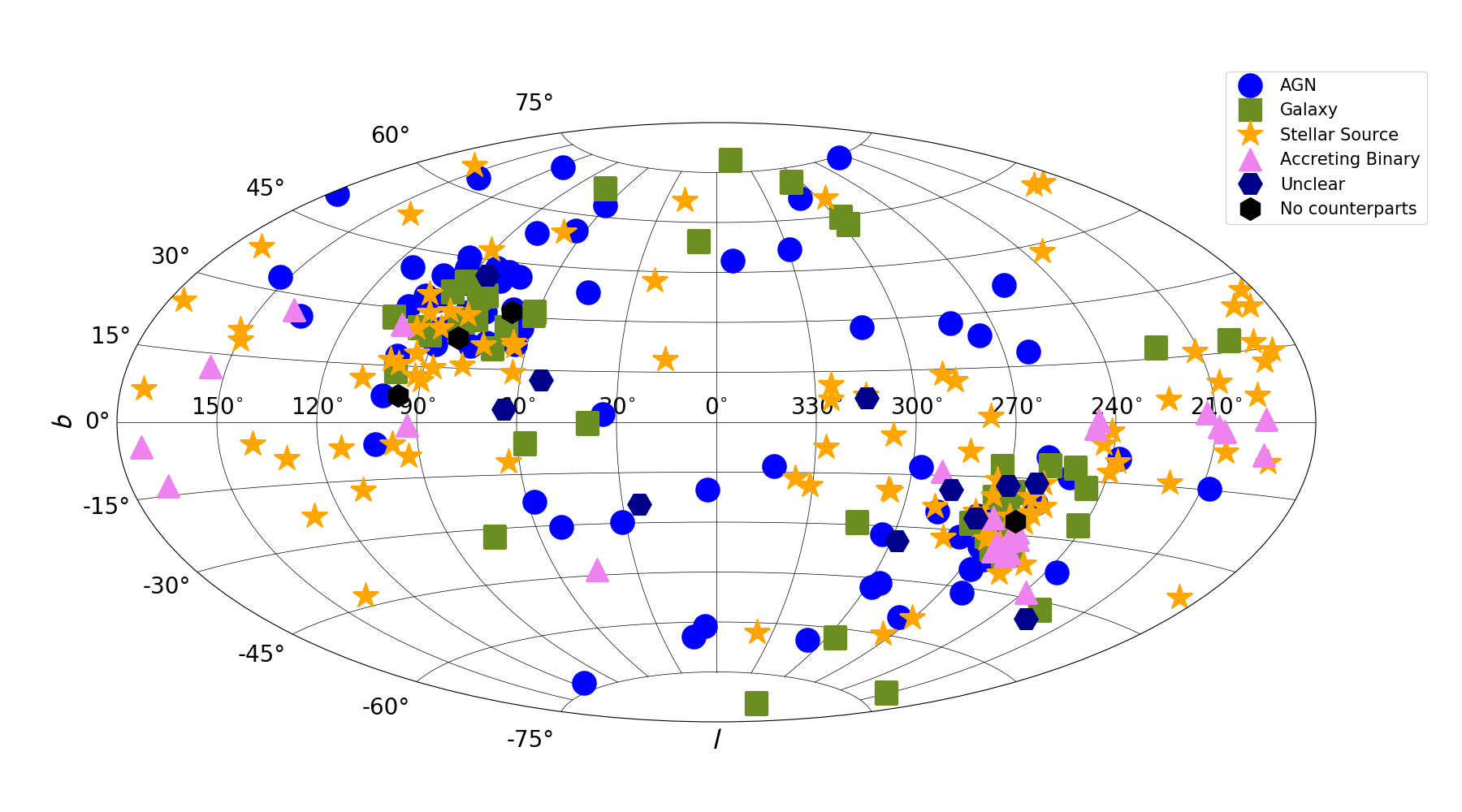}
    \caption{XMMSL2 variable sources in Galactic coordinates. Meanings of the markers and colours are indicated in the top-right corner. The identification of these variable sources is described in Section~\ref{sec:identification} and Section~\ref{sec:results}. }
    \label{fig:distribution}
\end{figure*}

Given that the spurious fraction of soft band sources in the XMMSL2 clean catalogue is expected to be around 4 per cent, we examined the slew images of these 295 sources by eye, and excluded 29 sources from the sample where the detection might be caused by a cosmic ray or variations in the background. We also found an anomaly that XMMSL2\,J051935.7-323932 and XMMSL2\,J051935.4-323929 were the same source but had a different unique source name in the XMMSL2 catalogue. We kept the source named XMMSL2\,J051935.7-323932 as it had the most significant detection. As a result, we had a sample of 265 sources, which is hereafter referred to as the variable sample. The variable sample objects comprise 2.5 per cent of the 10463 XMMSL2 clean sources meeting our selection criteria. Distribution of these sources on the sky in Galactic coordinates is shown in Figure~\ref{fig:distribution}. The sources are concentrated around the ecliptic poles where the slew paths predominantly pass. 

\section{Source identification}\label{sec:identification}

In the public XMMSL2 catalogue, 107 and 10 of our variable sample sources were identified by correlating the XMMSL2 coordinates with SIMBAD and NED archives, respectively.
In this paper, we undertake a more extensive identification process on the whole variable sample regardless of the XMMSL2 identification, in order to identify more of the sources and to obtain consistent identifications across the sample. 

\begin{figure}
    \centering
    \includegraphics[width=\columnwidth]{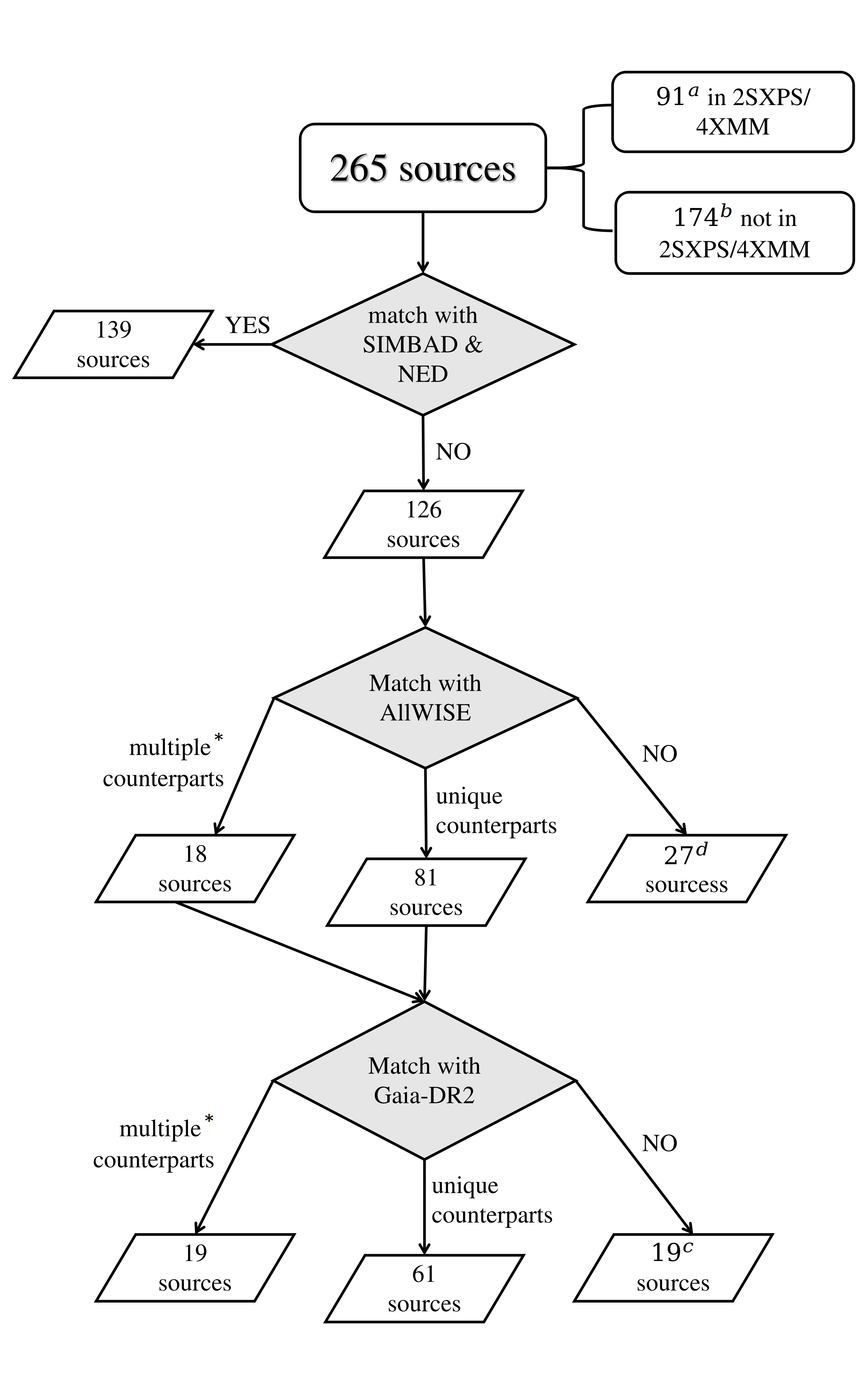}
    \caption{Identification process of the XMMSL2 variables adopted in this work.\\
    $^a$ These sources were cross-matched with the astronomical data bases and catalogues with positions and matching radii of 90 per cent position errors given in the 2SXPS/4XMM catalogue.\\
    $^b$ These sources were cross-matched with the astronomical data bases and catalogues with the XMMSL2 positions and a matching radius of 12”.\\
    $^c$ Identifications of these sources were given only based on their infrared colours.\\
    $^d$ The identification process was repeated for these sources with a larger radii, please refer to Section~\ref{sec:multiwave} for more details. \\
    $^*$ Identifications were given when the multiple counterparts of the XMMSL2 source were of the same type.}
    \label{fig:flowchart}
\end{figure}

The identification process is described in detail in the following subsections and is summarized here and in Figure~\ref{fig:flowchart}. Firstly, to reduce the astrometric uncertainty, all sources were correlated with the second \Swift-XRT Point Source Catalogue (2SXPS; \citealp{evans2020}) and 4XMM-DR9 to get better position accuracy. Secondly, the sources were correlated with SIMBAD and NED with the enhanced positions, or XMMSL2 positions if they were not in the point source catalogues, and the identifications would be adopted when specific\footnote{The `specific' here means that the category given in SIMBAD or NED provides information about the nature of the source, e.g. X-ray binary, binary, star, galaxy or AGN, rather than only the electromagnetic band in which the source was detected, e.g. IR source, X-ray source.} counterparts were found. Thirdly, for the sources with no specific counterparts in the data bases, multiwavelength properties would be used for the identification. 

\subsection{Cross-match with SIMBAD and NED}\label{sec:xmm_pointed}
All the XMMSL2 variable sources were correlated with 2SXPS and 4XMM-DR9 with an offset radius of 12\,arcsec, the same as that originally used in the correlation of the XMMSL2 catalogue with SIMBAD, NED and 3XMM. As a result, 79 and 49 sources were found in the 2SXPS and 4XMM-DR9 catalogues, respectively (with 37 sources in common). For sources with counterparts in both the 2SXPS and 4XMM-DR9 catalogues, the 4XMM positions and error circles were used in the identification. With better position accuracy, these 91 sources were first correlated with SIMBAD using the enhanced positions and matching radii of the 90 per cent confidence error on the position for each; 63\footnote{For XMMSL2\,J053527.7-691611, the unique entry is a cluster of stars in SIMBAD, and we did not take it as sources with unique counterpart at the moment.} XMMSL2 sources had a unique entry, while 7 sources had multiple entries in the SIMBAD database (see Table~\ref{tab:mul_swiftxmm} in Appendix~\ref{sec:app_mul}). We checked that for 6 out of the 7 sources, the multiple entries are associated with a single source and in fact only one source (XMMSL2\,J173616.9-444400 in Table~\ref{tab:mul_swiftxmm}) has multiple counterparts. This latter resides in a star cluster and we classified the source as “star”.

The remaining 21 sources with no SIMBAD counterparts, or where the nature of the counterparts was unclear (e.g. classified only as X-ray, UV, or IR sources), were then matched with NED with the same position and matching radii as used for SIMBAD, resulting in 3 further identifications. Consequently, 73 out of the 91 sources had been identified by cross-matching with SIMBAD and NED, and 18 sources remained unidentified.

\begin{figure}
	\includegraphics[width=\columnwidth]{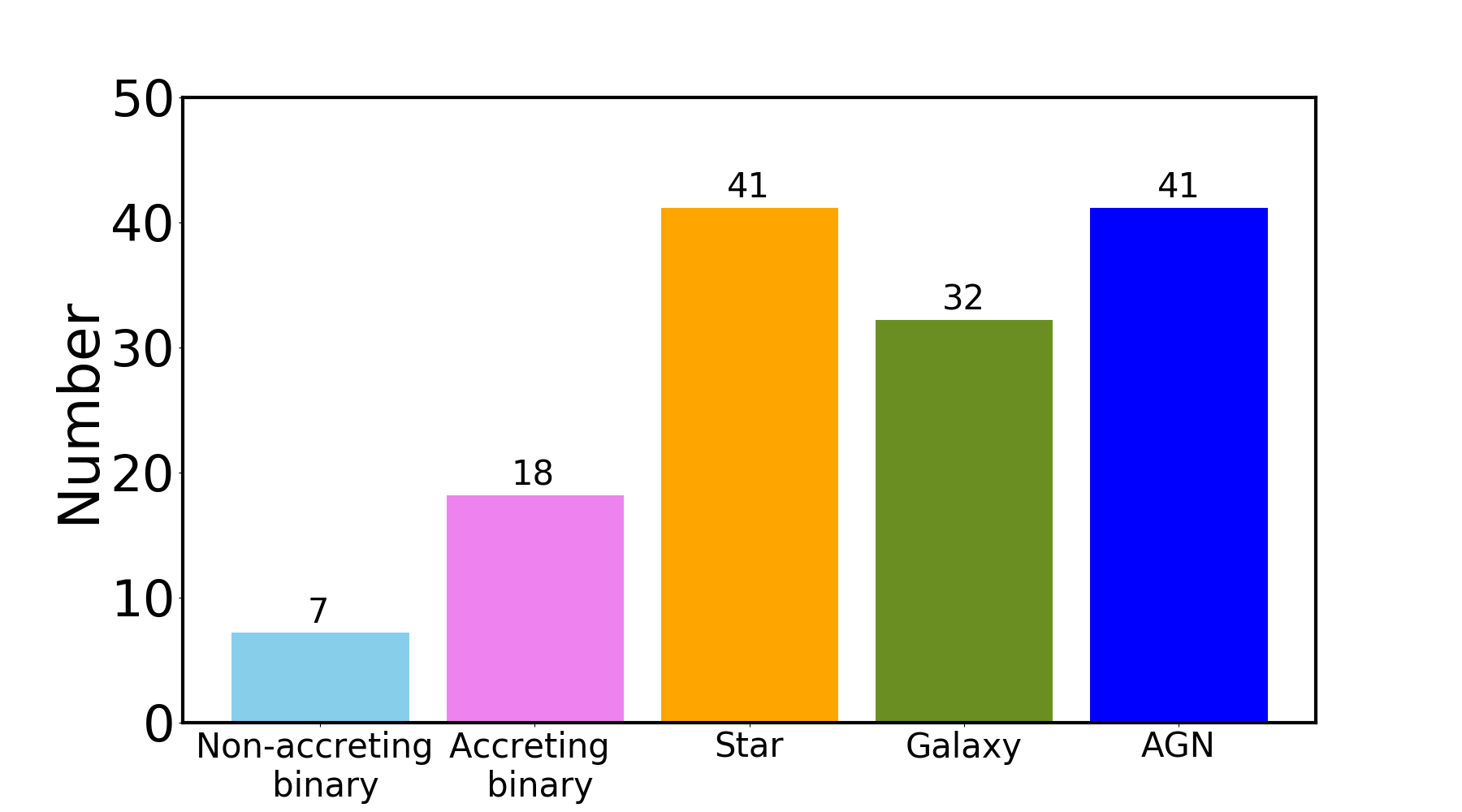}
    \caption{Category distribution of the sources identified according to their classifications in SIMBAD and NED as described in Section~\ref{sec:xmm_pointed}}.
    \label{fig:category_simbad}
\end{figure}

For the remaining 174 sources that did not have counterparts in the 4XMM-DR9 or 2SXPS catalogues, we cross-matched the XMMSL2 coordinates with the SIMBAD and NED archives by assuming a matching radius of 12\,arcmin. We obtained the identification for 66 sources (52 and 14 sources in SIMBAD and NED, respectively). Ff these, 20 had more than one entry in the SIMBAD/NED database (see Table~\ref{tab:mul_noxmmswift} in Appendix~\ref{sec:app_mul}), but only six sources were actually associated to multiple counterparts (XMMSL2\,J052443.6-613005, XMMSL2\,J092031.0+582243, XMMSL2\,J110535.5+095616, XMMSL2\,J123924.8-724407, XMMSL2\,J171238.5+584013, XMMSL2\,J173449.8+560256 in Table~\ref{tab:mul_noxmmswift}). We note however that the multiple entries associated with each of these six sources belong to the same category (e.g. galaxy or star). For this reason, while we can not associate the right counterpart to these XMMSLs sources, we can assign to them a classification. In summary, we obtained the identification for 139 XMMSL2 sources (see Figure~\ref{fig:category_simbad}), 131 of which have a unique counterpart in the SIMBAD/NED data bases.

\subsubsection{Characteristics with AllWISE and {\it Gaia} DR2}\label{sec:chara}

Infrared and optical properties of the 131 XMMSL2 variable sources with a unique counterpart in SIMBAD and NED were investigated by cross-matching with the AllWISE (i.e. the combination of WISE
and NEOWISE: \citealp{Wright2010} ; \citealp{mainzer2011,mainzer2014}) and the second {\it Gaia} data release ({\it Gaia} DR2; \citealp{gaia2018}) using the coordinates given in SIMBAD/NED with a matching radius of 3\,arcsec. 
Of these sources, 123 and 117 had a unique entry in AllWISE and {\it Gaia} DR2, respectively (with 114 in common).  

\begin{figure*}
    \centering
    \begin{multicols}{2}
        \includegraphics[width=\linewidth]{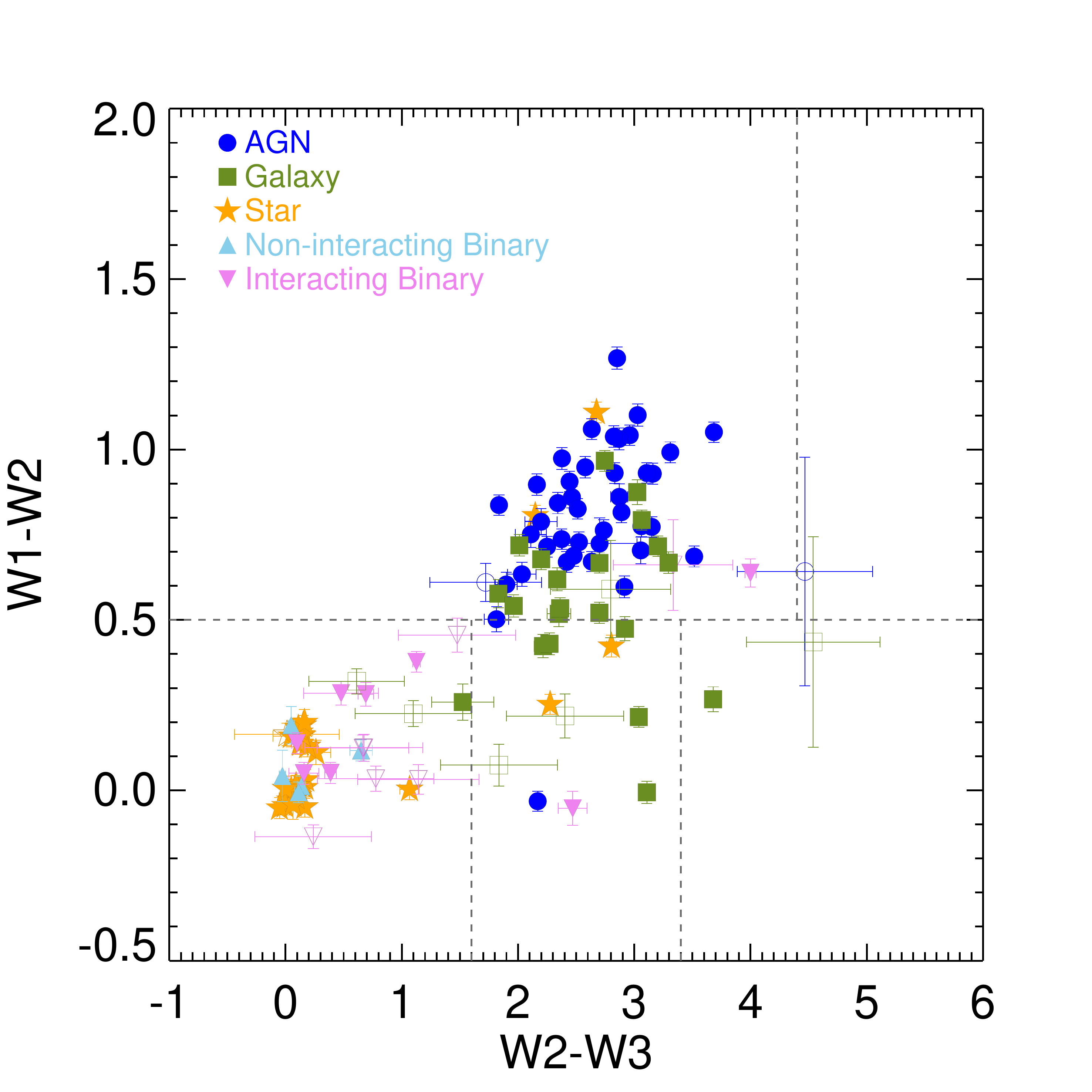}\par
        \includegraphics[width=\linewidth]{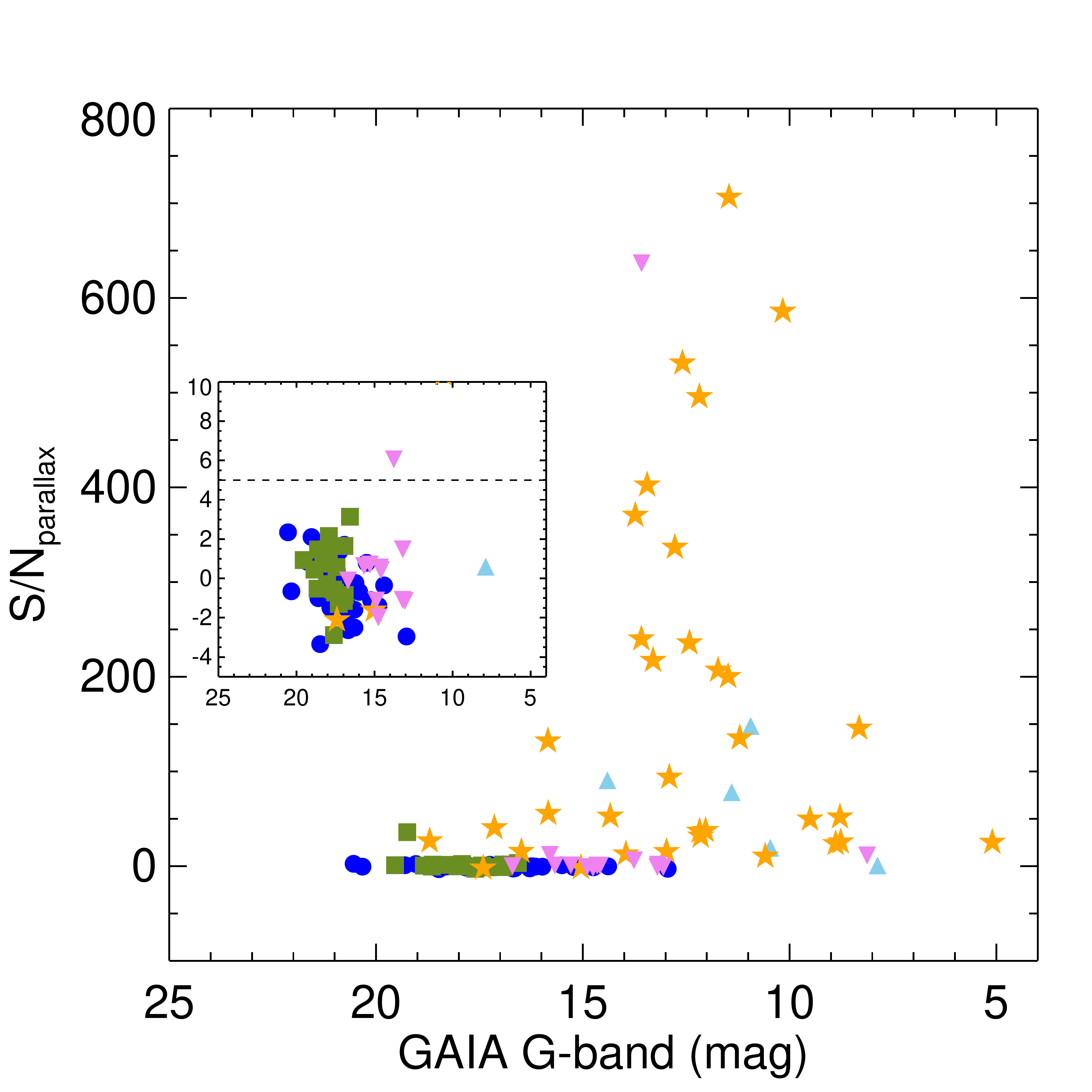}\par
    \end{multicols}
    \caption{ {\em left-hand panel:} \wise\ colours of the sources identified with SIMBAD and NED in Section~\ref{sec:xmm_pointed}. Dashed lines indicate boundaries proposed in \citet{mingo2016} (see Table~\ref{tab:wise_boundary}). The meaning of each symbol is shown in the top left-hand corner; sources that lack a reliable detection in W3 band (S/N<3) are represented with open symbols. {\em right-hand panel:} signal-to-noise ratio (S/N) measurements of the parallax vs. G-band magnitude of the sources identified with SIMBAD and NED. The symbols have the same meaning as in the {\em left-hand panel}. The inset plot shows the zoomed image for the
    sources with $\rm S/N_{parallax}$ in the range of -5 to 10, where dashed line indicates S/N$_{\rm parallax}$ = 5.}
    \label{fig:simbad_variables}
\end{figure*}

The left-hand panel of Figure~\ref{fig:simbad_variables} shows the locations of these sources in the \wise\ colour-colour diagram. The dashed lines mark the loci proposed by \citet{mingo2016} to classify sources based on their \wise\ colours (see also Table~\ref{tab:wise_boundary}). We can see that only one AGN lies below the line separating AGN and normal galaxies. In the region where W2-W3$<$1.6, we find most of the stars and three galaxies, consistent with the expectations of overlap between elliptical galaxies and stars for \wise\ shown in Figure 12 of \citet{Wright2010}.
Four of the stars have galaxy-like colours with W2-W3$>$1.6\footnote{Sources which are not located in the region expected for their category are, stars: J1924+5014 is a symbiotic star, J0534-6520 is a long period variable star, J2032+6712 and J0536-0017 are young stellar objects; AGN: J1115+1806 (NGC3599), the optical spectra indicated that it is a low-luminosity Seyfert/low-ionization nuclear emission-line region (LINER).}. There are also some galaxies located in the AGN region in the \wise\ colour-colour diagram; these sources are likely to be active but lack the optical spectra needed for confirmation.

 \begin{table}
	\centering
	\caption{Boundaries of the \wise\ colours used in the identification of variable sources, adopted from \citet{mingo2016}.} 
	\label{tab:wise_boundary}
	\begin{tabular}{ll} 
		\hline
		Label & {\it WISE} colour selection \\
		\hline
		Elliptical & $W1-W2 \le 0.5$; $0 < W2-W3 < 1.6$\\
		Spiral     & $W1-W2 < 0.5$; $1.6 \leq W2-W3 < 3.4$\\
		Starburst  & $W1-W2 < 0.5$; $W2-W3 \geq 3.4$\\
		AGN/QSO    & $W1-W2 \geq 0.5$; $W2-W3 \le 4.4$ \\
		\hline
	\end{tabular}
\end{table}

\begin{figure}
	\includegraphics[width=\columnwidth]{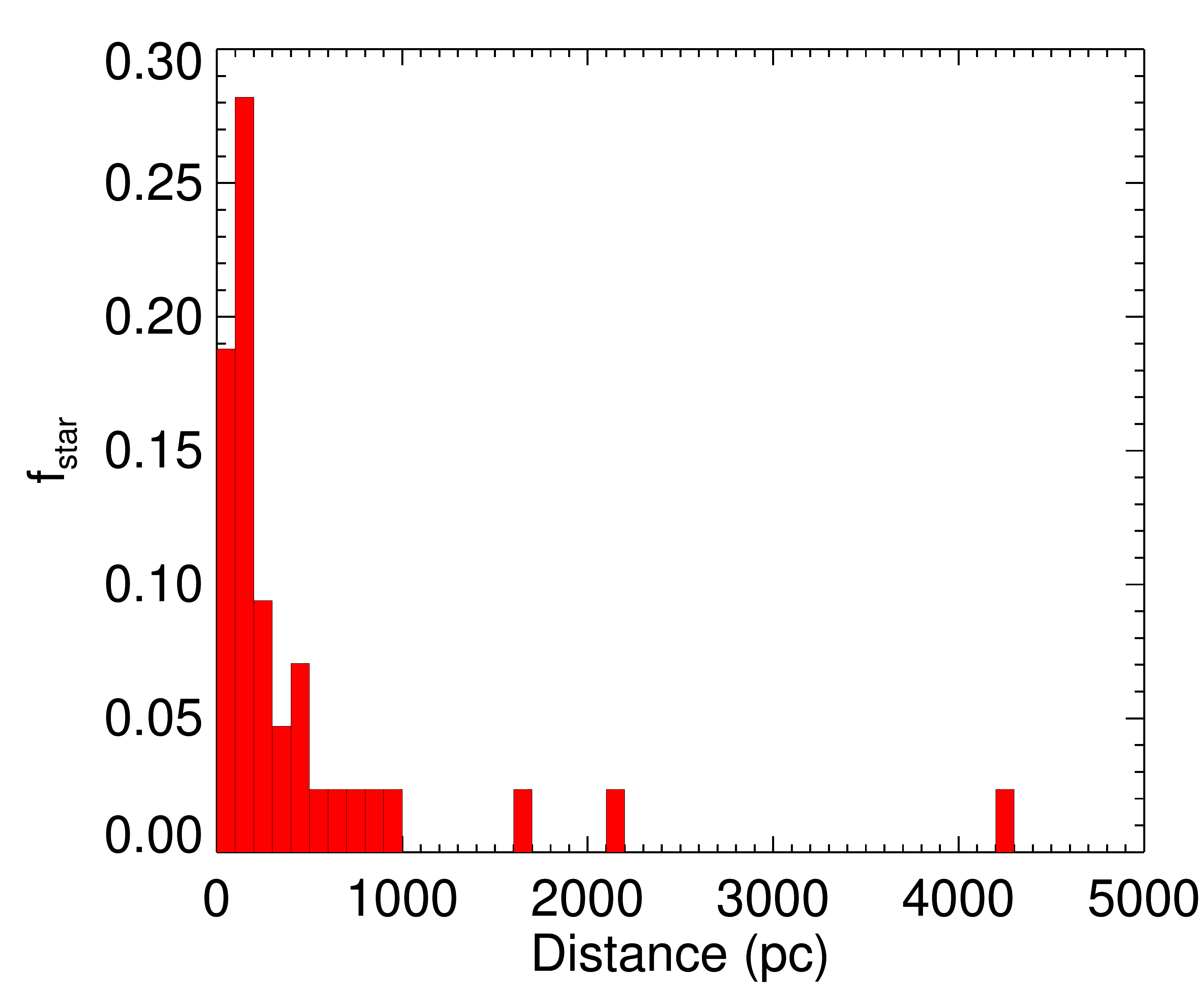}
    \caption{Distance distribution of the stars identified in SIMBAD and NED, as derived from {\it Gaia} parallax. The number of stars in each distance bin is normalized by the total number of stars.}
    \label{fig:gaia_distance}
\end{figure}

In the right panel of Figure~\ref{fig:simbad_variables}, we show the signal-to-noise ratio measurements of the parallax, $\rm S/N_{parallax}$, versus $G$-band magnitude from {\it Gaia} DR2 of the 117 XMMSL2 sources with a unique {\it Gaia} counterpart. The inset panel shows clearly that all of the extragalactic sources (galaxies and AGN) had $\rm S/N_{parallax} < 5$. There were also some stellar objects, like faint or distant stars and X-ray binaries, that had small $\rm S/N_{parallax}$. 
Figure~\ref{fig:gaia_distance} shows the distance distribution of the stars as derived from parallax given in {\it Gaia} GR2 and less than 10 per cent of stars having distances greater than 2000\,pc.
Based on the characteristics of the sources identified in SIMBAD or NED, the criteria used in the classification of the remaining sources using optical and IR properties are as follows: {\bf (a)} Sources with $\rm S/N_{parallax} \ge 5$ would be taken as Galactic stellar sources, {\bf (b)} sources with $\rm S/N_{parallax}<5$ and W1-W2$\ge $0.5 would be classified as AGN, and {\bf (c)} sources with $\rm S/N_{parallax}<5$, W2-W3$\ge$1.6 and W1-W2$<$0.5 would be (spiral or starburst) galaxies. The nature of sources with $\rm S/N_{parallax}<5$, W2-W3$<$1.6 and W1-W2$<$0.5 remains unclear in our identification work at this stage. If an X-ray source had one counterpart as a star but with distance larger than 2000\,pc, and another one as an extragalactic source, we would prefer the extragalactic source as the real counterpart. 

\subsection{Identification using multiwavelength properties}\label{sec:multiwave}
 
After correlating with SIMBAD and NED, 126 sources remained unidentified, of which 18 had {\it Swift/XMM} pointed observations. These sources were then classified based on their multiwavelength properties. 
First, the sources were cross-matched with AllWISE using XMMSL2 positions and a matching radius of 12\,arcsec, or 4XMM/2SXPS positions with 90 per cent position errors if available. 
Secondly, all the AllWISE counterparts were cross-matched with {\it Gaia} DR2 with a radius of 3\,arcsec. The identification process and number of counterparts in each step are given in Figure~\ref{fig:flowchart}. 

\begin{figure}
	\includegraphics[width=\columnwidth]{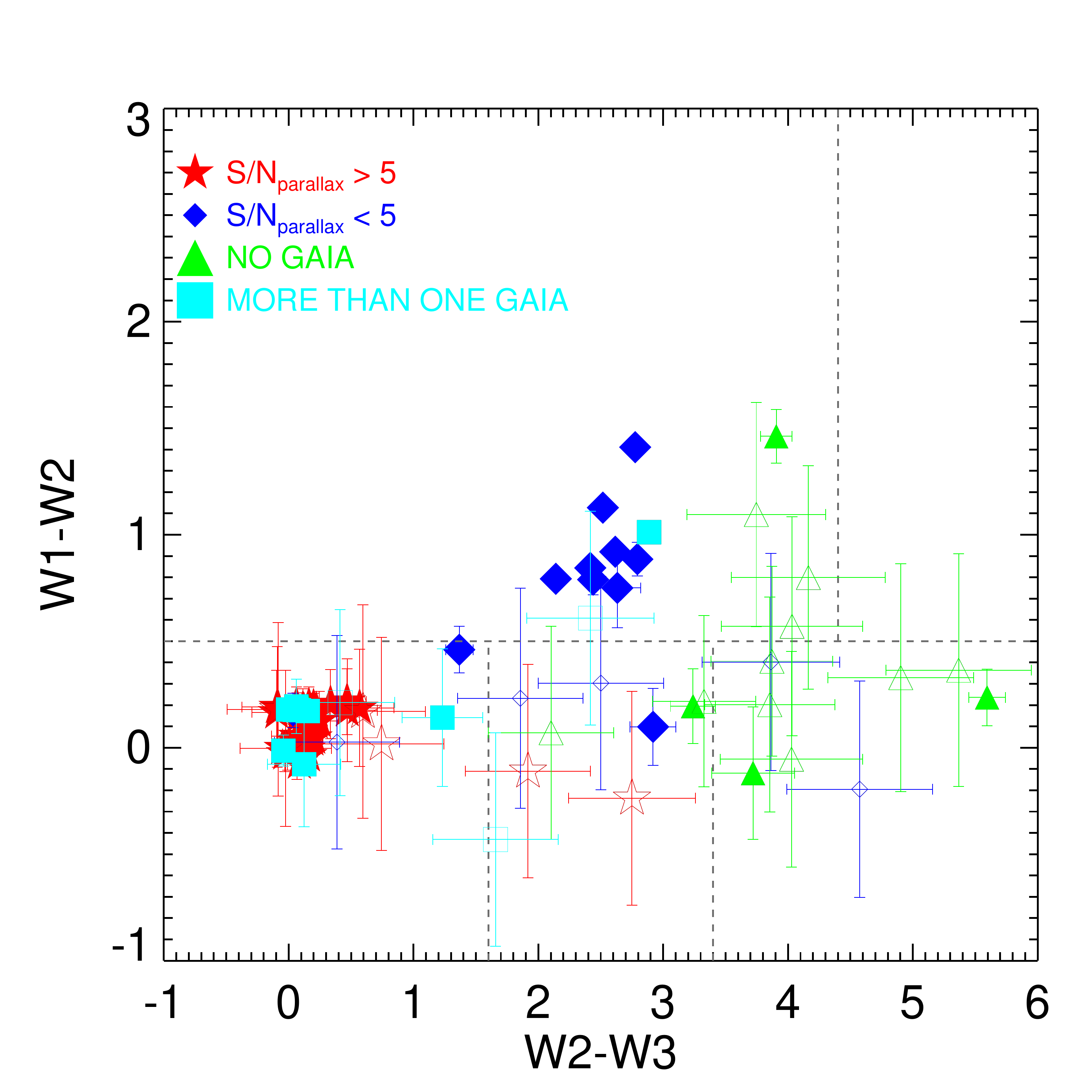}
    \caption{Locations of the sources with an AllWISE unique counterpart on the \wise\ colour-colour diagram. The dashed lines are the \wise\ colour boundaries used to give the classification as shown in Table \ref{tab:wise_boundary} adopted from \citet{mingo2016}. Sources that lack a reliable detection in W3 band (S/N <3) are represented by open symbols.}
    \label{fig:wise_one_gaia_one_combine}
\end{figure}

Classifications of each IR/optical counterpart were determined according to the criteria given in Section~\ref{sec:chara}. For sources that had AllWISE but no {\it Gaia} DR2 counterparts, their classifications were given merely based on the IR colours\footnote{We note that colour of galaxies span a relatively large range in the \wise\ colour-colour diagram, and there is overlap between Galactic sources and galaxies in Figure~\ref{fig:simbad_variables}. The identification of galaxies may not, therefore, be that robust compared to sources of other categories, and we will examine the results in detail in Section~\ref{sec:galaxy}.}. The classifications for all 81 sources with a unique AllWISE counterpart are 48 stars, 14 AGN and 19 galaxies\footnote{The relationship between {\wise} $W1$ magnitude and flux at 0.5--2\,keV from \citet{salvato2018} (Eq.~\ref{eq:x-w}), used to separate AGN and galaxies from stars, was also imposed as a cross-confirmation in the identification of galaxies.
 \begin{equation}
    [W1]=-1.625*{\rm logF_{(0.5-2\,keV)}}-8.8.
	\label{eq:x-w}
\end{equation}}. Locations of these sources on the \wise\ colour-colour diagram are shown in Figure~\ref{fig:wise_one_gaia_one_combine}. 

Table~\ref{tab:wise_mul} summarizes the properties of the AllWISE and {\it Gaia} counterparts for the 18 sources that had multiple AllWISE counterparts. The identifications were clear when all the counterparts were of the same type. For sources with one stellar and one extragalactic (AGN or galaxy) counterpart, the distance of the star would be taken into consideration.
As a consequence, these sources were identified as 2 stars, 7 galaxies, 2 AGN, and 7 sources were unidentified.

\begin{table*}
    \caption{List of sources with multiple AllWISE counterparts.}
    \begin{threeparttable}
    \begin{tabular}{ccccccc}
    \hline
    No. & XMMSL2 name    & AllWISE No.           & IR classification& S/N$_{\rm parallax}$ & Classification & Result \\
    \hline
       1.& XMMSL2 J030747.2+294208 & 1 & AGN     & <5                  & AGN     & Unclear \\
       & & 2 & Galaxy  & No                  & Galaxy  & \\
    \hline
       2.& XMMSL2 J044857.8-712734 & 1 & Galaxy  & Two<5               & Galaxy  & Galaxy \\
       & & 2 & Galaxy  & <5                  & Galaxy  & \\
    \hline
       3. & XMMSL2 J051627.4-650426 & 1 & Unclear & <5                  & Unclear & Galaxy$^a$ \\
       && 2 & Galaxy  & No                  & Galaxy &\\
    \hline
       4. & XMMSL2 J052953.5-655646 & 1 & Galaxy  & <5                  & Galaxy  & Galaxy \\
        & & 2 & Galaxy  & No                  & Galaxy  & \\
    \hline
       5.& XMMSL2 J064902.3-435452 & 1 & Star    & >5                  & Star & Unclear \\
       &   & 2 & Galaxy  & No                  & Galaxy &  \\
    \hline
       6. & XMMSL2 J065734.8-624934 & 1 & Galaxy  & No                  & Galaxy  & Galaxy \\
       &  & 2 & Galaxy  & No                  & Galaxy  &\\
    \hline
       7. & XMMSL2 J073252.9-610956 & 1 & Galaxy  & No                  & Galaxy  & Galaxy\\
       & & 2 & Galaxy  & No                  & Galaxy  &\\
    \hline
       8. & XMMSL2 J073808.3-583539 & 1 & Star    & One >5 and One NULL & Unclear & Unclear \\
       & & 2 & Unclear & >5                  & Star & \\
      & & 3 & Unclear & No                  & Unclear &\\
       & & 4 & Galaxy  & No                  & Galaxy &\\
    \hline
       9.& XMMSL2 J081713.9-651231 & 1 & Star    & >5                  & Star    & Unclear\\
       & & 2 & Galaxy  & No                  & Galaxy  &\\
    \hline
       10.& XMMSL2 J132438.1+611646 & 1 & AGN     & NO                  & AGN     & AGN\\
       & & 2 & AGN     & NO                  & AGN     &  \\
    \hline
       11. & XMMSL2 J161807.5+633132 & 1 & AGN  & No                     & AGN     & Unclear 
       \\
       & & 2 & Galaxy  & No                  & Galaxy  & \\
    \hline
       12.& XMMSL2 J162754.3-304651 & 1 & Star    & >5                  & Star& Star\\
       &  & 2 & Star    & >5                  & Star& \\
    \hline
       13. & XMMSL2 J165331.9+784514 & 1 & AGN     & No                  & AGN     & AGN\\
      & & 2 & AGN     & No                  & AGN     &\\
    \hline
       14. & XMMSL2 J170827.9+665244 & 1 & Galaxy  & <5                  & Galaxy  & Galaxy\\
      & & 2 & Galaxy  & No                  & Galaxy  & \\
    \hline
       15.& XMMSL2 J193437.2+490919 & 1 & Galaxy  & >5                  & Star& Galaxy$^b$\\
       & & 2 & Galaxy  & <5                  & Galaxy  & \\
    \hline
       16.& XMMSL2 J211111.7+215347 & 1 & Unclear & >5                  & Star& Star\\
       & & 2 & Unclear & >5                  & Star& \\
    \hline
       17. & XMMSL2 J234136.9-661053 & 1 & Unclear & <5                  & Unclear  & Unclear \\
       & & 2 & Star    & >5                  & Star & \\
    \hline
       18.& XMMSL2 J234412.1+653357 & 1 & Unclear & Two <5              & Unclear & Unclear\\
       & & 2 & Unclear & <5                  & Unclear \\
    \hline
    \end{tabular}
    {Column 2: unique XMMSL2 source name; column 3: numbering of AllWISE counterparts; column 4: \wise\ colour classification of each counterpart; column 5: $\rm S/N_{parallax}$ of the {\it Gaia} counterparts for each AllWISE counterpart; column 6: identification of each AllWISE counterpart based on the \wise\ colour and {\it Gaia} $\rm S/N_{parallax}$; column 7: identification of each XMMSL2 variable source.\\
    $^a$ The counterpart classified as `unclear' based on \wise\ infrared colour has negative parallax in {\it Gaia} DR2, and we preferred this source as galaxy.\\
    $^b$ Parallax of the counterpart classified as `star' was 0.49, corresponding to a distance greater than 2000\,pc, which was too large for a variable star in XMMSL2 according to Figure~\ref{fig:gaia_distance}. So we suggested the galaxy as the counterpart of this XMMSL2 source. \\}
    \end{threeparttable}
    \label{tab:wise_mul}
\end{table*}

After the above identification process, There were 8 (19) XMMSL2 sources with (without) {\XMM} or {\it Swift} pointed observations that did not have AllWISE counterparts within the matching radius. The statistical localization of the 8 sources was typically very small, and did not reflect systematic uncertainties, leading to the need to expand the cross-matching radius until a counterpart was found in SIMBAD. Table~\ref{tab:simbad_unknown} gives the details of these 8 sources.

\begin{table*}
\centering
\caption{List of sources with counterparts in 2SXPS or 4XMM-DR9 but did not have AllWISE counterparts within 90 per cent positional error circle.}
\begin{threeparttable}
    \begin{tabular}{cccccc}
    \hline
    No. & XMMSL2 name & Pos. err.(\,arcsec) & Sep.(\,arcsec)& Object name & Object type  \\
    \hline
    1   & XMMSL2\,J050521.3-684543 & 0.92 & 7.58 & 2MASS J05052138-6845332 & LP* \\
    \hline
    2   & XMMSL2\,J053041.6-665430 & 1.17 & 1.92 & 2MASS J05304215-6654303 & IrS$^a$ \\
    \hline
    3   & XMMSL2\,J053113.1-660706 & 0.87 & 2.07 & RX J0531.2-6609 & HMXB \\
    \hline
    4   & XMMSL2\,J054134.4-682549 & 0.49 & 0.86 & UCAC2 2073373 & HMXB \\
    \hline
    5   & XMMSL2\,J074007.3-853929 & 0.33 & 0.47 & 2MASX J07400785-8539307 & GiG \\
    \hline
    6   & XMMSL2\,J161243.5-522531 & 2.77 & 12.80 & V* QX Nor & LMXB \\
    \hline
    7   & XMMSL2\,J170144.0-405132 & 2.00 & 3.75 & 4U 1701-407 & LMXB \\
    \hline
    8   & XMMSL2\,J181921.7-252429 & 1.70 & 3.89 & V* V4641 Sgr & HMXB \\
    \hline
    \end{tabular}
    Column 3: 90 per cent confidence level in the position error, in unit of arcsecond; column 4: separation between the source position given in 2SXPS or 4XMM-DR9 and
    the position of the counterpart in SIMBAD, in unit of arcsecond; column 5: object names of the counterparts; column 6: Object type of the source given in SIMBAD.  LP*: Long-period variable star. IrS: Infrared source. HMXB: High mass X-ray binary. GIC: Galaxy in groups. LMXB: Low mass X-ray binary.\\
    $^a$ This source is a star based on {\it Gaia} parallax measurement.
    \end{threeparttable}
    \label{tab:simbad_unknown}
\end{table*}

\begin{figure}
	\includegraphics[width=\columnwidth]{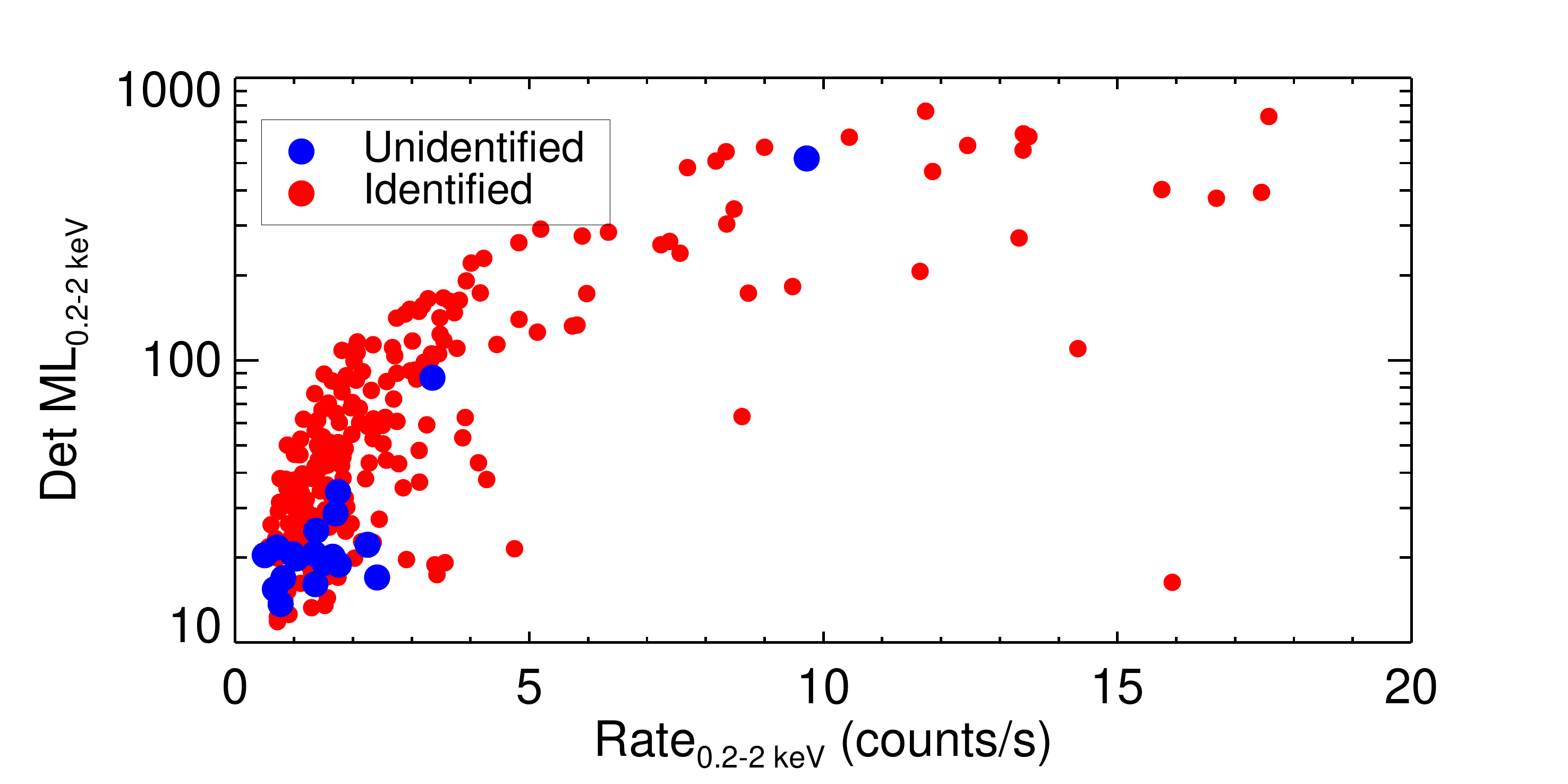}
    \caption{Distribution of soft X-ray count rates and detection likelihoods of the identified sources (red circles) and unidentified ones (blue circles) after identification via the procedure described in Section \ref{sec:xmm_pointed}.}
    \label{fig:rate_det}
\end{figure}

Figure~\ref{fig:rate_det} compares the soft band count rate and detection likelihood distribution of the 19 unidentified sources with that of the identified sources. It shows clearly that the soft band detection likelihoods of these 19 sources were generally lower than those of the identified ones. According to \citet{saxton2008}, although the 1$\sigma$ position uncertainty of the sources was 8\,arcsec, there was a long tail in the histogram of the distribution of the angular separation of the slew sources from their SIMBAD counterparts. To cater for this possibility, we repeated the identification procedure for the 19 sources, namely cross-matching with different data bases and catalogues, with a larger matching radius of 20\,arcsec. This led to 2 sources, XMMSL2\,J070307.8-703233 and J154212.6+253352, being classified as stars based on their SIMBAD counterparts and 3 as AGN and 6 as galaxies from their IR colour and parallax measurements. Four of the remaining sources had counterparts but with an
unclear classification and 4 sources did not have any AllWISE counterparts within the matching radius.

\section{results}\label{sec:results}

Of the sample of 265 highly variable sources, 94.3 per cent have been identified; 131 based on their SIMBAD classification, 18 from NED, and 101 sources from their multiwavelength properties.
We have compared our identifications with the work of \citet{salvato2018}, who used the Bayesian cross-matching algorithm NWAY to identify AllWISE counterparts to XMMMSL2 sources with |b|$>$15$^\circ$. 
There were 76 sources in common between the \citet{salvato2018} sample and those which we have identified based on their IR colours. Most of these sources have
the same assigned counterpart in the two surveys, with 5 (XMMSL2\,J181345.8+652137, XMMSL2\,J181046.9+700440, XMMSL2\,J175743.5+561159, XMMSL2\,J070728.4-565912, and XMMSL2\,J060915.4-550443) having a different association. For each of these 5, the \wise\ counterpart given in \citet{salvato2018} was the second nearest one, was located more than 12\,arcsec\ (the cross-matching radius used in our identification) away from the XMMSL2 position, and was much brighter than the nearest AllWISE counterpart. For a further 6 sources in our sample that had an AllWISE counterpart but whose nature remained unclear, \citet{salvato2018} find multiple possible counterparts and assign a probability for each. 

The original XMMSL2 catalogue contains identifications, based on cross-matching with SIMBAD and NED, for 117 of the objects in our sample. Of these, 4 sources have a different identification to this work (Table~\ref{tab:diff_id}). The differences may be due to subsequent updates in the SIMBAD classifications or to the fact that SIMBAD sometimes provides multiple types for the same object. For example, the main type of XMMSL2\,J113107.3-625651 is HMXB in SIMBAD now, but it is also a Be star according to other object types given in SIMBAD. Using this information, we were able to update our identification for XMMSL2\,J175328.3-012708 from binary star to LMXB (see Table~\ref{tab:diff_id} for details). 

\begin{table*}
\caption{Sources with different identifications in the XMMSL2 catalogue and this work.}
    \begin{threeparttable}
    \centering 
    \begin{tabular}{ccccc}
    \hline
    No. & XMMSL2 name & XMMSL2 ID & This work & ID\\
    \hline
    1   & XMMSL2\,J053527.7-691611 & BlueSG* &  SN* & SN*\\
    \hline
    2   & XMMSL2\,J113107.3-625651 & Be*     & HMXB & HMXB \\
    \hline
    3   & XMMSL2\,J175328.3-012708 & LMXB    & **   & LMXB \\
    \hline
    4   & XMMSL2\,J195402.8+104145 & WV*     & Ro* & Ro*\\
    \hline
 \end{tabular}
Column 2: Unique source name in the XMMSL2 catalog; column 3: Identification in the initial XMMSL2 catalog,  BlueSG*: blue supergiant star. Be*: Be star. WV*: Variable star of W Vir type; column 4: identification in SIMBAD or
    NED obtained from the identification process in this work. SN*: supernova. **: double or multiple star. Ro*: Rotationally variable star; column 5: Identification adopted. Reasons for the difference for each source is as follows.\\
    1. The blue super giant is the nearest one to the XMMSL2 position while cross-matching with SIMBAD. By checking the X-ray light curves, we found that the X-ray emission was from the supernova 1987A. \\
    2. The main type of the source is HXMB in SIMBAD.\\
    3. J1753-0127 was in the 4XMM catalogue and the 90 per cent position uncertainty was 1.95\,arcsec. The separation between the LMXB and the 4XMM position is 1.99\,arcsec\ as given in SIMBAD. While in NED, there was a binary located 1.6\,arcsec\ away from the 4XMM position. We concluded that the binary in NED and the LMXB in SIMBAD are the same source.\\
    4. The main type was rotationally variable star in SIMBAD when we did the cross-match.
    \end{threeparttable}
    \label{tab:diff_id}
\end{table*}

The full sample of variable sources have been broadly split into categories of normal galaxies, AGN, and stellar sources. In the following sections we study the properties of each category.

\subsection{Galaxies}
\label{sec:galaxy}

Sixty five of the variable sources were classified as galaxies after our identification work. We further subdivide these into three types according to their soft X-ray light curves, using data from XMMSL2, {\Swift} and {\xmm} pointed observations, as follows:

\begin{description}
\item[{\it 1. Flaring source.}] The flux ratio between the highest XMMSL2 flux and the lowest subsequent flux (or upper limit) was greater than 10. Fifteen galaxies were classified into this type. 

\item[{\it  2. Non-flaring source.}] Defined by a light curve which has more than one detection after the RASS measurement and where the ratio between the highest and lowest post-RASS fluxes was smaller than 10 over a time range longer than 1 year. Sixteen sources had this classification.

\item[{\it 3. Unknown.}] The remaining 34 galaxies did not fit into the first two definitions. Either they had only one XMMSL2 detection, and the flux ratio between the detection and following upper limits was smaller than 10, or despite having more than one detection, the flux ratio between the highest and lowest detection was smaller than 10 over a period of less than 1 year. 

\end{description}

\begin{figure}
	\includegraphics[width=\columnwidth]{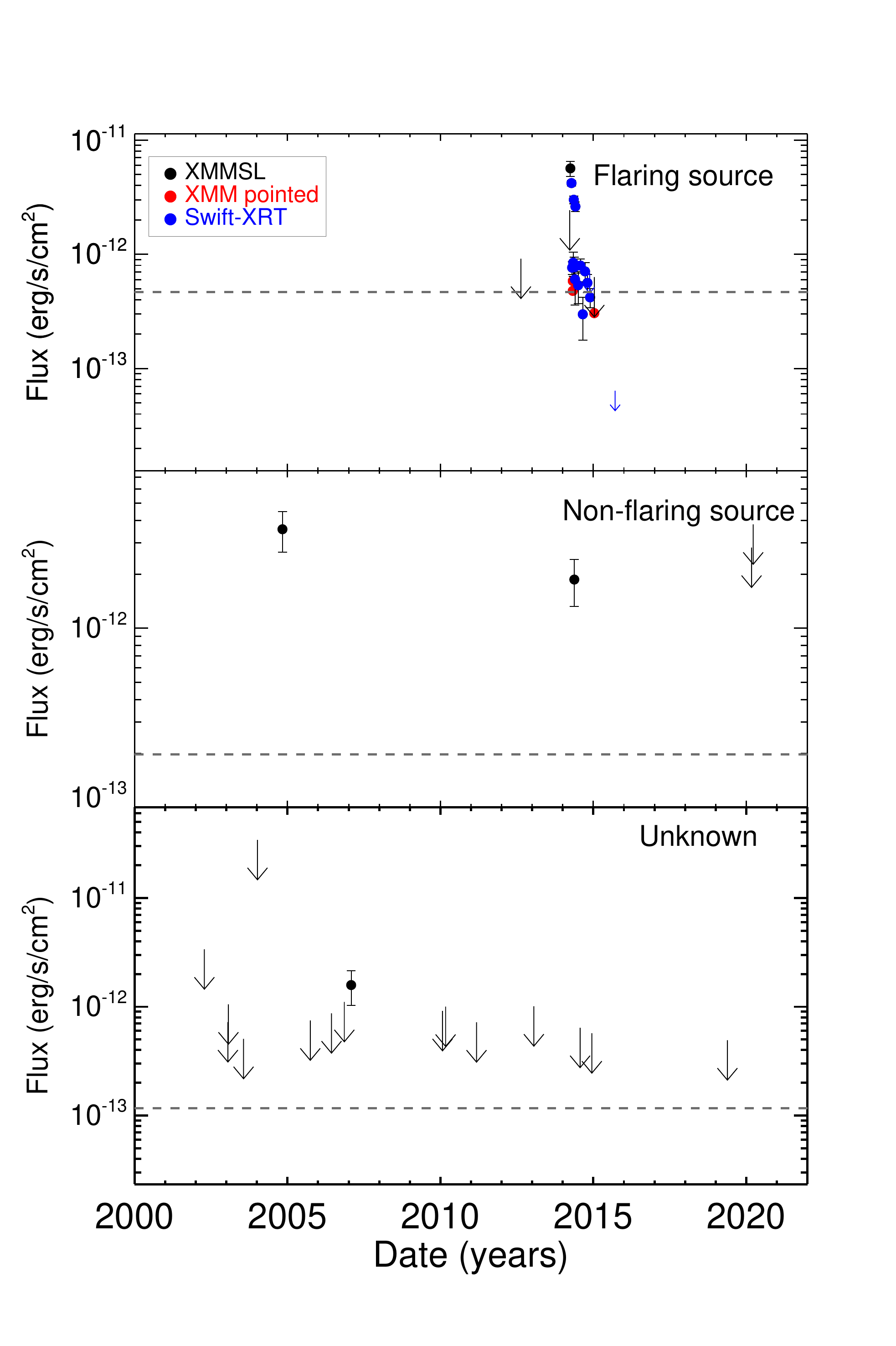}
    \caption{Example light curves for each type of galaxy (described in Section \ref{sec:galaxy}). The meaning of symbols are described in the top left-hand corner. The dashed grey line in each panel shows the flux (or upper limit) of each source in the RASS.}
    \label{fig:lc_galaxy_sample}
\end{figure}

Figure~\ref{fig:lc_galaxy_sample} shows an example light curve for each type. For the galaxy candidates classified as {\it flaring source}, 
4 have already been suggested as TDEs\footnote{We note that the XMMSL2/RASS ratio was 9.97 for the TDE XMMSL2\,J131952.3+225958 found in \citet{li2020}, which thus was not included in this sample.}: XMMSL2\,J132342.0+482700 \citep{esquej2008}, XMMSL2\,J120135.4+300305 \citep{saxton2012}, XMMSL2\,J074007.3-853929 \citep{saxton2017}, and XMMSL2\,J061927.5-655310 \citep{saxton2014}\footnote{In this paper, they did not give a conclusive statement, and the flare in this galaxy has either been caused by a tidal disruption event or by an increase in the accretion rate of a persistent active galactic nucleus.}.
For a further two, XMMSL2\,J045106.4-694800 and XMMSL2\,J162638.0-515635, a closer inspection of the sources shows their counterparts to be X-ray binaries instead of galaxies. XMMSL2\,J045106.4-694800 was included in the 4XMM-DR9 with a position error of 0.61\,arcsec, and had a HMXB located 1.5\,arcsec\ away. For XMMSL2\,J162638.0-515635, there was a HMXB located 15.6\,arcsec\ away from the XMMSL2 position. This explains why these two sources were missed out in the cross-correlation with SIMBAD and NED. More details of these two HMXB are listed in Table~\ref{tab:xb} 
in Section~\ref{sec:xb}. 

Some of the 16 galaxies classified as {\it non-flaring source} may be previously unknown AGN and actually 9 of them are located in the AGN region in the \wise\ colour-colour diagram. 

For the remaining 43 galaxies (9 {\it flaring} sources and 34 {\it unknown}), 6 are located in the AGN region of the \wise\ colour-colour diagram.  More information is needed to reveal their variability mechanism. 
One of the {\it flaring source} galaxy candidates, XMMSL2\,J045314.4-691132, is located in the outskirts of the Large Magellanic Cloud (LMC), and the X-ray variability here could be from a variable source in the LMC or from a background galaxy.

\subsection{AGN}

Sixty sources were identified as AGN in our XMMSL2 variable sample, several of which have been extensively monitored, giving us greater insight into their variability mechanisms. 
These include a changing-look AGN (XMMSL2\,J113851.0-232135, known as HE 1136-2304) whose large amplitude variability was most likely due to an increase in the accretion rate \citep{parker2016} and XMMSL2\,J143622.6+584740 (Mrk 817), a well-monitored Sy1.5 galaxy whose X-ray variability strongly correlates with X-ray power law index, while no correlation is found with simultaneous optical/UV photometry \citep{morales2019}. XMMSL2\,J111527.3+180636, also known as NGC 3599, 
was one of the first candidate TDEs found in XMMSL \citep{esquej2008} although 
it is possible that the variability was actually due to a disc instability \citep{saxton2015}.
The sharp X-ray variability in XMMSL2\,J051935.7-323932 (ESO 362-18) was caused by the supernova SN2010JR \citep{immler2010}. XMMSL2\,J164924.9+523515 and XMMSL2\,J200112.7+435248 \citep{bassani2009} are BL Lac objects, in which rapid X-ray variability could be linked to jet activity. 
We found two previously unknown AGN, XMMSL2\,J073649.5-692547 and XMMSL2\,J154214.0+671656, fortuitously located in the field of view of the {\Swift} monitoring of other objects, which showed strong variability on a timescale of several months. The soft X-ray flux of J0736-6925 dropped by a factor of 20 within one year, while J1542+6716 showed variability throughout the wave bands from infrared to X-ray. 

We have compared our sample of variable AGN with the remaining AGN in XMMSL2 that displayed far less long-term variability.
We selected 1832 sources identified as AGN, from the initial 10463  XMMSL2 clean sources that meet the criteria specified in Section~\ref{sec:selection}, and with an XMMSL2/RASS flux ratio in the range of 0.3--3. These sources were defined as the `low-variability AGN' sample. 
Both samples were cross-matched with the Veron catalogue \citep{veron2010} with a matching radius of 5\,arcsec\ using the coordinates of the identifiers. The low-variability AGN sample contained 1324 objects with a counterpart in the Veron catalogue (1286 with redshift measurements), of which 1148 sources had a unique counterpart in the Two-Micron All Sky Survey (2MASS; \citealp{skrutskie2006}) Point Source Catalogue within 5\,arcsec. The variable sample has 23 sources in the Veron AGN catalogue, all with redshift measurements and 22 with 2MASS counterparts.

\begin{figure}
	\includegraphics[width=\columnwidth]{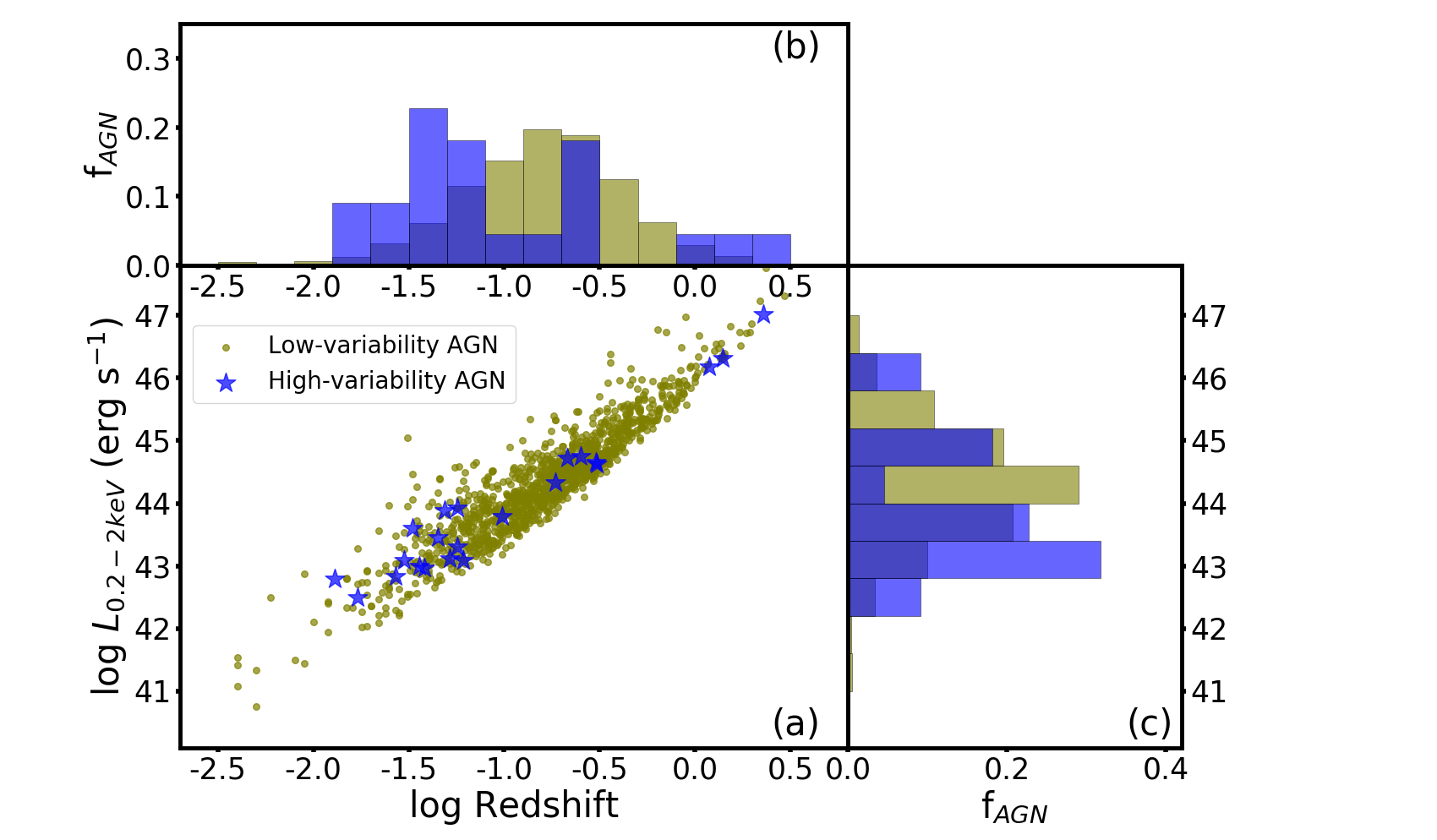}
    \caption{Panel (a): redshift and unabsorbed soft X-ray luminosity distribution of the high- (blue stars) and low-variability (green circles) AGN. Panel (b): histogram of the redshift of high- (blue bars) and low-variability (green bars) AGN. Panel (c): luminosity histogram for the high- and low-variability AGN samples.  The histograms in panels (b) and (c) have been normalized by the number of sources in each sample.}
    \label{fig:agn_z_lumi_bh}
\end{figure}

\begin{figure}
	\includegraphics[width=\columnwidth]{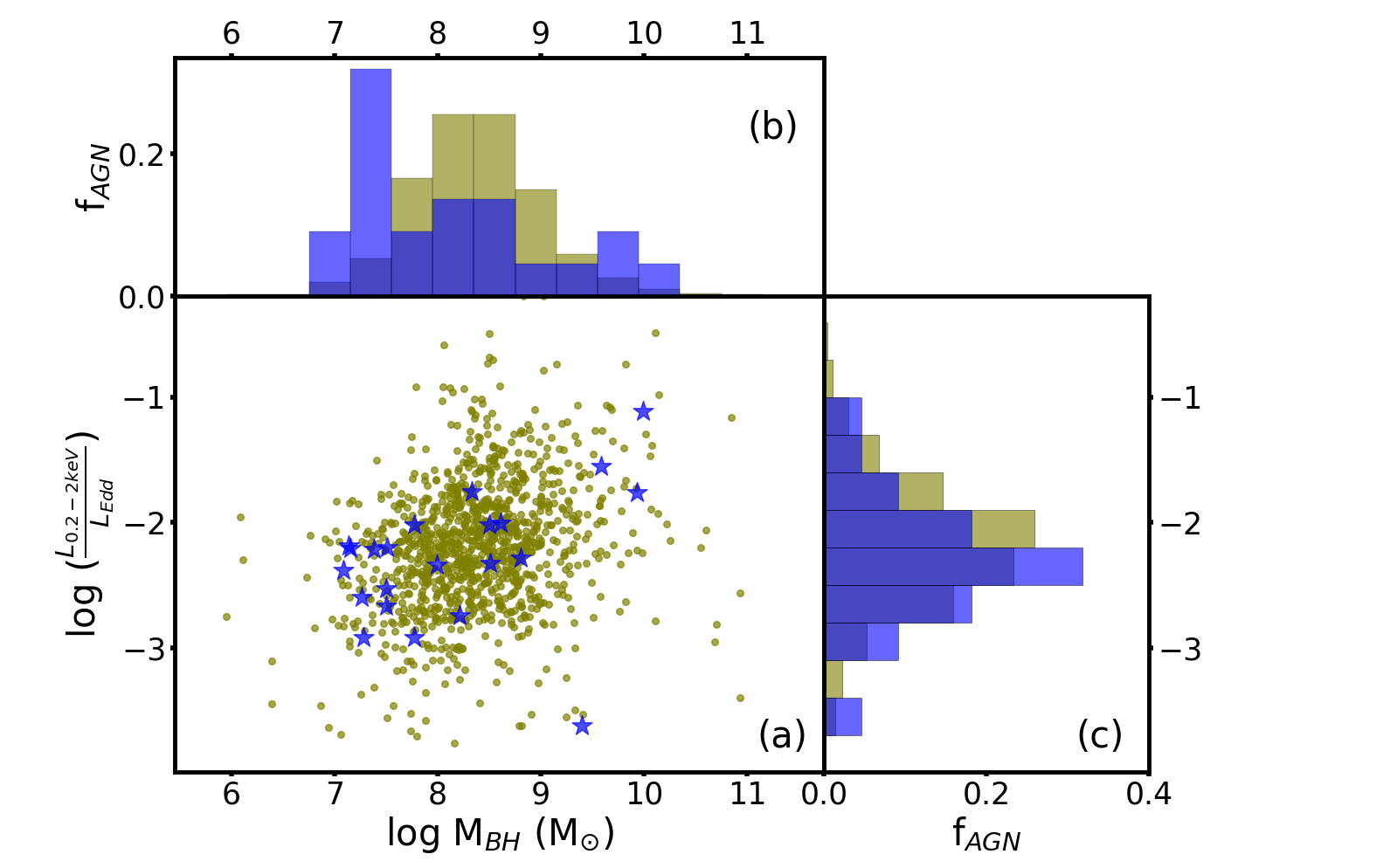}
    \caption{Panel (a): Eddington ratio and black hole mass distribution of the high- (blue stars) and low-variability (green circles) AGN. Panel (b): histogram of black hole mass for high- (blue bars) and low-variability (green bars) AGN. Panel (c): Eddington ratio distribution for the high- and low-variability AGN samples.  The histograms in panel (b) and (c) have been normalized by the number of sources in each sample.}
    \label{fig:agn_edd_bh}
\end{figure}

Figure~\ref{fig:agn_z_lumi_bh} compares the redshift and unabsorbed soft X-ray luminosity (de-absorbed for Galactic absorption only) distributions of the high- and low-variability AGN samples. Here, luminosity refers to the average of the XMMSL2 and RASS values. We can see that the highly variable AGN have relatively lower values for both quantities. We have examined whether the high- and low-variability sources were drawn randomly from the same sample using the Kolmogorov-Smirnov test (KS test). For the redshift, this hypothesis has a probability lower than 3$\times$10$^{-4}$, while for the soft X-ray luminosity, it is 0.02. We also examined the black hole masses of these AGN using the empirical relationship between bulge K$_{s}$ band luminosity and black hole mass described in \citet{marconi2003}. This method has a scatter of 0.3 dex. For all of the AGN, the K$_s$ band luminosities have been corrected to the bulge luminosities by adding 0.8 mag, a correction that corresponds to a lenticular host galaxy which lies in between the correction for an elliptical and a spiral galaxy. The K-correction, for the redshift of the K$_s$ band luminosity, has been performed using the SED of an elliptical galaxy with stellar population of 2\,Gyr. We note that, although the absolute value may not be accurate for any individual source, the relative ensemble values can be compared with confidence. Figure~\ref{fig:agn_edd_bh} shows the distribution of Eddington ratios and black hole masses for both AGN samples, where the Eddington ratio is defined as $L_{0.2-2 \rm keV}$/$L_{\rm Edd}$ and $L_{0.2-2 \rm keV}$ is an average of the XMMSL2 and RASS values. The sample of highly variable AGN tend to have lower black hole masses, with a KS test probability that the high- and low-variability AGN were drawn randomly from the same sample of 0.008. We find no significant difference in the Eddington ratio distributions, with a KS test probability of 0.4. We thus conclude that highly variable AGN tend to have lower black hole masses, redshifts, and luminosities but note that larger samples will be necessary to confirm this result. 

\subsection{Stellar sources}

There were 127 stellar objects (101 stars and 26\footnote{Two of them, XMMSL2\,J045106.4-694800 and XMMSL2\,J162638.0-515635, were identified as HMXB after examination of the `galaxies' as described in Section~\ref{sec:galaxy}.} accreting binaries) in the variable sample. Here we give some basic information and further analysis for each sub-category.

\subsubsection{Stars}

The sample of 101 stars contains 6\footnote{2 spectroscopic binaries, 3 eclipsing binaries and 1 RSCVn.} known non-accreting binaries and 2 (XMMSL2\,J053628.7-001721 and XMMSL2\,J203232.6+671219) candidate young stellar objects. A further one, XMMSL2\,J203232.6+671219, is located in a star cluster with the X-ray emission from a LMXB. The X-rays seen from XMMSL2\,J053527.7-691611 were in fact found to be from the nearby supernova remnant 1987A. These 10 sources were excluded in the following statistical analysis of the properties of single stars.

\begin{figure}
	\includegraphics[width=\columnwidth]{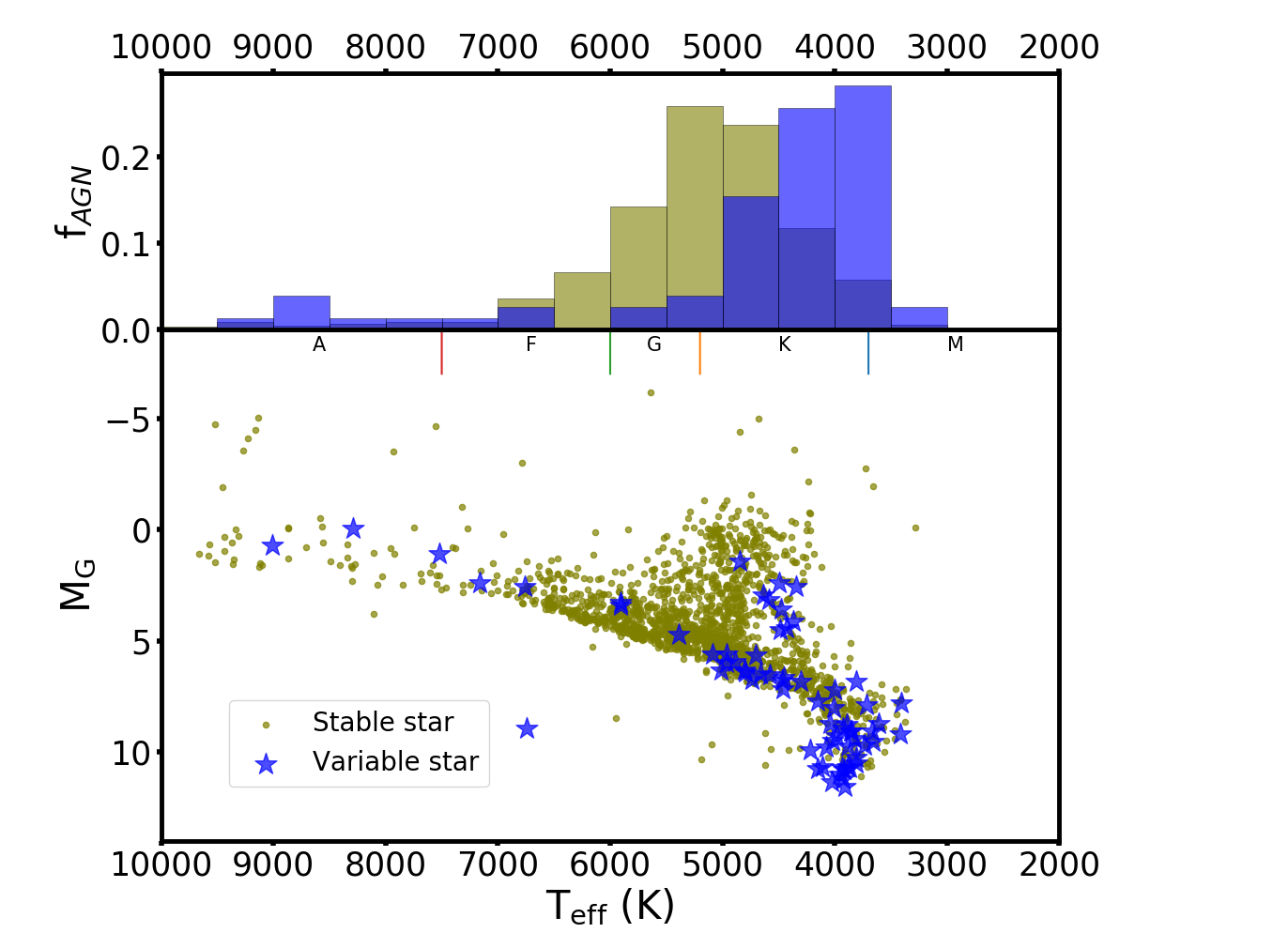}
    \caption{{\em Lower panel:} Hertzsprung-Russell diagram of the variable (blue stars) and stable (green circles) stars with known parallax and effective temperature in {\it Gaia} DR2, with the x-axis being the effective temperature given in {\it Gaia} DR2 and the y-axis being the G band absolute magnitude.
    {\em Upper panel:} effective temperature distribution of the variable (blue bars) and stable (green bars) stars.}
   \label{fig:teff_star}
\end{figure}

Among the 10463 XMMSL2 sources that meet our selection criteria given in Section~\ref{sec:selection}, 2181 sources were identified as single stars with XMMSL2/RASS flux ratios in the range of 0.3-3, and are hereafter referred to as the stable star sample. Both variable and stable stars were cross-matched with {\it Gaia} DR2 with a matching radius of 3\,arcsec. Temperatures, parallax, and absolute magnitudes derived using a conversion of parallax into distance were found for 70 variable and 1763 stable stars.
A comparison between the properties of the samples (Figure~\ref{fig:teff_star}) shows that the variable star population consists of stars that are relatively cooler than those of the stable sample; a characteristic seen in previous studies \citep{mcquillan2014}. We use the Morgan-Keenan system \citep{morgan1943} to extract the spectral type from the effective temperature and find that $>80$ per cent of the variable stars, with {\it Gaia} parallax and effective temperature measurements, are late-type K and M dwarfs. 

\subsubsection{Accreting binaries}\label{sec:xb}

\begin{figure}
	\includegraphics[width=\columnwidth]{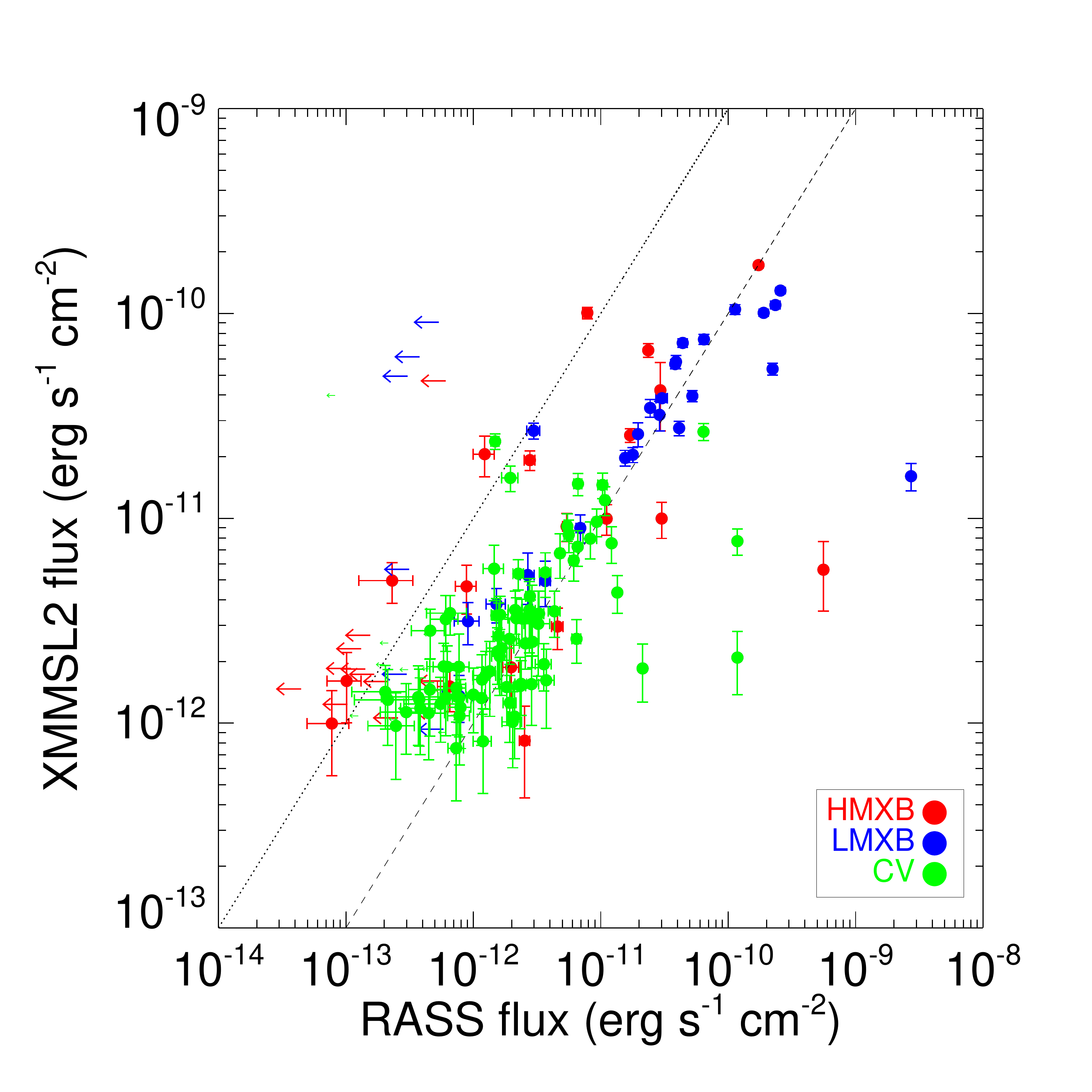}
    \caption{RASS vs. XMMSL2 fluxes of accreting binaries in the full XMMSL2 clean catalogue. The black dotted line indicates a flux difference of a factor of 10, from above which our variable sample is drawn. The black dashed line indicates the equality in flux for reference.}
   \label{fig:interacting_binaries}
\end{figure}

Accreting binaries comprised 27 of our identified sample (including XMMSL2\,J0451-6948, J1626-5156, and J1736-4444), in which X-ray emission is likely attributable to an accretion flow or strong shocks. Our variable sample contains 6 cataclysmic variables (CVs), 5 low-mass X-ray binaries (LMXBs) and 16 high-mass X-ray binaries (HMXBs). In Figure~\ref{fig:interacting_binaries} we show the RASS versus XMMSL2 fluxes of accreting binaries from the XMMSL2 catalogue (after application of the criteria listed in Section~\ref{sec:selection}). It is clear from the dividing line, delineating the objects in our sample which varied by a factor of more than 10, that the majority of LMXBs were more or less stable between these two epochs while a high fraction of HMXBs show strong variability. Due to the survey cadence and exposure time of XMMSL, we tend to preferentially select short-term transients with a greater duty cycle. In Table~\ref{tab:xb} we show that most of the HMXB selected in this work are in Be X-ray binary systems, which are known to show frequent outbursts \citep[e.g.,][]{reig2011}. Low mass X-ray binaries have a duty cycle of 0.01-0.1, with an average of 0.03 \citep{yan2015} and are hence under-represented. 
CVs represent the largest class of accreting binaries within XMMSL2, mirroring the result of RXTE studies \citep{sazonov2005a}. Only 4 per cent of the XMMSL2 CVs show flux variability in excess of a factor of 10, which may indicate that those few were caught in an outburst state. More information about each of the CVs and X-ray binaries is given in Tables~\ref{tab:cv} and ~\ref{tab:xb}, respectively.

\begin{table*}
\caption{List of sources identified as CVs.}
\begin{threeparttable}
    \centering
    \begin{tabular}{ccccc}
    \hline
    No. & XMMSL2 name & Identifier$^a$ & Type$^b$ & Reference \\
    \hline
    1   & XMMSL2\,J012838.3+184537  & CRTS\,J012838.3+184536  & CV? & \citealp{drake2014} \\
    2   & XMMSL2\,J060636.5-694932  & XMMSL1\,J060636.2-694933 & Nova & \citealp{read2009} \\
    3  & XMMSL2\,J063045.7-603110  & XMMSL1\,J063045.9-603110 & Nova & \citealp{oliveira2017}\\
    4  & XMMSL2\,J070542.7-381442  & XMMSL1\,J070542.7-381442 & Nova & \citealp{read2008} \\
    5  & XMMSL2\,J172700.3+181422  & ASASSN-13ag    & CV   & \citealp{shappee2013} \\
    6  & XMMSL2\,J181613.0+495207  & V* AM Her & AMher& \citealp{hiroi2013} \\
    \hline
    \end{tabular}
    $^a$ Name of the identifier adopted in SIMBAD. \\
    $^b$ Subtype of the CVs. Nova, where the eruption is caused by explosive burning in degenerate matter at the bottom of an accreted envelope on a WD. AMher means CV of AM Her type, a class of polar.
    \end{threeparttable}
    \label{tab:cv}
\end{table*}

\begin{table*}
\caption{List of sources identified as X-ray binaries.}
    \begin{threeparttable}
    \centering
    \begin{tabular}{ccccccc}
    \hline
    No. & XMMSL2 name & Identifier$^a$ & Category$^b$& Spec.$^c$ & Type$^d$ & Reference\\
    \hline
    1 & XMMSL2\,J011705.4-732644 & SMC X-1 & HMXB &  B0Ie & T,P & 1-3 \\
    2 & XMMSL2\,J045106.4-694800 & Swift J045106.8$-$694803  & HMXB & B0-1III-V & P & 4-5\\
    3 & XMMSL2\,J050045.8-704436 &  IGR J0500-7047 & HMXB & B2IIIe  & T,P & 6-8 \\
    4 &XMMSL2\,J051328.1-654721 &  Swift J0513.4-6547 & HMXB & B1Ve & P & 9-10\\
    5 &XMMSL2\,J052948.3-655643 & RX J0529.7-6556 & HMXB & B2e & T,P & 11 \\
    6 &XMMSL2\,J053011.9-655125 & RX J0530.1-6551 & HXMB &  & P & 12\\
    7 &XMMSL2\,J053113.1-660706 & EXO 053109-6609.2 & HMXB & B0.7Ve? & T,P & 13-14 \\
    8 & XMMSL2\,J053231.8-655132 & RX J0532.5-6551   & HMXB & B0II&  & 15-16 \\
    9 & XMMSL2\,J054134.4-682549 & XMMU J054134.7$-$682550 & HMXB &   & P & 17 \\
    10 & XMMSL2\,J054405.2-710052 & RX J0544.1-7100 & HMXB & B0Ve & T,P & 18-19\\
    11 &XMMSL2\,J113107.3-625651 & IGR J11305-6256 & HMXB & B0IIIe & T,P? & 20-121 \\
    12 & XMMSL2\,J114359.5-610738 & IGR J11435-6109  & HMXB & B3e & T,P & 22-23\\
    13 & XMMSL2\,J154223.4-522306 & 4U 1538-52       & HMXB & B0Iab & P & 24-26\\
    14 & XMMSL2\,J161243.5-522531 & 4U1608-52        & LMXB &        & T & 27-28\\
    15 & XMMSL2\,J162638.0-515635 & Swift J1626.6-5156& HMXB & B0Ve   & T, P & 29-31\\
    16 & XMMSL2\,J170144.0-405132 & XTE~1701-407     & LMXB &        & T & 32\\
    17 & XMMSL2\,J173616.9-444400 & IGRJ17361-4441 & LMXB &  & T & 33\\
    18 & XMMSL2\,J175328.3$-$012708 & Swift J1753.5$-$0127 & LMXB$^{BH}$ & & T & 34 \\
    19 & XMMSL2\,J181921.7$-$252429 & SAX J1819.3$-$2525 & HMXB$^{BH}$ & B9III & & 35-36 \\
    20 & XMMSL2\,J190008.6-245510 & HETE J1900.1-2455 & LMXB & & T,P & 37\\
    21 & XMMSL2\,J210336.0+454506 & SAX J2103.5+4545  & HMXB & B0Ve & T,P & 38-39\\
    \hline
    \end{tabular}
    $^a$ X-ray name of the identifier, with rough information on its sky location. The prefix `J' indicates that the epoch of the coordinates is J2000, otherwise the 1950 coordinates are used in the name. The first part of name is a string identifying the satellite by which the source was listed or detected (except for SMC X-1). \\
    $^b$ Category of the X-ray source. LMXB for low mass X-ray binary and HMXB for high mass X-ray binary. BH in the superscript means the compact object in the binary is a black hole, otherwise it is a neutron star.\\
    $^c$ The spectral type of the optical counterpart. \\
    $^d$ The X-ray source type. T: transient X-ray source. P: X-ray pulsar.\\
    {\bf Reference:} 1. \citealp{schreier1972}, 2. \citealp{webster1972}, 3. \citealp{lucke1976}, 4. \citealp{Beardmore2009}, 5. \citealp{Massey2002},
    6. \citealp{sazonov2005b}, 7. \citealp{masetti2006}, 8. \citealp{vasilopoulos2016}, 9. \citealp{krimm2009}, 10. \citealp{coe2015}, 11. \citealp{haberl1997}, 12. \citealp{liu2005}, 13. \citealp{pakull1985}, 14.\citealp{mcgowan2002}, 15. \citealp{haberl1995b}, 16. \citealp{negueruela2002}, 17. \citealp{shtykovskiy2005}, 18. \citealp{haberl1998}, 19. \citealp{haberl1999}, 20. \citealp{produit2004}, 21. \citealp{garrison1977}, 22. \citealp{grebenev2004}, 23. \citealp{negueruela2007}, 24. \citealp{davison1977}, 25. \citealp{becker1977}, 26. \citealp{reynolds1992}, 27. \citealp{grindlay1976}, 28. \citealp{tananbaum1976}, 29. \citealp{krimm2005}, 30. \citealp{palmer2005b}, 31. \citealp{negueruela2006}, 32. \citealp{markwardt2008}, 33. \citealp{gibaud2011}, 34. \citealp{palmer2005a}, 35. \citealp{intzand1999}, 36. \citealp{markwardt1999}, 37. \citealp{vanderspek2005}, 38. \citealp{hulleman1998}, 39. \citealp{reig2004}
    \end{threeparttable}
    \label{tab:xb}
\end{table*}

\subsection{Summary of Classification}

\begin{figure}
	\includegraphics[width=\columnwidth]{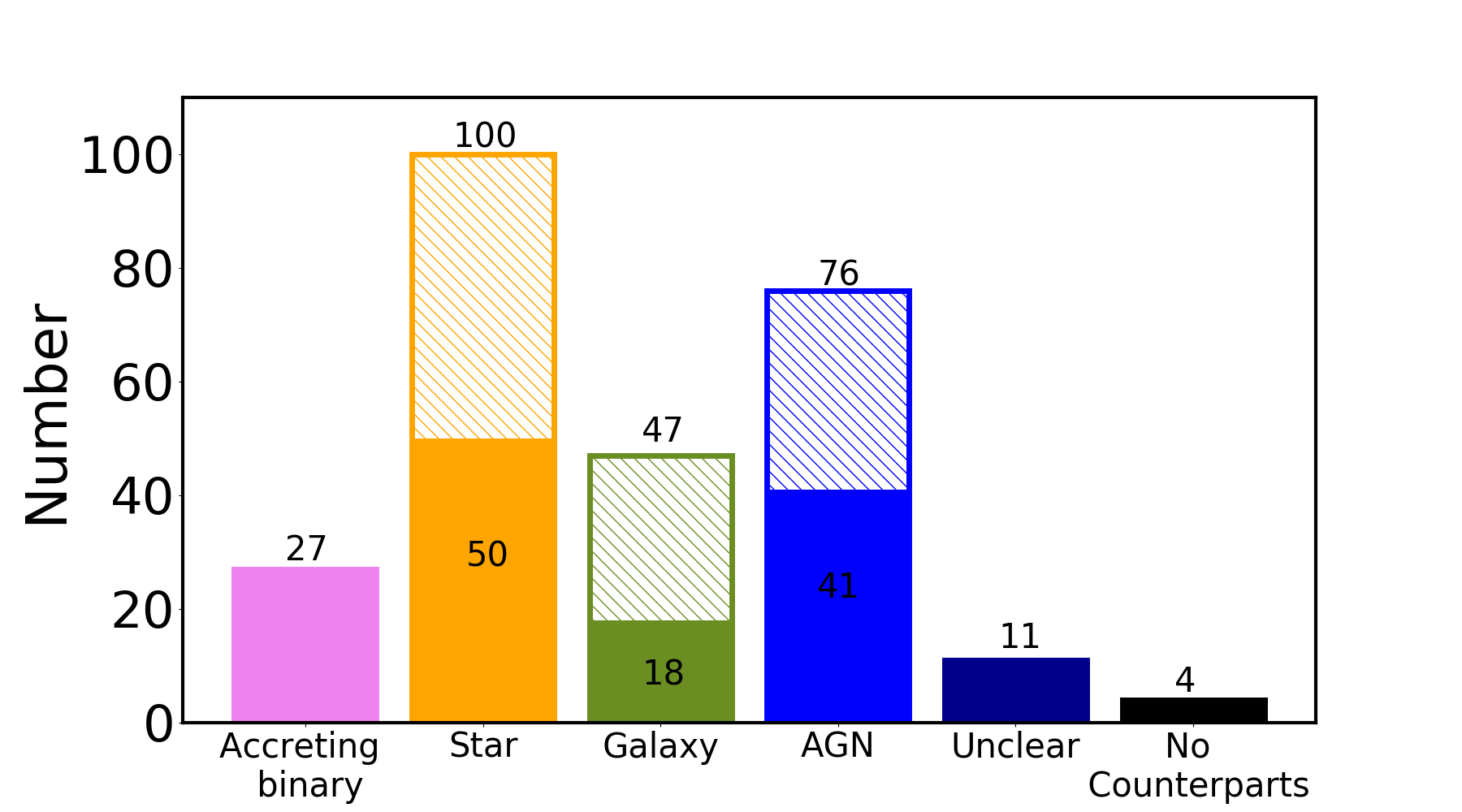}
    \caption{Updated category distribution of all the XMMSL2 variable sources after checking the properties for each sub-category. The filled bars indicate the number of sources identified with SIMBAD and NED, and the hatched  for sources identified using IR/optical properties. The classifications for some sources, mainly the galaxies, have been updated as described in Section~\ref{sec:galaxy}.}
    \label{fig:distribution_update}
\end{figure}

The identification of a few sources changed after carefully checking the light curves and images as described above. The updated category distribution of the variable sources is shown in Figure~\ref{fig:distribution_update}, where the 16 galaxies classified as {\it non-flaring source} based on their light curves were taken as AGN. The stars include single stars, stars in non-accreting binaries and the SN 1987A. Accreting binaries include cataclysmic variable stars, HMXB and LMXB. 
Of the 250 (94.3 per cent of the XMMSL2 variable sample) identified sources, 127 were stellar with 27 being accreting binaries, 47 were galaxies and 76 were AGN. 
The distribution of different types of variable sources in Galactic coordinates has been shown in Figure~\ref{fig:distribution}.
There was no obvious difference among the sky distributions of different types of variables. A catalogue of the XMMSL2 variable sources has been made available in the Paperdata of National Astronomical Data Center
at (\url{http://paperdata.china-vo.org/Li.Dongyue/xmmsl2_variables_final_upload.csv}). For each source, information about the XMMSL2 detections and identifications has been given,  and the redshift of galaxies and AGN has been provided when available. For galaxies classified as {\it flaring source}, the timescale when the fluxes declined by a factor of ten has been given. Please refer to Appendix~\ref{sec:data_release} for a further description of the released catalogue.

\section{Discussion}\label{sec:discussion}

\subsection{Variable content of the XMMSL2}

Our study revealed widespread variability among different types of sources detected in the XMMSL2 data, among which, sources identified as galaxies are of particular interest as inactive galaxies are normally considered less-variable in X-rays. One possibility is that these galaxies are indeed active but that low-level AGN activity is not picked up using the criteria in this work. Another possibility is that AGN whose infrared spectra are dominated by galaxy colours would not have been identified as AGN where classification is based on infrared colours only. It is therefore important to obtain optical spectroscopy to examine the nuclear activity and distinguish between persistent AGN accretion activity and extreme events, such as TDEs, AGN turn-on, or quasi-periodic eruptions (QPE) which have been suggested in some quiescent galaxies \citep[e.g.,][]{gezari2017, yan2019, arcodia2021}. The galaxies classified as {\it non-flaring sources} according to their soft X-ray light curves have been taken as AGN in this work, but note that the variability could also be caused by a long-decay mechanism such as a decade-long TDE \citep{lin2017}. multiwavelength observations are necessary to uncover the nature of these larger amplitude variabilities.

We have built a highly variable AGN sample based on their long-term X-ray variability and investigated the basic properties. 
Compared with the less-variable AGN from the same parent sample, the highly variable sample sources tend to have a lower redshift, soft X-ray luminosity, and black hole mass, but comparable Eddington ratio. Some of our findings are consistent with a previous study of AGN in the XMMSL. \citet{strotjohann2016} also studied the highly variable AGN in the XMMSL; some of their selection criteria differed from those used here, while there is some overlap between the sample objects. They also compared the basic properties of the variable AGN with the wider XMMSL AGN sample and found marginal evidence that highly variable AGN have a lower redshift and lower X-ray luminosity, while no significant difference was seen in the black hole mass distributions of the varying and non-varying sample.

The tentative anticorrelations between variability amplitudes and black hole masses, X-ray luminosities and redshifts that we find in our XMMSL2 variable AGN have been previously reported in both short-term and long-term light curve studies \citep[e.g.,][]{papadakis2004, oneill2005, papadakis2008, lanzuisi2014,pan2015}. There followed a suggestion that the anticorrelation between variability and luminosity may in fact be induced by a more fundamental relation with black hole mass \citep{papadakis2004, lanzuisi2014}.

The observed relationship between redshift and variability has two possible origins: an intrinsic evolutionary effect, or an observational bias. In a flux-limited catalogue like XMMSL2, at higher redshift a selection bias towards AGN with higher luminosity is likely at play, and a variability--luminosity anticorrelation, as is widely observed, would induce the variability--redshift anticorrelation. In addition, according to the structure function, AGN would be more variable on longer time scales \citep[e.g.][]{vagnetti2011, middei2017}. At higher redshift, due to the time dilation effect, the same observed time range corresponds to a shorter rest-frame interval by a factor of $1/(1+z)$, also resulting in lower observed variability. For these reasons we expect that selection effects play a significant role in the observed decrease in variability with redshift.

\citet{freund2018} previously found that the stellar population of the XMMSL2 catalogue has a disproportionately high X-ray to optical luminosity ratio, implying that this flux-limited survey preferentially selects stars in their flaring state while not detecting the same objects in their quiescent states.
Among the 100 variable stars, the majority are late-type which are particularly prone to X-ray flaring events due to their magnetic activity \citep{grindlay1970,BENZ2010,stelzer13}. Two of the sources in the variable sample were identified as young stellar objects, which are typically characterized by strong and variable X-ray emission occurring in the stellar magnetosphere, at the star-disc interface, or above the circumstellar disc \citep[e.g.,][]{feigelson1999,preibisch2005}.
Our variable stellar sample also includes 11 earlier type stars (A, F and G),  comprising 15.7 per cent of the 70 variable stars that have temperature and parallax measurements in {\it Gaia} DR2. We have also calculated the spectral types of the full sample of XMMSL2 stars in \citet{freund2018} using the effective temperature given in their catalogue, and found 34.4 per cent A, F and G stars. X-ray emission from these stars is due to coronal activity in their quiescent state \citep[e.g.,][]{lucy1980, owocki1988} with luminous X-ray flares being much less common than for later-type stars, although a number were observed in ROSAT data \citep{schaefer2000}.
Apparent X-ray variability in these objects can also be due to an unseen companion. A late-type stellar companion can cause X-ray flares while remaining unnoticed in an optical survey \citep[e.g.][]{ramsay2003}. We also found 6 non-accreting binaries showing large amplitude variability, one of which is a RS CVn system. RSCVn systems consist of a close binary pair of F, G or K stars and have been detected in {\xmm} observations \citep{Lopez2012,ramsay2003}. Enhanced chromospheric activity is responsible for producing characteristic strong flares in these systems. Ultimately an optical spectrum will be necessary to make progress in determining the nature of the object in each case.
We note that our sample contains no very massive stars, consistent with the view that isolated massive stars typically show low levels of X-ray variability \citep{testa2010}.  

There were 4 sources that did not have IR or optical counterparts within 20\,arcsec around their XMMSL2 positions. We have examined the slew images of these sources; although the possibility that these 4 sources were caused by spurious detection could not be totally ruled out, we suggest that at least 3 of them were real detections. They may be extreme eruptions taking place in distant galaxies, like gamma-ray bursts \citep[e.g.,][]{kouveliotou1993}, with host galaxies too faint to be detected in current optical/IR surveys.

\begin{figure}
	\includegraphics[width=\columnwidth]{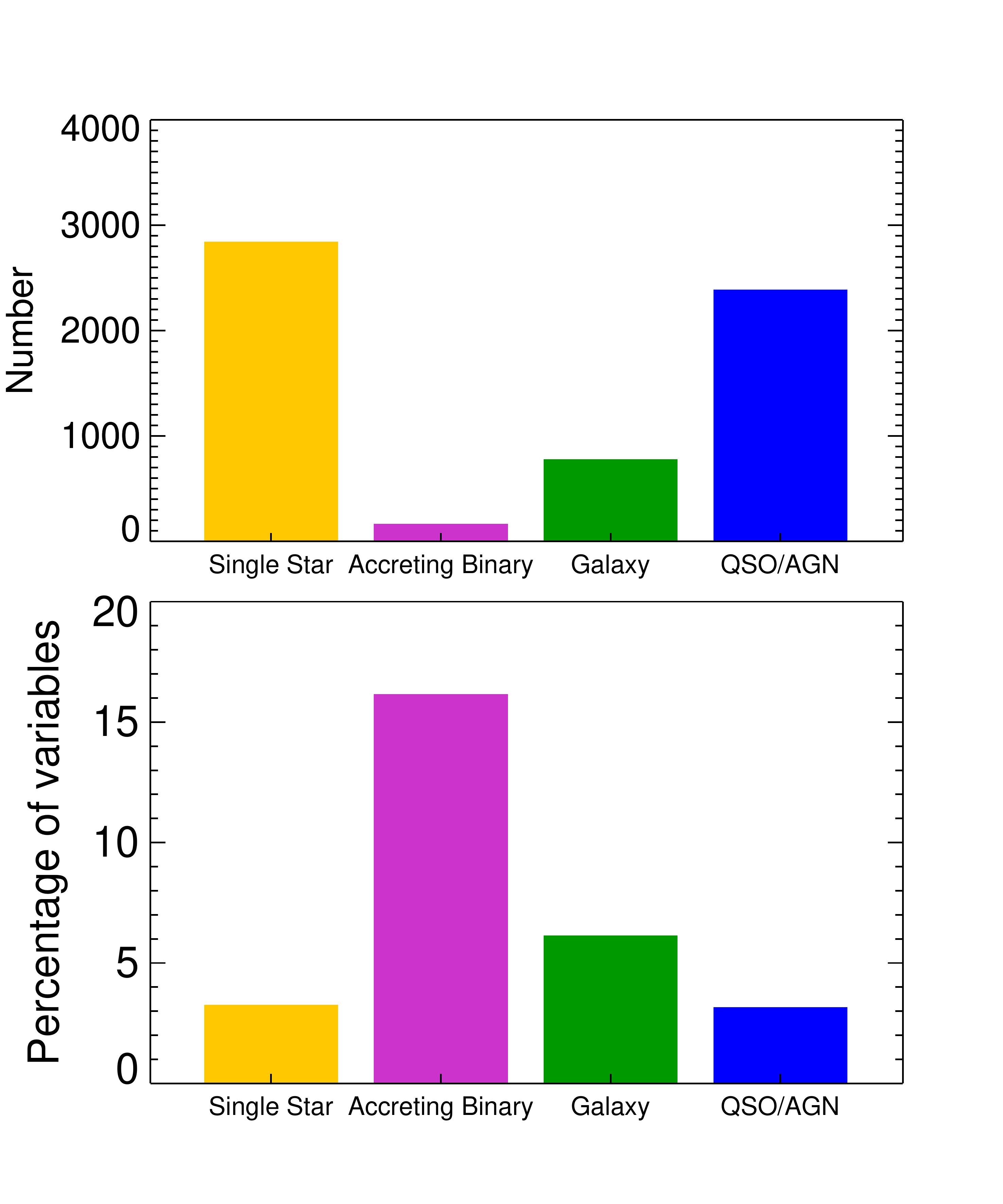}
    \caption{Upper panel: total number of sources identified in each category in the XMMSL2 clean catalogue. Lower panel: percentage of variables that met our variability criteria in each category.}
    \label{fig:distribution_whole_sample}
\end{figure}

As discussed earlier, identification of all the XMMSL2 sources has been carried out by cross-matching mainly with SIMBAD and NED. The number of sources of each category in the whole XMMSL2 clean sample is shown in Figure~\ref{fig:distribution_whole_sample}, along with the variable proportion for each category. The variable proportions of stars and AGN are comparable ($\sim$3 per cent). In the whole sample, the number of AGN is about three times that of galaxies, while the variable proportion of galaxies ($\sim$5 per cent) is higher than expected. One reason may be that some variable sources identified as galaxies are actually AGN or that a galaxy is preferentially detected in XMMSL2 during a flaring event.

\subsection{Comparison with other surveys}\label{sec:comparison}

In a systematic study of X-ray variability in the RASS, \citet{fuhrmeister2003} generated light curves covering several days for about 30000 X-ray point sources, 
finding 1207 variable objects ($\sim 4$ per cent). These variable sources were optically identified with counterparts in SIMBAD, USNO-A2.0 and NED. They found that 63.5 per cent of the identified variable sources were classified as stars while only 9.8 per cent were of extragalactic origin. The difference in their extragalactic fraction, from that which we find in XMMSL2, is likely due to the different time range of data used to search for variables, as extragalactic sources are known to become increasingly variable on longer time-scales \citep{mchardy2006}.

\citet{traulsen2020} studied long-term variability in the
4XMM-DR9s catalogue of sources found in overlapping {\it XMM-Newton} pointed observations. They found that 3.75 per cent of all multiply observed sources were likely long-term variables. These sources, classified by cross-matching with SIMBAD and SDSS-DR12 \citep{alam2015}, consisted of Galactic and extragalactic objects in similar proportions. For many sources, the identifications only contained information about the region of the electromagnetic spectrum in which the sources have been detected, e.g. X-ray source or $\gamma$-ray source. Such objects are more likely to be of extragalactic origin given the depth of {\xmm} pointed observations.

The extended Roentgen Survey with an Imaging Telescope Array (eROSITA; \citealp{merloni2012}, \citealp{predehl2017}) is expected to detect more faint X-ray variables, especially extragalactic sources such as variable AGNs and TDEs.
During the four-year all-sky survey, most of the sky will be revisited approximately every 6 months, and each location in the sky will be visited (at least) eight times during the eROSITA survey phase. 
For a given sky location, each of these visits will consist of a few 30 s long scans, separated by 4 hours, amounting to a total exposure of $\sim$ 250s.
With this pattern, eROSITA will be sensitive to a wide range of variability time-scales, from (tens of) seconds to months and years. 
In the eROSITA Final Equatorial-Depth Survey (eFEDS), by using reliable spectra, {\it Gaia} parallaxes, and/or multiwavelength properties, \citet{salvato2021} classified the sources as 2822 Galactic and 21952 extragalactic. 
For the eFEDS, the FoV passing time of a given source is about 300 seconds. Each position is typically covered in 3 consecutive visits separated by about 2000 seconds.
This scanning strategy is especially sensitive to stellar flares, which decay on a time scale of a few thousands seconds. \citet{boller2021} performed a variability test on the sources, and identified 65 sources as significantly variable in the soft band. Most of these variable sources resemble stellar flare events, and only 18 per cent are of extragalactic origin. 

The differences among the variable contents found in the surveys of XMMSL2, 4XMM-DR9s, RASS and eFEDS show clearly that the different sensitivity and cadence of surveys will sample predominantly different populations of variable sources. 

The above surveys are made with sensitive X-ray telescopes with a small  FoV of 1-2 degrees or less. 
For all-sky monitor (ASM) type instruments, such as the Monitor of All-sky X-ray Image (MAXI, \citealp{matsuoka2009}) and the Burst Alert Telescope (BAT) on {\Swift} \citep{barthelmy2005} , the situation is quite different.
These monitors have very large instantaneous FoVs of hundreds or thousands of square degrees, but much shallower sensitivity compared to the narrow-FoV instruments by several orders of magnitude.
As such, MAXI (GSC: 2-30 keV) and {\Swift} (15-150 keV) detect only the brightest X-ray variables or transients, mostly Galactic X-ray binary systems, stellar super-flares, GRBs, a few of the brightest AGN/blazars and relativistic, jetted TDEs.     
Covering a broad range of sampling cadence (from sub-seconds to years), these ASM surveys are complementary to the narrow-FoV but deep surveys in the parameter space of sensitivity and cadence, revealing different populations and properties of variable X-ray sources.  

Future missions in the field of time-domain X-ray astronomy include the Einstein Probe (EP; \citealp{yuan2016a, 2018SPIE10699E..25Y}), which adopts the lobster-eye micro-pore X-ray focusing optics and extends the energy band to the soft X-rays in 0.5-4 keV.
While having a FoV of $3600$ square degrees, its sensitivity is expected to improve significantly upon those of the previous and current X-ray ASM (assuming a typical power-law spectrum for cosmic X-ray sources).
For a typical exposure of about 20 minutes, a sensitivity of a few times $10^{-11}$ erg s$^{-1}$ cm$^{-2}$ in 0.5-4 keV is expected.
The observing cadence is typically several observations per day for a given position in the night sky. 
In this way, a weekly monitoring sensitivity of several times $10^{-12}$ erg s$^{-1}$ cm$^{-2}$ can be achieved, which is, though still less than, comparable to that of the XMMSL. Therefore the results presented in this work can be used as a reference in predicting the survey outcome for variables around these flux levels. 

\section{Summary}\label{sec:summary}
The XMMSL is an ideal survey for studying the long-term variability of various types of X-ray sources over a large part of the sky, thanks to the large and nearly uniform sky coverage and the long operation duration of the {\xmm} mission.
We have built a sample of highly variable X-ray sources comprising 265 objects drawn from the XMMSL2 catalogue, whose fluxes varied by a factor of more than ten when compared with those measured in the RASS about 10-25 years earlier. 
We performed careful identification for the sample sources using archival multiwavelength data.  
Among them, 250 (94.3 per cent) sources have been identified according to their SIMBAD, NED counterparts and their multiwavelength properties, yielding 100 stars (6 in known non-accreting binaries), 27 accreting binaries, 76 AGN and 47 galaxies.
One source associated with an AGN and four with galaxies were found to be TDEs in previous studies using XMMSL data.
For the remaining sources classified as galaxies, further observations are needed to understand the origin of their X-ray emission and variability. 
Our results show that the highly variable AGN tend to have lower black hole masses, redshifts, and luminosities compared to those with smaller X-ray variation amplitudes in the XMMSL2 sample. No significant difference is found in the distribution of the Eddington ratio between the high- and low-variability AGN.
The majority of the variable stars are magnetically active late-type stars, while the remaining tens are early-type and young stellar objects, where the X-ray flares can be caused by winds or accretion.
The statistical properties of this flux-limited sample of highly variable X-ray sources provide a useful reference for predicting the survey outcome of future X-ray time-domain survey missions. 

\section*{Acknowledgements}
We thank the anonymous referee who has helped to greatly improve the paper. We thank Sarah Casewell, Wenda Zhang, Julian Osborne, and Heyang Liu for useful discussions. DL acknowledges support from the National Natural Science Foundation of China (Nos. 12073037 and 11803047), a Chinese Academy of Sciences award to visit the University of Leicester, and ESA through the Faculty of the European Space Astronomy Centre (ESAC) - Funding reference 464/2017. RLCS acknowledges support from STFC. H-WP acknowledges support from National Natural Science Foundation of China (Nos. U1838202 and 11873054). WY acknowledges support by the Strategic Priority Research Program of the Chinese Academy of Sciences (No. XDA15052100). This work is mainly based on observations obtained with {\xmm}, an ESA science mission with instruments and contributions directly funded by ESA Member States and NASA. This research has made use of the NASA/IPAC Extragalactic Database (NED), which is funded by the National Aeronautics and Space Administration and operated by the California Institute of Technology. This work makes use of data supplied by the UK {\Swift} Science Data Centre at the University of Leicester, and data obtained from the 4XMM serendipitous source catalogue compiled by the 10 institutes of the {\xmm} Survey Science Centre selected by ESA. Thiswork has made use of data from the European Space Agency (ESA) mission Gaia (\url{https://www.cosmos.esa.int /gaia} ), processed by the Gaia Data Processing and Analysis Consortium (DPAC, \url{https:// www.cosmos.esa.int/web/gaia/ dpac/consortium}). Funding for theDPAC has been provided by national institutions, in particular the institutions participating in the Gaia Multilateral Agreement. This work also makes use of data products from {\it WISE} and 2MASS. {\it WISE} is a joint project of the University of California, Los Angeles, and the Jet Propulsion Laboratory/California Institute of Technology, funded by the National Aeronautics and Space Administration. 2MASS is a joint project of the University of Massachusetts and the Infrared Processing and Analysis Center/California Institute of Technology, funded by the National Aeronautics and Space Administration and the National Science Foundation.

\section*{DATA AVAILABILITY}

The XMMSL2 clean catalogue is available at (\url{https://www.cosmos.esa.int/web/xmm-newton/xsa#download}). The RASS2RXS catalogue is available at (\url{https://heasarc.gsfc.nasa.gov/W3Browse/rosat/rass2rxs.html}). The AllWISE, GAIA, and 2MASS data used in this work is available through the NASA/IPAC infrared archive
(\url{https://irsa.ipac.caltech.edu/applications/Gator/}). The 2SXPS catalogue is available at (\url{https://www.swift.ac.uk/2SXPS/}) and the 4XMM-DR9 catalogue can be downloaded at (\url{http://xmmssc.irap.omp.eu/Catalogue/4XMM-DR9/4XMM_DR9.html}). The catalogue of XMMSL2 variable sources generated in this work is available in the Paperdata of National Astronomical Data Center at (\url{http://paperdata.china-vo.org/Li.Dongyue/xmmsl2_variables_final_upload.csv}).




\bibliographystyle{mnras}
\bibliography{ref} 

\begin{thebibliography}{}
\makeatletter
\relax
\def\mn@urlcharsother{\let\do\@makeother \do\$\do\&\do\#\do\^\do\_\do\%\do\~}
\def\mn@doi{\begingroup\mn@urlcharsother \@ifnextchar [ {\mn@doi@}
  {\mn@doi@[]}}
\def\mn@doi@[#1]#2{\def\@tempa{#1}\ifx\@tempa\@empty \href
  {http://dx.doi.org/#2} {doi:#2}\else \href {http://dx.doi.org/#2} {#1}\fi
  \endgroup}
\def\mn@eprint#1#2{\mn@eprint@#1:#2::\@nil}
\def\mn@eprint@arXiv#1{\href {http://arxiv.org/abs/#1} {{\tt arXiv:#1}}}
\def\mn@eprint@dblp#1{\href {http://dblp.uni-trier.de/rec/bibtex/#1.xml}
  {dblp:#1}}
\def\mn@eprint@#1:#2:#3:#4\@nil{\def\@tempa {#1}\def\@tempb {#2}\def\@tempc
  {#3}\ifx \@tempc \@empty \let \@tempc \@tempb \let \@tempb \@tempa \fi \ifx
  \@tempb \@empty \def\@tempb {arXiv}\fi \@ifundefined
  {mn@eprint@\@tempb}{\@tempb:\@tempc}{\expandafter \expandafter \csname
  mn@eprint@\@tempb\endcsname \expandafter{\@tempc}}}

\bibitem[\protect\citeauthoryear{{Alam} et~al.,}{{Alam}
  et~al.}{2015}]{alam2015}
{Alam} S.,  et~al., 2015, \mn@doi [\apjs] {10.1088/0067-0049/219/1/12}, \href
  {https://ui.adsabs.harvard.edu/abs/2015ApJS..219...12A} {219, 12}

\bibitem[\protect\citeauthoryear{{Arcodia} et~al.,}{{Arcodia}
  et~al.}{2021}]{arcodia2021}
{Arcodia} R.,  et~al., 2021, \mn@doi [\nat] {10.1038/s41586-021-03394-6}, \href
  {https://ui.adsabs.harvard.edu/abs/2021Natur.592..704A} {592, 704}

\bibitem[\protect\citeauthoryear{{Barthelmy} et~al.,}{{Barthelmy}
  et~al.}{2005}]{barthelmy2005}
{Barthelmy} S.~D.,  et~al., 2005, \mn@doi [\ssr] {10.1007/s11214-005-5096-3},
  \href {https://ui.adsabs.harvard.edu/abs/2005SSRv..120..143B} {120, 143}

\bibitem[\protect\citeauthoryear{{Bassani}, {Landi}, {Masetti}, {Parisi},
  {Bazzano}  \& {Ubertini}}{{Bassani} et~al.}{2009}]{bassani2009}
{Bassani} L.,  {Landi} R.,  {Masetti} N.,  {Parisi} P.,  {Bazzano} A.,
  {Ubertini} P.,  2009, \mn@doi [\mnras] {10.1111/j.1745-3933.2009.00683.x},
  \href {https://ui.adsabs.harvard.edu/abs/2009MNRAS.397L..55B} {397, L55}

\bibitem[\protect\citeauthoryear{{Beardmore}, {Coe}, {Markwardt}, {Osborne},
  {Baumgartner}, {Tueller}  \& {Gehrels}}{{Beardmore}
  et~al.}{2009}]{Beardmore2009}
{Beardmore} A.~P.,  {Coe} M.~J.,  {Markwardt} C.,  {Osborne} J.~P.,
  {Baumgartner} W.~H.,  {Tueller} J.,   {Gehrels} N.,  2009, The Astronomer's
  Telegram, \href {https://ui.adsabs.harvard.edu/abs/2009ATel.1901....1B}
  {1901, 1}

\bibitem[\protect\citeauthoryear{{Becker}, {Swank}, {Boldt}, {Holt}, {Pravdo},
  {Saba}  \& {Serlemitsos}}{{Becker} et~al.}{1977}]{becker1977}
{Becker} R.~H.,  {Swank} J.~H.,  {Boldt} E.~A.,  {Holt} S.~S.,  {Pravdo} S.~H.,
   {Saba} J.~R.,   {Serlemitsos} P.~J.,  1977, \mn@doi [\apjl]
  {10.1086/182498}, \href
  {https://ui.adsabs.harvard.edu/abs/1977ApJ...216L..11B} {216, L11}

\bibitem[\protect\citeauthoryear{{Benz} \& {G{\"u}del}}{{Benz} \&
  {G{\"u}del}}{2010}]{BENZ2010}
{Benz} A.~O.,  {G{\"u}del} M.,  2010, \mn@doi [\araa]
  {10.1146/annurev-astro-082708-101757}, \href
  {https://ui.adsabs.harvard.edu/abs/2010ARA&A..48..241B} {48, 241}

\bibitem[\protect\citeauthoryear{{Boella}, {Butler}, {Perola}, {Piro}, {Scarsi}
   \& {Bleeker}}{{Boella} et~al.}{1997}]{boella1997}
{Boella} G.,  {Butler} R.~C.,  {Perola} G.~C.,  {Piro} L.,  {Scarsi} L.,
  {Bleeker} J.~A.~M.,  1997, \mn@doi [\aaps] {10.1051/aas:1997136}, \href
  {https://ui.adsabs.harvard.edu/abs/1997A&AS..122..299B} {122, 299}

\bibitem[\protect\citeauthoryear{{Boller}, {Freyberg}, {Tr{\"u}mper}, {Haberl},
  {Voges}  \& {Nandra}}{{Boller} et~al.}{2016}]{boller2016}
{Boller} T.,  {Freyberg} M.~J.,  {Tr{\"u}mper} J.,  {Haberl} F.,  {Voges} W.,
  {Nandra} K.,  2016, \mn@doi [\aap] {10.1051/0004-6361/201525648}, \href
  {https://ui.adsabs.harvard.edu/abs/2016A&A...588A.103B} {588, A103}

\bibitem[\protect\citeauthoryear{{Boller} et~al.,}{{Boller}
  et~al.}{2021}]{boller2021}
{Boller} T.,  et~al., 2021, arXiv e-prints, \href
  {https://ui.adsabs.harvard.edu/abs/2021arXiv210614523B} {p. arXiv:2106.14523}

\bibitem[\protect\citeauthoryear{{Brunner} et~al.,}{{Brunner}
  et~al.}{2021}]{brunner2021}
{Brunner} H.,  et~al., 2021, arXiv e-prints, \href
  {https://ui.adsabs.harvard.edu/abs/2021arXiv210614517B} {p. arXiv:2106.14517}

\bibitem[\protect\citeauthoryear{{Coe}, {Finger}, {Bartlett}  \&
  {Udalski}}{{Coe} et~al.}{2015}]{coe2015}
{Coe} M.~J.,  {Finger} M.,  {Bartlett} E.~S.,   {Udalski} A.,  2015, \mn@doi
  [\mnras] {10.1093/mnras/stu2562}, \href
  {https://ui.adsabs.harvard.edu/abs/2015MNRAS.447.1630C} {447, 1630}

\bibitem[\protect\citeauthoryear{{Davison}}{{Davison}}{1977}]{davison1977}
{Davison} P.~J.~N.,  1977, \mn@doi [\mnras] {10.1093/mnras/179.1.35P}, \href
  {https://ui.adsabs.harvard.edu/abs/1977MNRAS.179P..35D} {179, 35P}

\bibitem[\protect\citeauthoryear{{Drake} et~al.,}{{Drake}
  et~al.}{2014}]{drake2014}
{Drake} A.~J.,  et~al., 2014, \mn@doi [\mnras] {10.1093/mnras/stu639}, \href
  {https://ui.adsabs.harvard.edu/abs/2014MNRAS.441.1186D} {441, 1186}

\bibitem[\protect\citeauthoryear{{Esquej} et~al.,}{{Esquej}
  et~al.}{2008}]{esquej2008}
{Esquej} P.,  et~al., 2008, \mn@doi [\aap] {10.1051/0004-6361:200810110}, \href
  {https://ui.adsabs.harvard.edu/abs/2008A&A...489..543E} {489, 543}

\bibitem[\protect\citeauthoryear{{Evans} et~al.,}{{Evans}
  et~al.}{2020}]{evans2020}
{Evans} P.~A.,  et~al., 2020, \mn@doi [\apjs] {10.3847/1538-4365/ab7db9}, \href
  {https://ui.adsabs.harvard.edu/abs/2020ApJS..247...54E} {247, 54}

\bibitem[\protect\citeauthoryear{{Feigelson} \& {Montmerle}}{{Feigelson} \&
  {Montmerle}}{1999}]{feigelson1999}
{Feigelson} E.~D.,  {Montmerle} T.,  1999, \mn@doi [\araa]
  {10.1146/annurev.astro.37.1.363}, \href
  {https://ui.adsabs.harvard.edu/abs/1999ARA&A..37..363F} {37, 363}

\bibitem[\protect\citeauthoryear{{Freund}, {Robrade}, {Schneider}  \&
  {Schmitt}}{{Freund} et~al.}{2018}]{freund2018}
{Freund} S.,  {Robrade} J.,  {Schneider} P.~C.,   {Schmitt} J.~H.~M.~M.,  2018,
  \mn@doi [\aap] {10.1051/0004-6361/201732009}, \href
  {https://ui.adsabs.harvard.edu/abs/2018A&A...614A.125F} {614, A125}

\bibitem[\protect\citeauthoryear{{Fuhrmeister} \& {Schmitt}}{{Fuhrmeister} \&
  {Schmitt}}{2003}]{fuhrmeister2003}
{Fuhrmeister} B.,  {Schmitt} J.~H.~M.~M.,  2003, \mn@doi [\aap]
  {10.1051/0004-6361:20030303}, \href
  {https://ui.adsabs.harvard.edu/abs/2003A&A...403..247F} {403, 247}

\bibitem[\protect\citeauthoryear{{Gaia Collaboration} et~al.,}{{Gaia
  Collaboration} et~al.}{2018}]{gaia2018}
{Gaia Collaboration} et~al., 2018, \mn@doi [\aap]
  {10.1051/0004-6361/201833051}, \href
  {https://ui.adsabs.harvard.edu/abs/2018A&A...616A...1G} {616, A1}

\bibitem[\protect\citeauthoryear{{Garrison}, {Hiltner}  \& {Schild}}{{Garrison}
  et~al.}{1977}]{garrison1977}
{Garrison} R.~F.,  {Hiltner} W.~A.,   {Schild} R.~E.,  1977, \mn@doi [\apjs]
  {10.1086/190468}, \href
  {https://ui.adsabs.harvard.edu/abs/1977ApJS...35..111G} {35, 111}

\bibitem[\protect\citeauthoryear{{Gezari} et~al.,}{{Gezari}
  et~al.}{2017}]{gezari2017}
{Gezari} S.,  et~al., 2017, \mn@doi [\apj] {10.3847/1538-4357/835/2/144}, \href
  {https://ui.adsabs.harvard.edu/abs/2017ApJ...835..144G} {835, 144}

\bibitem[\protect\citeauthoryear{{Gibaud} et~al.,}{{Gibaud}
  et~al.}{2011}]{gibaud2011}
{Gibaud} L.,  et~al., 2011, The Astronomer's Telegram, \href
  {https://ui.adsabs.harvard.edu/abs/2011ATel.3565....1G} {3565, 1}

\bibitem[\protect\citeauthoryear{{Grebenev}, {Ubertini}, {Chenevez}, {Mowlavi},
  {Roques}, {Gehrels}  \& {Kuulkers}}{{Grebenev} et~al.}{2004}]{grebenev2004}
{Grebenev} S.~A.,  {Ubertini} P.,  {Chenevez} J.,  {Mowlavi} N.,  {Roques}
  J.~P.,  {Gehrels} N.,   {Kuulkers} E.,  2004, The Astronomer's Telegram,
  \href {https://ui.adsabs.harvard.edu/abs/2004ATel..350....1G} {350, 1}

\bibitem[\protect\citeauthoryear{{Grindlay}}{{Grindlay}}{1970}]{grindlay1970}
{Grindlay} J.~E.,  1970, \mn@doi [\apj] {10.1086/150645}, \href
  {https://ui.adsabs.harvard.edu/abs/1970ApJ...162..187G} {162, 187}

\bibitem[\protect\citeauthoryear{{Grindlay} \& {Gursky}}{{Grindlay} \&
  {Gursky}}{1976}]{grindlay1976}
{Grindlay} J.,  {Gursky} H.,  1976, \mn@doi [\apjl] {10.1086/182268}, \href
  {https://ui.adsabs.harvard.edu/abs/1976ApJ...209L..61G} {209, L61}

\bibitem[\protect\citeauthoryear{{Haberl} \& {Pietsch}}{{Haberl} \&
  {Pietsch}}{1999}]{haberl1999}
{Haberl} F.,  {Pietsch} W.,  1999, \aap, \href
  {https://ui.adsabs.harvard.edu/abs/1999A&A...344..521H} {344, 521}

\bibitem[\protect\citeauthoryear{{Haberl}, {Pietsch}  \& {Dennerl}}{{Haberl}
  et~al.}{1995}]{haberl1995b}
{Haberl} F.,  {Pietsch} W.,   {Dennerl} K.,  1995, \aap, \href
  {https://ui.adsabs.harvard.edu/abs/1995A&A...303L..49H} {303, L49}

\bibitem[\protect\citeauthoryear{{Haberl}, {Dennerl}, {Pietsch}  \&
  {Reinsch}}{{Haberl} et~al.}{1997}]{haberl1997}
{Haberl} F.,  {Dennerl} K.,  {Pietsch} W.,   {Reinsch} K.,  1997, \aap, \href
  {https://ui.adsabs.harvard.edu/abs/1997A&A...318..490H} {318, 490}

\bibitem[\protect\citeauthoryear{{Haberl}, {Pietsch}  \& {Filipovic}}{{Haberl}
  et~al.}{1998}]{haberl1998}
{Haberl} F.,  {Pietsch} W.,   {Filipovic} M.~D.,  1998, \iaucirc, \href
  {https://ui.adsabs.harvard.edu/abs/1998IAUC.6867....1H} {6867, 1}

\bibitem[\protect\citeauthoryear{{Hiroi} et~al.,}{{Hiroi}
  et~al.}{2013}]{hiroi2013}
{Hiroi} K.,  et~al., 2013, \mn@doi [\apjs] {10.1088/0067-0049/207/2/36}, \href
  {https://ui.adsabs.harvard.edu/abs/2013ApJS..207...36H} {207, 36}

\bibitem[\protect\citeauthoryear{{Hulleman}, {in 't Zand}  \&
  {Heise}}{{Hulleman} et~al.}{1998}]{hulleman1998}
{Hulleman} F.,  {in 't Zand} J.~J.~M.,   {Heise} J.,  1998, \aap, \href
  {https://ui.adsabs.harvard.edu/abs/1998A&A...337L..25H} {337, L25}

\bibitem[\protect\citeauthoryear{{Immler}, {Russell}, {Roming}  \&
  {Pooley}}{{Immler} et~al.}{2010}]{immler2010}
{Immler} S.,  {Russell} B.~R.,  {Roming} P.~W.~A.,   {Pooley} D.,  2010, The
  Astronomer's Telegram, \href
  {https://ui.adsabs.harvard.edu/abs/2010ATel.3045....1I} {3045, 1}

\bibitem[\protect\citeauthoryear{{Kanner}, {Baker}, {Blackburn}, {Camp},
  {Mooley}, {Mushotzky}  \& {Ptak}}{{Kanner} et~al.}{2013}]{kanner2013}
{Kanner} J.,  {Baker} J.,  {Blackburn} L.,  {Camp} J.,  {Mooley} K.,
  {Mushotzky} R.,   {Ptak} A.,  2013, \mn@doi [\apj]
  {10.1088/0004-637X/774/1/63}, \href
  {https://ui.adsabs.harvard.edu/abs/2013ApJ...774...63K} {774, 63}

\bibitem[\protect\citeauthoryear{{Komossa}}{{Komossa}}{2017}]{komossa2017}
{Komossa} S.,  2017, \mn@doi [Astronomische Nachrichten]
  {10.1002/asna.201713339}, \href
  {https://ui.adsabs.harvard.edu/abs/2017AN....338..256K} {338, 256}

\bibitem[\protect\citeauthoryear{{Komossa} \& {Bade}}{{Komossa} \&
  {Bade}}{1999}]{KomossaBade1999}
{Komossa} S.,  {Bade} N.,  1999, \aap, \href
  {http://adsabs.harvard.edu/abs/1999A%26A...343..775K} {343, 775}

\bibitem[\protect\citeauthoryear{{K{\"o}nig} et~al.,}{{K{\"o}nig}
  et~al.}{2022}]{koenig2022}
{K{\"o}nig} O.,  et~al., 2022, \mn@doi [Astronomy and Computing]
  {10.1016/j.ascom.2021.100529}, \href
  {https://ui.adsabs.harvard.edu/abs/2022A&C....3800529K} {38, 100529}

\bibitem[\protect\citeauthoryear{{Kouveliotou}, {Meegan}, {Fishman}, {Bhat},
  {Briggs}, {Koshut}, {Paciesas}  \& {Pendleton}}{{Kouveliotou}
  et~al.}{1993}]{kouveliotou1993}
{Kouveliotou} C.,  {Meegan} C.~A.,  {Fishman} G.~J.,  {Bhat} N.~P.,  {Briggs}
  M.~S.,  {Koshut} T.~M.,  {Paciesas} W.~S.,   {Pendleton} G.~N.,  1993,
  \mn@doi [\apjl] {10.1086/186969}, \href
  {https://ui.adsabs.harvard.edu/abs/1993ApJ...413L.101K} {413, L101}

\bibitem[\protect\citeauthoryear{{Krimm}, {Barthelmy}, {Capalbi}, {Gehrels},
  {Gronwall}  \& {Palmer}}{{Krimm} et~al.}{2005}]{krimm2005}
{Krimm} H.,  {Barthelmy} S.,  {Capalbi} M.,  {Gehrels} N.,  {Gronwall} C.,
  {Palmer} D.,  2005, GRB Coordinates Network, \href
  {https://ui.adsabs.harvard.edu/abs/2005GCN..4361....1K} {4361, 1}

\bibitem[\protect\citeauthoryear{{Krimm} et~al.,}{{Krimm}
  et~al.}{2009}]{krimm2009}
{Krimm} H.~A.,  et~al., 2009, The Astronomer's Telegram, \href
  {https://ui.adsabs.harvard.edu/abs/2009ATel.2011....1K} {2011, 1}

\bibitem[\protect\citeauthoryear{{Lanzuisi} et~al.,}{{Lanzuisi}
  et~al.}{2014}]{lanzuisi2014}
{Lanzuisi} G.,  et~al., 2014, \mn@doi [\apj] {10.1088/0004-637X/781/2/105},
  \href {https://ui.adsabs.harvard.edu/abs/2014ApJ...781..105L} {781, 105}

\bibitem[\protect\citeauthoryear{{Li} et~al.,}{{Li} et~al.}{2020}]{li2020}
{Li} D.,  et~al., 2020, \mn@doi [\apj] {10.3847/1538-4357/ab744a}, \href
  {https://ui.adsabs.harvard.edu/abs/2020ApJ...891..121L} {891, 121}

\bibitem[\protect\citeauthoryear{{Lin} et~al.,}{{Lin} et~al.}{2017}]{lin2017}
{Lin} D.,  et~al., 2017, \mn@doi [Nature Astronomy] {10.1038/s41550-016-0033},
  \href {https://ui.adsabs.harvard.edu/abs/2017NatAs...1E..33L} {1, 0033}

\bibitem[\protect\citeauthoryear{{Liu}, {van Paradijs}  \& {van den
  Heuvel}}{{Liu} et~al.}{2005}]{liu2005}
{Liu} Q.~Z.,  {van Paradijs} J.,   {van den Heuvel} E.~P.~J.,  2005, \mn@doi
  [\aap] {10.1051/0004-6361:20053718}, \href
  {https://ui.adsabs.harvard.edu/abs/2005A&A...442.1135L} {442, 1135}

\bibitem[\protect\citeauthoryear{{L{\'o}pez-Santiago}, {Stelzer}  \&
  {Saxton}}{{L{\'o}pez-Santiago} et~al.}{2012}]{Lopez2012}
{L{\'o}pez-Santiago} J.,  {Stelzer} B.,   {Saxton} R.,  2012, \mn@doi [\pasp]
  {10.1086/666647}, \href
  {https://ui.adsabs.harvard.edu/abs/2012PASP..124..682L} {124, 682}

\bibitem[\protect\citeauthoryear{{Lucke}, {Yentis}, {Friedman}, {Fritz}  \&
  {Shulman}}{{Lucke} et~al.}{1976}]{lucke1976}
{Lucke} R.,  {Yentis} D.,  {Friedman} H.,  {Fritz} G.,   {Shulman} S.,  1976,
  \mn@doi [\apjl] {10.1086/182125}, \href
  {https://ui.adsabs.harvard.edu/abs/1976ApJ...206L..25L} {206, L25}

\bibitem[\protect\citeauthoryear{{Lucy} \& {White}}{{Lucy} \&
  {White}}{1980}]{lucy1980}
{Lucy} L.~B.,  {White} R.~L.,  1980, \mn@doi [\apj] {10.1086/158342}, \href
  {https://ui.adsabs.harvard.edu/abs/1980ApJ...241..300L} {241, 300}

\bibitem[\protect\citeauthoryear{{Mainzer} et~al.,}{{Mainzer}
  et~al.}{2011}]{mainzer2011}
{Mainzer} A.,  et~al., 2011, \mn@doi [\apj] {10.1088/0004-637X/743/2/156},
  \href {https://ui.adsabs.harvard.edu/abs/2011ApJ...743..156M} {743, 156}

\bibitem[\protect\citeauthoryear{{Mainzer} et~al.,}{{Mainzer}
  et~al.}{2014}]{mainzer2014}
{Mainzer} A.,  et~al., 2014, \mn@doi [\apj] {10.1088/0004-637X/792/1/30}, \href
  {https://ui.adsabs.harvard.edu/abs/2014ApJ...792...30M} {792, 30}

\bibitem[\protect\citeauthoryear{{Marconi} \& {Hunt}}{{Marconi} \&
  {Hunt}}{2003}]{marconi2003}
{Marconi} A.,  {Hunt} L.~K.,  2003, \mn@doi [\apjl] {10.1086/375804}, \href
  {https://ui.adsabs.harvard.edu/abs/2003ApJ...589L..21M} {589, L21}

\bibitem[\protect\citeauthoryear{{Markwardt}, {Swank}  \&
  {Marshall}}{{Markwardt} et~al.}{1999}]{markwardt1999}
{Markwardt} C.~B.,  {Swank} J.~H.,   {Marshall} F.~E.,  1999, \iaucirc, \href
  {https://ui.adsabs.harvard.edu/abs/1999IAUC.7120....1M} {7120, 1}

\bibitem[\protect\citeauthoryear{{Markwardt}, {Pereira}  \&
  {Swank}}{{Markwardt} et~al.}{2008}]{markwardt2008}
{Markwardt} C.~B.,  {Pereira} D.,   {Swank} J.~H.,  2008, The Astronomer's
  Telegram, \href {https://ui.adsabs.harvard.edu/abs/2008ATel.1569....1M}
  {1569, 1}

\bibitem[\protect\citeauthoryear{{Masetti} et~al.,}{{Masetti}
  et~al.}{2006}]{masetti2006}
{Masetti} N.,  et~al., 2006, \mn@doi [\aap] {10.1051/0004-6361:20054332}, \href
  {https://ui.adsabs.harvard.edu/abs/2006A&A...449.1139M} {449, 1139}

\bibitem[\protect\citeauthoryear{{Massey}}{{Massey}}{2002}]{Massey2002}
{Massey} P.,  2002, \mn@doi [\apjs] {10.1086/338286}, \href
  {https://ui.adsabs.harvard.edu/abs/2002ApJS..141...81M} {141, 81}

\bibitem[\protect\citeauthoryear{{Matsuoka} et~al.,}{{Matsuoka}
  et~al.}{2009}]{matsuoka2009}
{Matsuoka} M.,  et~al., 2009, \mn@doi [\pasj] {10.1093/pasj/61.5.999}, \href
  {https://ui.adsabs.harvard.edu/abs/2009PASJ...61..999M} {61, 999}

\bibitem[\protect\citeauthoryear{{McGowan} \& {Charles}}{{McGowan} \&
  {Charles}}{2002}]{mcgowan2002}
{McGowan} K.~E.,  {Charles} P.~A.,  2002, \mn@doi [\mnras]
  {10.1046/j.1365-8711.2002.05675.x}, \href
  {https://ui.adsabs.harvard.edu/abs/2002MNRAS.335..941M} {335, 941}

\bibitem[\protect\citeauthoryear{{McHardy}, {Koerding}, {Knigge}, {Uttley}  \&
  {Fender}}{{McHardy} et~al.}{2006}]{mchardy2006}
{McHardy} I.~M.,  {Koerding} E.,  {Knigge} C.,  {Uttley} P.,   {Fender} R.~P.,
  2006, \mn@doi [\nat] {10.1038/nature05389}, \href
  {https://ui.adsabs.harvard.edu/abs/2006Natur.444..730M} {444, 730}

\bibitem[\protect\citeauthoryear{{McQuillan}, {Mazeh}  \&
  {Aigrain}}{{McQuillan} et~al.}{2014}]{mcquillan2014}
{McQuillan} A.,  {Mazeh} T.,   {Aigrain} S.,  2014, \mn@doi [\apjs]
  {10.1088/0067-0049/211/2/24}, \href
  {https://ui.adsabs.harvard.edu/abs/2014ApJS..211...24M} {211, 24}

\bibitem[\protect\citeauthoryear{{Melia}}{{Melia}}{2009}]{melia2009}
{Melia} F.,  2009, {High-Energy Astrophysics}

\bibitem[\protect\citeauthoryear{{Merloni} et~al.,}{{Merloni}
  et~al.}{2012}]{merloni2012}
{Merloni} A.,  et~al., 2012, arXiv e-prints, \href
  {https://ui.adsabs.harvard.edu/abs/2012arXiv1209.3114M} {p. arXiv:1209.3114}

\bibitem[\protect\citeauthoryear{{Middei}, {Vagnetti}, {Bianchi}, {La Franca},
  {Paolillo}  \& {Ursini}}{{Middei} et~al.}{2017}]{middei2017}
{Middei} R.,  {Vagnetti} F.,  {Bianchi} S.,  {La Franca} F.,  {Paolillo} M.,
  {Ursini} F.,  2017, \mn@doi [\aap] {10.1051/0004-6361/201629940}, \href
  {https://ui.adsabs.harvard.edu/abs/2017A&A...599A..82M} {599, A82}

\bibitem[\protect\citeauthoryear{{Mingo} et~al.,}{{Mingo}
  et~al.}{2016}]{mingo2016}
{Mingo} B.,  et~al., 2016, \mn@doi [\mnras] {10.1093/mnras/stw1826}, \href
  {https://ui.adsabs.harvard.edu/abs/2016MNRAS.462.2631M} {462, 2631}

\bibitem[\protect\citeauthoryear{{Miniutti}, {Saxton},
  {Rodr{\'\i}guez-Pascual}, {Read}, {Esquej}, {Colless}, {Dobbie}  \&
  {Spolaor}}{{Miniutti} et~al.}{2013}]{miniutti2013}
{Miniutti} G.,  {Saxton} R.~D.,  {Rodr{\'\i}guez-Pascual} P.~M.,  {Read} A.~M.,
   {Esquej} P.,  {Colless} M.,  {Dobbie} P.,   {Spolaor} M.,  2013, \mn@doi
  [\mnras] {10.1093/mnras/stt850}, \href
  {https://ui.adsabs.harvard.edu/abs/2013MNRAS.433.1764M} {433, 1764}

\bibitem[\protect\citeauthoryear{{Morales}, {Miller}, {Cackett}, {Reynolds}  \&
  {Zoghbi}}{{Morales} et~al.}{2019}]{morales2019}
{Morales} A.~M.,  {Miller} J.~M.,  {Cackett} E.~M.,  {Reynolds} M.~T.,
  {Zoghbi} A.,  2019, \mn@doi [\apj] {10.3847/1538-4357/aaeff9}, \href
  {https://ui.adsabs.harvard.edu/abs/2019ApJ...870...54M} {870, 54}

\bibitem[\protect\citeauthoryear{{Morgan}, {Keenan}  \& {Kellman}}{{Morgan}
  et~al.}{1943}]{morgan1943}
{Morgan} W.~W.,  {Keenan} P.~C.,   {Kellman} E.,  1943, {An atlas of stellar
  spectra, with an outline of spectral classification}

\bibitem[\protect\citeauthoryear{{Negueruela} \& {Coe}}{{Negueruela} \&
  {Coe}}{2002}]{negueruela2002}
{Negueruela} I.,  {Coe} M.~J.,  2002, \mn@doi [\aap]
  {10.1051/0004-6361:20020139}, \href
  {https://ui.adsabs.harvard.edu/abs/2002A&A...385..517N} {385, 517}

\bibitem[\protect\citeauthoryear{{Negueruela} \& {Marco}}{{Negueruela} \&
  {Marco}}{2006}]{negueruela2006}
{Negueruela} I.,  {Marco} A.,  2006, The Astronomer's Telegram, \href
  {https://ui.adsabs.harvard.edu/abs/2006ATel..739....1N} {739, 1}

\bibitem[\protect\citeauthoryear{{Negueruela}, {Torrej{\'o}n}  \&
  {McBride}}{{Negueruela} et~al.}{2007}]{negueruela2007}
{Negueruela} I.,  {Torrej{\'o}n} J.~M.,   {McBride} V.,  2007, The Astronomer's
  Telegram, \href {https://ui.adsabs.harvard.edu/abs/2007ATel.1239....1N}
  {1239, 1}

\bibitem[\protect\citeauthoryear{{O'Neill}, {Nandra}, {Papadakis}  \&
  {Turner}}{{O'Neill} et~al.}{2005}]{oneill2005}
{O'Neill} P.~M.,  {Nandra} K.,  {Papadakis} I.~E.,   {Turner} T.~J.,  2005,
  \mn@doi [\mnras] {10.1111/j.1365-2966.2005.08860.x}, \href
  {https://ui.adsabs.harvard.edu/abs/2005MNRAS.358.1405O} {358, 1405}

\bibitem[\protect\citeauthoryear{{Oliveira}, {Rodrigues}, {Cieslinski},
  {Jablonski}, {Silva}, {Almeida}, {Rodr{\'\i}guez-Ardila}  \&
  {Palhares}}{{Oliveira} et~al.}{2017}]{oliveira2017}
{Oliveira} A.~S.,  {Rodrigues} C.~V.,  {Cieslinski} D.,  {Jablonski} F.~J.,
  {Silva} K.~M.~G.,  {Almeida} L.~A.,  {Rodr{\'\i}guez-Ardila} A.,   {Palhares}
  M.~S.,  2017, \mn@doi [\aj] {10.3847/1538-3881/aa610d}, \href
  {https://ui.adsabs.harvard.edu/abs/2017AJ....153..144O} {153, 144}

\bibitem[\protect\citeauthoryear{{Owocki}, {Castor}  \& {Rybicki}}{{Owocki}
  et~al.}{1988}]{owocki1988}
{Owocki} S.~P.,  {Castor} J.~I.,   {Rybicki} G.~B.,  1988, \mn@doi [\apj]
  {10.1086/166977}, \href
  {https://ui.adsabs.harvard.edu/abs/1988ApJ...335..914O} {335, 914}

\bibitem[\protect\citeauthoryear{{Pakull}, {Brunner}, {Pietsch}, {Staubert},
  {Beuermann}, {van der Klis}  \& {Bonnet-Bidaud}}{{Pakull}
  et~al.}{1985}]{pakull1985}
{Pakull} M.,  {Brunner} H.,  {Pietsch} W.,  {Staubert} A.,  {Beuermann} K.,
  {van der Klis} M.,   {Bonnet-Bidaud} J.~M.,  1985, \mn@doi [\ssr]
  {10.1007/BF00179844}, \href
  {https://ui.adsabs.harvard.edu/abs/1985SSRv...40..379P} {40, 379}

\bibitem[\protect\citeauthoryear{{Palmer}, {Barthelmey}, {Cummings}, {Gehrels},
  {Krimm}, {Markwardt}, {Sakamoto}  \& {Tueller}}{{Palmer}
  et~al.}{2005a}]{palmer2005a}
{Palmer} D.~M.,  {Barthelmey} S.~D.,  {Cummings} J.~R.,  {Gehrels} N.,  {Krimm}
  H.~A.,  {Markwardt} C.~B.,  {Sakamoto} T.,   {Tueller} J.,  2005a, The
  Astronomer's Telegram, \href
  {https://ui.adsabs.harvard.edu/abs/2005ATel..546....1P} {546, 1}

\bibitem[\protect\citeauthoryear{{Palmer}, {Barthelmy}, {Cummings}, {Gehrels},
  {Kennea}, {Krimm}, {Markwardt}  \& {Tueller}}{{Palmer}
  et~al.}{2005b}]{palmer2005b}
{Palmer} D.,  {Barthelmy} S.,  {Cummings} J.,  {Gehrels} N.,  {Kennea} J.,
  {Krimm} H.,  {Markwardt} C.~B.,   {Tueller} J.,  2005b, The Astronomer's
  Telegram, \href {https://ui.adsabs.harvard.edu/abs/2005ATel..678....1P} {678,
  1}

\bibitem[\protect\citeauthoryear{{Pan}, {Yuan}, {Zhou}, {Dong}  \& {Liu}}{{Pan}
  et~al.}{2015}]{pan2015}
{Pan} H.-W.,  {Yuan} W.,  {Zhou} X.-L.,  {Dong} X.-B.,   {Liu} B.,  2015,
  \mn@doi [\apj] {10.1088/0004-637X/808/2/163}, \href
  {https://ui.adsabs.harvard.edu/abs/2015ApJ...808..163P} {808, 163}

\bibitem[\protect\citeauthoryear{{Papadakis}}{{Papadakis}}{2004}]{papadakis2004}
{Papadakis} I.~E.,  2004, \mn@doi [\mnras] {10.1111/j.1365-2966.2004.07351.x},
  \href {https://ui.adsabs.harvard.edu/abs/2004MNRAS.348..207P} {348, 207}

\bibitem[\protect\citeauthoryear{{Papadakis}, {Chatzopoulos}, {Athanasiadis},
  {Markowitz}  \& {Georgantopoulos}}{{Papadakis} et~al.}{2008}]{papadakis2008}
{Papadakis} I.~E.,  {Chatzopoulos} E.,  {Athanasiadis} D.,  {Markowitz} A.,
  {Georgantopoulos} I.,  2008, \mn@doi [\aap] {10.1051/0004-6361:200809572},
  \href {https://ui.adsabs.harvard.edu/abs/2008A&A...487..475P} {487, 475}

\bibitem[\protect\citeauthoryear{{Parker} et~al.,}{{Parker}
  et~al.}{2016}]{parker2016}
{Parker} M.~L.,  et~al., 2016, \mn@doi [\mnras] {10.1093/mnras/stw1449}, \href
  {https://ui.adsabs.harvard.edu/abs/2016MNRAS.461.1927P} {461, 1927}

\bibitem[\protect\citeauthoryear{{Predehl}}{{Predehl}}{2017}]{predehl2017}
{Predehl} P.,  2017, \mn@doi [Astronomische Nachrichten]
  {10.1002/asna.201713324}, \href
  {https://ui.adsabs.harvard.edu/abs/2017AN....338..159P} {338, 159}

\bibitem[\protect\citeauthoryear{{Preibisch} et~al.,}{{Preibisch}
  et~al.}{2005}]{preibisch2005}
{Preibisch} T.,  et~al., 2005, \mn@doi [\apjs] {10.1086/432891}, \href
  {https://ui.adsabs.harvard.edu/abs/2005ApJS..160..401P} {160, 401}

\bibitem[\protect\citeauthoryear{{Produit}, {Ballet}  \& {Mowlavi}}{{Produit}
  et~al.}{2004}]{produit2004}
{Produit} N.,  {Ballet} J.,   {Mowlavi} N.,  2004, The Astronomer's Telegram,
  \href {https://ui.adsabs.harvard.edu/abs/2004ATel..278....1P} {278, 1}

\bibitem[\protect\citeauthoryear{{Ramsay}, {Harra}  \& {Kay}}{{Ramsay}
  et~al.}{2003}]{ramsay2003}
{Ramsay} G.,  {Harra} L.,   {Kay} H.,  2003, \mn@doi [\mnras]
  {10.1046/j.1365-8711.2003.06517.x}, \href
  {https://ui.adsabs.harvard.edu/abs/2003MNRAS.341.1388R} {341, 1388}

\bibitem[\protect\citeauthoryear{{Read} et~al.,}{{Read}
  et~al.}{2008}]{read2008}
{Read} A.~M.,  et~al., 2008, \mn@doi [\aap] {10.1051/0004-6361:200809456},
  \href {https://ui.adsabs.harvard.edu/abs/2008A&A...482L...1R} {482, L1}

\bibitem[\protect\citeauthoryear{{Read} et~al.,}{{Read}
  et~al.}{2009}]{read2009}
{Read} A.~M.,  et~al., 2009, \mn@doi [\aap] {10.1051/0004-6361/200912082},
  \href {https://ui.adsabs.harvard.edu/abs/2009A&A...506.1309R} {506, 1309}

\bibitem[\protect\citeauthoryear{{Rees}}{{Rees}}{1988}]{rees1988}
{Rees} M.~J.,  1988, \mn@doi [\nat] {10.1038/333523a0}, \href
  {https://ui.adsabs.harvard.edu/abs/1988Natur.333..523R} {333, 523}

\bibitem[\protect\citeauthoryear{{Reig}}{{Reig}}{2011}]{reig2011}
{Reig} P.,  2011, \mn@doi [\apss] {10.1007/s10509-010-0575-8}, \href
  {https://ui.adsabs.harvard.edu/abs/2011Ap&SS.332....1R} {332, 1}

\bibitem[\protect\citeauthoryear{{Reig}, {Negueruela}, {Fabregat}, {Chato},
  {Blay}  \& {Mavromatakis}}{{Reig} et~al.}{2004}]{reig2004}
{Reig} P.,  {Negueruela} I.,  {Fabregat} J.,  {Chato} R.,  {Blay} P.,
  {Mavromatakis} F.,  2004, \mn@doi [\aap] {10.1051/0004-6361:20035786}, \href
  {https://ui.adsabs.harvard.edu/abs/2004A&A...421..673R} {421, 673}

\bibitem[\protect\citeauthoryear{{Remillard} \& {McClintock}}{{Remillard} \&
  {McClintock}}{2006}]{remillard2006}
{Remillard} R.~A.,  {McClintock} J.~E.,  2006, \mn@doi [\araa]
  {10.1146/annurev.astro.44.051905.092532}, \href
  {https://ui.adsabs.harvard.edu/abs/2006ARA&A..44...49R} {44, 49}

\bibitem[\protect\citeauthoryear{{Reynolds}, {Bell}  \& {Hilditch}}{{Reynolds}
  et~al.}{1992}]{reynolds1992}
{Reynolds} A.~P.,  {Bell} S.~A.,   {Hilditch} R.~W.,  1992, \mn@doi [\mnras]
  {10.1093/mnras/256.3.631}, \href
  {https://ui.adsabs.harvard.edu/abs/1992MNRAS.256..631R} {256, 631}

\bibitem[\protect\citeauthoryear{{Salvato} et~al.,}{{Salvato}
  et~al.}{2018}]{salvato2018}
{Salvato} M.,  et~al., 2018, \mn@doi [\mnras] {10.1093/mnras/stx2651}, \href
  {https://ui.adsabs.harvard.edu/abs/2018MNRAS.473.4937S} {473, 4937}

\bibitem[\protect\citeauthoryear{{Salvato} et~al.,}{{Salvato}
  et~al.}{2021}]{salvato2021}
{Salvato} M.,  et~al., 2021, arXiv e-prints, \href
  {https://ui.adsabs.harvard.edu/abs/2021arXiv210614520S} {p. arXiv:2106.14520}

\bibitem[\protect\citeauthoryear{{Saxton}, {Read}, {Esquej}, {Freyberg},
  {Altieri}  \& {Bermejo}}{{Saxton} et~al.}{2008}]{saxton2008}
{Saxton} R.~D.,  {Read} A.~M.,  {Esquej} P.,  {Freyberg} M.~J.,  {Altieri} B.,
   {Bermejo} D.,  2008, \mn@doi [\aap] {10.1051/0004-6361:20079193}, \href
  {https://ui.adsabs.harvard.edu/abs/2008A&A...480..611S} {480, 611}

\bibitem[\protect\citeauthoryear{{Saxton}, {Read}, {Esquej}, {Komossa},
  {Dougherty}, {Rodriguez-Pascual}  \& {Barrado}}{{Saxton}
  et~al.}{2012}]{saxton2012}
{Saxton} R.~D.,  {Read} A.~M.,  {Esquej} P.,  {Komossa} S.,  {Dougherty} S.,
  {Rodriguez-Pascual} P.,   {Barrado} D.,  2012, \mn@doi [\aap]
  {10.1051/0004-6361/201118367}, \href
  {https://ui.adsabs.harvard.edu/abs/2012A&A...541A.106S} {541, A106}

\bibitem[\protect\citeauthoryear{{Saxton} et~al.,}{{Saxton}
  et~al.}{2014}]{saxton2014}
{Saxton} R.~D.,  et~al., 2014, \mn@doi [\aap] {10.1051/0004-6361/201424347},
  \href {https://ui.adsabs.harvard.edu/abs/2014A&A...572A...1S} {572, A1}

\bibitem[\protect\citeauthoryear{{Saxton}, {Motta}, {Komossa}  \&
  {Read}}{{Saxton} et~al.}{2015}]{saxton2015}
{Saxton} R.~D.,  {Motta} S.~E.,  {Komossa} S.,   {Read} A.~M.,  2015, \mn@doi
  [\mnras] {10.1093/mnras/stv2160}, \href
  {https://ui.adsabs.harvard.edu/abs/2015MNRAS.454.2798S} {454, 2798}

\bibitem[\protect\citeauthoryear{{Saxton}, {Read}, {Komossa}, {Lira},
  {Alexander}  \& {Wieringa}}{{Saxton} et~al.}{2017}]{saxton2017}
{Saxton} R.~D.,  {Read} A.~M.,  {Komossa} S.,  {Lira} P.,  {Alexander} K.~D.,
  {Wieringa} M.~H.,  2017, \mn@doi [\aap] {10.1051/0004-6361/201629015}, \href
  {https://ui.adsabs.harvard.edu/abs/2017A&A...598A..29S} {598, A29}

\bibitem[\protect\citeauthoryear{{Saxton}, {Komossa}, {Auchettl}  \&
  {Jonker}}{{Saxton} et~al.}{2021}]{saxton2021}
{Saxton} R.,  {Komossa} S.,  {Auchettl} K.,   {Jonker} P.~G.,  2021, \mn@doi
  [\ssr] {10.1007/s11214-020-00759-7}, \href
  {https://ui.adsabs.harvard.edu/abs/2021SSRv..217...18S} {217, 18}

\bibitem[\protect\citeauthoryear{{Saxton} et~al.,}{{Saxton}
  et~al.}{2022}]{saxton2022}
{Saxton} R.~D.,  et~al., 2022, \mn@doi [Astronomy and Computing]
  {10.1016/j.ascom.2021.100531}, \href
  {https://ui.adsabs.harvard.edu/abs/2022A&C....3800531S} {38, 100531}

\bibitem[\protect\citeauthoryear{{Sazonov}, {Churazov}, {Revnivtsev},
  {Vikhlinin}  \& {Sunyaev}}{{Sazonov} et~al.}{2005}]{sazonov2005b}
{Sazonov} S.,  {Churazov} E.,  {Revnivtsev} M.,  {Vikhlinin} A.,   {Sunyaev}
  R.,  2005, \mn@doi [\aap] {10.1051/0004-6361:200500205}, \href
  {https://ui.adsabs.harvard.edu/abs/2005A&A...444L..37S} {444, L37}

\bibitem[\protect\citeauthoryear{{Sazonov}, {Revnivtsev}, {Gilfanov},
  {Churazov}  \& {Sunyaev}}{{Sazonov} et~al.}{2006}]{sazonov2005a}
{Sazonov} S.,  {Revnivtsev} M.,  {Gilfanov} M.,  {Churazov} E.,   {Sunyaev} R.,
   2006, \mn@doi [\aap] {10.1051/0004-6361:20054297}, \href
  {https://ui.adsabs.harvard.edu/abs/2006A&A...450..117S} {450, 117}

\bibitem[\protect\citeauthoryear{{Schaefer}, {King}  \&
  {Deliyannis}}{{Schaefer} et~al.}{2000}]{schaefer2000}
{Schaefer} B.~E.,  {King} J.~R.,   {Deliyannis} C.~P.,  2000, \mn@doi [\apj]
  {10.1086/308325}, \href
  {https://ui.adsabs.harvard.edu/abs/2000ApJ...529.1026S} {529, 1026}

\bibitem[\protect\citeauthoryear{{Schreier}, {Giacconi}, {Gursky}, {Kellogg}
  \& {Tananbaum}}{{Schreier} et~al.}{1972}]{schreier1972}
{Schreier} E.,  {Giacconi} R.,  {Gursky} H.,  {Kellogg} E.,   {Tananbaum} H.,
  1972, \mn@doi [\apjl] {10.1086/181086}, \href
  {https://ui.adsabs.harvard.edu/abs/1972ApJ...178L..71S} {178, L71}

\bibitem[\protect\citeauthoryear{{Seward} \& {Charles}}{{Seward} \&
  {Charles}}{2010}]{seward2010}
{Seward} F.~D.,  {Charles} P.~A.,  2010, {Exploring the X-ray Universe}

\bibitem[\protect\citeauthoryear{{Shappee} et~al.,}{{Shappee}
  et~al.}{2013}]{shappee2013}
{Shappee} B.~J.,  et~al., 2013, The Astronomer's Telegram, \href
  {https://ui.adsabs.harvard.edu/abs/2013ATel.5052....1S} {5052, 1}

\bibitem[\protect\citeauthoryear{{Shtykovskiy} \& {Gilfanov}}{{Shtykovskiy} \&
  {Gilfanov}}{2005}]{shtykovskiy2005}
{Shtykovskiy} P.,  {Gilfanov} M.,  2005, \mn@doi [\aap]
  {10.1051/0004-6361:20041074}, \href
  {https://ui.adsabs.harvard.edu/abs/2005A&A...431..597S} {431, 597}

\bibitem[\protect\citeauthoryear{{Skrutskie} et~al.,}{{Skrutskie}
  et~al.}{2006}]{skrutskie2006}
{Skrutskie} M.~F.,  et~al., 2006, \mn@doi [\aj] {10.1086/498708}, \href
  {https://ui.adsabs.harvard.edu/abs/2006AJ....131.1163S} {131, 1163}

\bibitem[\protect\citeauthoryear{{Stelzer}, {Marino}, {Micela},
  {L{\'o}pez-Santiago}  \& {Liefke}}{{Stelzer} et~al.}{2013}]{stelzer13}
{Stelzer} B.,  {Marino} A.,  {Micela} G.,  {L{\'o}pez-Santiago} J.,   {Liefke}
  C.,  2013, \mn@doi [\mnras] {10.1093/mnras/stt225}, \href
  {https://ui.adsabs.harvard.edu/abs/2013MNRAS.431.2063S} {431, 2063}

\bibitem[\protect\citeauthoryear{{Strotjohann}, {Saxton}, {Starling}, {Esquej},
  {Read}, {Evans}  \& {Miniutti}}{{Strotjohann} et~al.}{2016}]{strotjohann2016}
{Strotjohann} N.~L.,  {Saxton} R.~D.,  {Starling} R.~L.~C.,  {Esquej} P.,
  {Read} A.~M.,  {Evans} P.~A.,   {Miniutti} G.,  2016, \mn@doi [\aap]
  {10.1051/0004-6361/201628241}, \href
  {https://ui.adsabs.harvard.edu/abs/2016A&A...592A..74S} {592, A74}

\bibitem[\protect\citeauthoryear{{Tananbaum}, {Chaisson}, {Forman}, {Jones}  \&
  {Matilsky}}{{Tananbaum} et~al.}{1976}]{tananbaum1976}
{Tananbaum} H.,  {Chaisson} L.~J.,  {Forman} W.,  {Jones} C.,   {Matilsky}
  T.~A.,  1976, \mn@doi [\apjl] {10.1086/182282}, \href
  {https://ui.adsabs.harvard.edu/abs/1976ApJ...209L.125T} {209, L125}

\bibitem[\protect\citeauthoryear{{Testa}}{{Testa}}{2010}]{testa2010}
{Testa} P.,  2010, \mn@doi [Proceedings of the National Academy of Science]
  {10.1073/pnas.0913822107}, \href
  {https://ui.adsabs.harvard.edu/abs/2010PNAS..107.7158T} {107, 7158}

\bibitem[\protect\citeauthoryear{{Traulsen} et~al.,}{{Traulsen}
  et~al.}{2020}]{traulsen2020}
{Traulsen} I.,  et~al., 2020, \mn@doi [\aap] {10.1051/0004-6361/202037706},
  \href {https://ui.adsabs.harvard.edu/abs/2020A&A...641A.137T} {641, A137}

\bibitem[\protect\citeauthoryear{{Vagnetti}, {Turriziani}  \&
  {Trevese}}{{Vagnetti} et~al.}{2011}]{vagnetti2011}
{Vagnetti} F.,  {Turriziani} S.,   {Trevese} D.,  2011, \mn@doi [\aap]
  {10.1051/0004-6361/201118072}, \href
  {https://ui.adsabs.harvard.edu/abs/2011A&A...536A..84V} {536, A84}

\bibitem[\protect\citeauthoryear{{Vanderspek}, {Morgan}, {Crew}, {Graziani}  \&
  {Suzuki}}{{Vanderspek} et~al.}{2005}]{vanderspek2005}
{Vanderspek} R.,  {Morgan} E.,  {Crew} G.,  {Graziani} C.,   {Suzuki} M.,
  2005, The Astronomer's Telegram, \href
  {https://ui.adsabs.harvard.edu/abs/2005ATel..516....1V} {516, 1}

\bibitem[\protect\citeauthoryear{{Vasilopoulos}, {Haberl}, {Delvaux}, {Sturm}
  \& {Udalski}}{{Vasilopoulos} et~al.}{2016}]{vasilopoulos2016}
{Vasilopoulos} G.,  {Haberl} F.,  {Delvaux} C.,  {Sturm} R.,   {Udalski} A.,
  2016, \mn@doi [\mnras] {10.1093/mnras/stw1408}, \href
  {https://ui.adsabs.harvard.edu/abs/2016MNRAS.461.1875V} {461, 1875}

\bibitem[\protect\citeauthoryear{{V{\'e}ron-Cetty} \&
  {V{\'e}ron}}{{V{\'e}ron-Cetty} \& {V{\'e}ron}}{2010}]{veron2010}
{V{\'e}ron-Cetty} M.~P.,  {V{\'e}ron} P.,  2010, \mn@doi [\aap]
  {10.1051/0004-6361/201014188}, \href
  {https://ui.adsabs.harvard.edu/abs/2010A&A...518A..10V} {518, A10}

\bibitem[\protect\citeauthoryear{{Voges} et~al.,}{{Voges}
  et~al.}{1999}]{voges1999}
{Voges} W.,  et~al., 1999, \aap, \href
  {https://ui.adsabs.harvard.edu/abs/1999A&A...349..389V} {349, 389}

\bibitem[\protect\citeauthoryear{{Warwick}, {Saxton}  \& {Read}}{{Warwick}
  et~al.}{2012}]{warwick2012}
{Warwick} R.~S.,  {Saxton} R.~D.,   {Read} A.~M.,  2012, \mn@doi [\aap]
  {10.1051/0004-6361/201118642}, \href
  {https://ui.adsabs.harvard.edu/abs/2012A&A...548A..99W} {548, A99}

\bibitem[\protect\citeauthoryear{{Webb} et~al.,}{{Webb}
  et~al.}{2020}]{webb2020}
{Webb} N.~A.,  et~al., 2020, \mn@doi [\aap] {10.1051/0004-6361/201937353},
  \href {https://ui.adsabs.harvard.edu/abs/2020A&A...641A.136W} {641, A136}

\bibitem[\protect\citeauthoryear{{Webster}, {Martin}, {Feast}  \&
  {Andrews}}{{Webster} et~al.}{1972}]{webster1972}
{Webster} B.~L.,  {Martin} W.~L.,  {Feast} M.~W.,   {Andrews} P.~J.,  1972,
  \mn@doi [Nature Physical Science] {10.1038/physci240183b0}, \href
  {https://ui.adsabs.harvard.edu/abs/1972NPhS..240R.183W} {240, 183}

\bibitem[\protect\citeauthoryear{{Wenger} et~al.,}{{Wenger}
  et~al.}{2000}]{wenger2000}
{Wenger} M.,  et~al., 2000, \mn@doi [\aaps] {10.1051/aas:2000332}, \href
  {https://ui.adsabs.harvard.edu/abs/2000A&AS..143....9W} {143, 9}

\bibitem[\protect\citeauthoryear{{Wright} et~al.,}{{Wright}
  et~al.}{2010}]{Wright2010}
{Wright} E.~L.,  et~al., 2010, \mn@doi [\aj] {10.1088/0004-6256/140/6/1868},
  \href {https://ui.adsabs.harvard.edu/abs/2010AJ....140.1868W} {140, 1868}

\bibitem[\protect\citeauthoryear{{Yan} \& {Yu}}{{Yan} \& {Yu}}{2015}]{yan2015}
{Yan} Z.,  {Yu} W.,  2015, \mn@doi [\apj] {10.1088/0004-637X/805/2/87}, \href
  {https://ui.adsabs.harvard.edu/abs/2015ApJ...805...87Y} {805, 87}

\bibitem[\protect\citeauthoryear{{Yan} et~al.,}{{Yan} et~al.}{2019}]{yan2019}
{Yan} L.,  et~al., 2019, \mn@doi [\apj] {10.3847/1538-4357/ab074b}, \href
  {https://ui.adsabs.harvard.edu/abs/2019ApJ...874...44Y} {874, 44}

\bibitem[\protect\citeauthoryear{{Yuan}, {Osborne}, {Watson}  \&
  {Komossa}}{{Yuan} et~al.}{2006}]{yuan2006}
{Yuan} W.,  {Osborne} J.~P.,  {Watson} M.~G.,   {Komossa} S.,  2006, \mn@doi
  [Advances in Space Research] {10.1016/j.asr.2005.03.107}, \href
  {https://ui.adsabs.harvard.edu/abs/2006AdSpR..38.1421Y} {38, 1421}

\bibitem[\protect\citeauthoryear{{Yuan} et~al.,}{{Yuan}
  et~al.}{2016}]{yuan2016a}
{Yuan} W.,  et~al., 2016, \mn@doi [\ssr] {10.1007/s11214-016-0274-z}, \href
  {https://ui.adsabs.harvard.edu/abs/2016SSRv..202..235Y} {202, 235}

\bibitem[\protect\citeauthoryear{{Yuan} et~al.,}{{Yuan}
  et~al.}{2018}]{2018SPIE10699E..25Y}
{Yuan} W.,  et~al., 2018, in {den Herder} J.-W.~A.,  {Nikzad} S.,   {Nakazawa}
  K.,  eds,  Society of Photo-Optical Instrumentation Engineers (SPIE)
  Conference Series Vol. 10699, Space Telescopes and Instrumentation 2018:
  Ultraviolet to Gamma Ray. p. 1069925, \mn@doi{10.1117/12.2313358}

\bibitem[\protect\citeauthoryear{{in 't Zand}, {Heise}, {Bazzano}, {Cocchi},
  {di Ciolo}  \& {Muller}}{{in 't Zand} et~al.}{1999}]{intzand1999}
{in 't Zand} J.,  {Heise} J.,  {Bazzano} A.,  {Cocchi} M.,  {di Ciolo} L.,
  {Muller} J.~M.,  1999, \iaucirc, \href
  {https://ui.adsabs.harvard.edu/abs/1999IAUC.7119....1I} {7119, 1}

\makeatother
\end{thebibliography}



\appendix\label{sec:app}

{ \section{Chance match estimation}\label{app:spurious}
In this section, we justify the matching radii adopted for correlating between different catalogues and estimate the
potential rate of matches which occur by chance. 
As an example, we have investigated the probability of finding a chance RASS2RXS counterpart to XMMSL2 as a function of angular separation. The positional accuracies of XMMSL2 and RASS2RXS are 8\,arcsec and 15\,arcsec, respectively. If we cross-match the catalogues out to a large radius of 4\,arcmin, we find 8037 associations for the 10463 XMMSL2 sources. The distribution of angular separations between the XMMSL2 sources and their RASS2RXS counterparts is shown in  Figure~\ref{fig:rass_spurious} (panel b). Next we shifted the XMMSL2 positions randomly by 5\,arcmin--26\,arcmin\ in right ascension and declination, and re-correlated the shifted sources with RASS2RXS using the same matching radius. This was repeated 10 times and the results averaged to estimate the probability of chance coincidence as a function of separation (panel a of Figure~\ref{fig:rass_spurious}). By multiplying these probabilities by the actual distribution of
separation distances we obtain the number of potential spurious (i.e. chance) matches in each separation bin (panel c of Figure~\ref{fig:rass_spurious}). As the number of spurious matches increases significantly outside 120\,arcsec\ and as more than 97 per cent of the counterparts are located inside, 120\,arcsec\ is taken as a valid search radius for RASS counterparts of XMMSL2 sources. 
\begin{figure}
    \centering
    \includegraphics[width=\columnwidth]{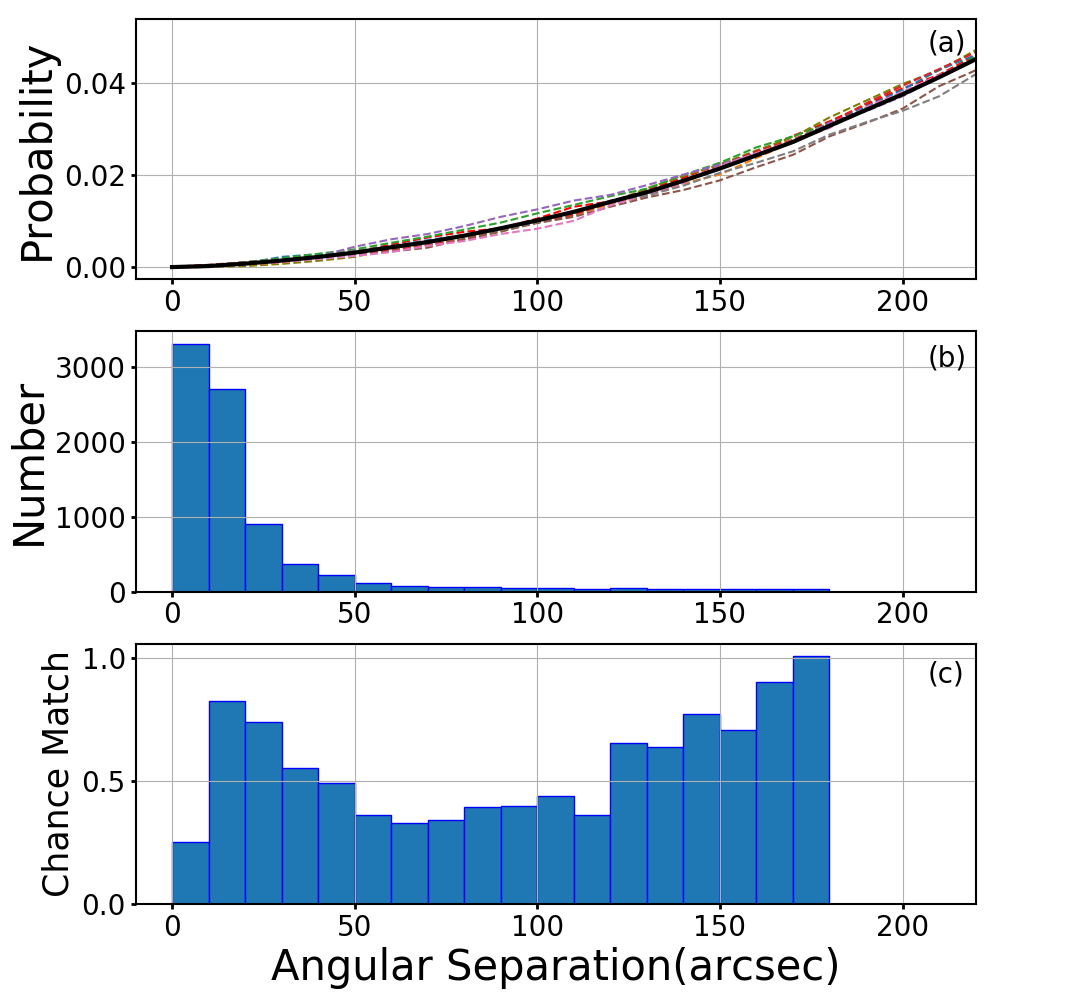}
    \caption{(a) The probability of finding a RASS2RXS counterpart within a given radius. The coloured dashed line shows the result of each trial and the black solid line shows the averaged result of all the trials.  (b)  Histogram of the angular separation between the true XMMSL2 sources and their RASS counterparts. (c) Number of matches of XMMSL2 sources which may have arisen by chance in each separation bin.}
    \label{fig:rass_spurious}
\end{figure}

We applied the same technique to the XMMSL2-AllWISE correlation finding the results shown in Figure~\ref{fig:allwise_spurious}. According to the distribution of angular separation and number of potential spurious matches, 12\,arcsec\ defines a reasonable boundary to search the AllWISE counterparts for XMMSL2 sources. 

\begin{figure}
    \centering
    \includegraphics[width=\columnwidth]{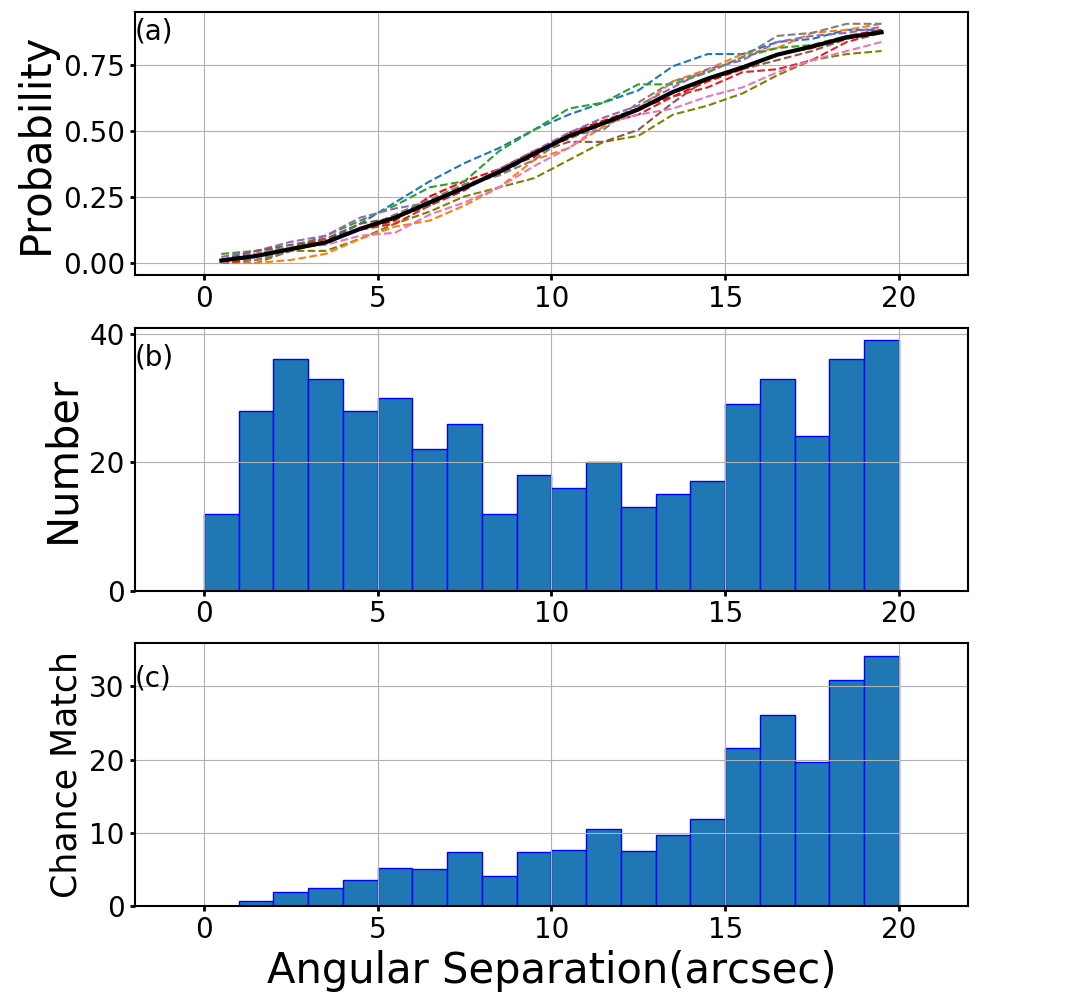}
    \caption{(a) The probability of finding an AllWISE counterpart within a given radius. The coloured dashed line shows the result of each trial and the black solid line shows the averaged result of all the trials.  (b)  Histogram of the angular separation between the true XMMSL2 sources and their AllWISE counterparts. (c) Number of matches of XMMSL2 sources which may have arisen by chance for different radii.}
    \label{fig:allwise_spurious}
\end{figure}

\begin{table}
\caption{The matching radii, the number of matches, and the estimated number of chance matches for each catalogue correlation$^{a}$.}
\begin{threeparttable}
    \centering
    \begin{tabular}{cccc}
    \hline
         Matched & Search radius & No. of & Chance \\
         catalogues & (\,arcsec) & counterparts & matches \\
    \hline
    XMMSL2-RASS2RXS & 120 & 7848 & 15.0 \\
    XMMSL2-AllWISE & 12 & 106 & 22.4\\
    AllWISE-Gaia & 3 & 87 & 1.7 \\
    AGN-2MASS$^{b}$ & 5 & 1170 & 4.5 \\
    \hline
    \end{tabular}
    $^{a}$ The number of chance matches in the correlation of XMMSL2-4XMM-DR9, XMMSL2-2SXPS, XMMSL2-SIMBAD, and AGN-Veron is very small.\\
    $^{b}$ The AGN sample includes both the low- and high-variability AGN.
\end{threeparttable}
\label{tab:spurious_summary}
\end{table}

We then adopted the method of \citet{warwick2012} to investigate the effect of limiting count rate or magnitude on the probability of finding an object within a matching radius.
The method is quite similar to that used to find the optimum matching radius, but with count rate or magnitude used instead of separation as the criterion for a chance match. The results of XMMSL-RASS2RXS and XMMSL-AllWISE correlations are shown in Figure~\ref{fig:rass_spurious_rate} and Figure~\ref{fig:allwise_spurious_mag}, respectively. 
In our work, 7848 XMMSL2 sources had RASS2RXS counterparts within 120\,arcsec, and the total estimated number of chance matches is 15.0 distributed predominantly over RASS count-rates of 0.01 to 0.1 c/s. The number of associations in the XMMSL-AllWISE correlation is 106 within the 12\,arcsec matching radius. The number of potential chance matches was 4.1 out of 65 for counterparts with W1$\le$15, and 18.3 out of 41 for counterparts with W1>15. The number of spurious matches in the other catalogue correlations was estimated using the same method and results are summarized in Table~\ref{tab:spurious_summary}. }

\begin{figure}
    \centering
    \includegraphics[width=\columnwidth]{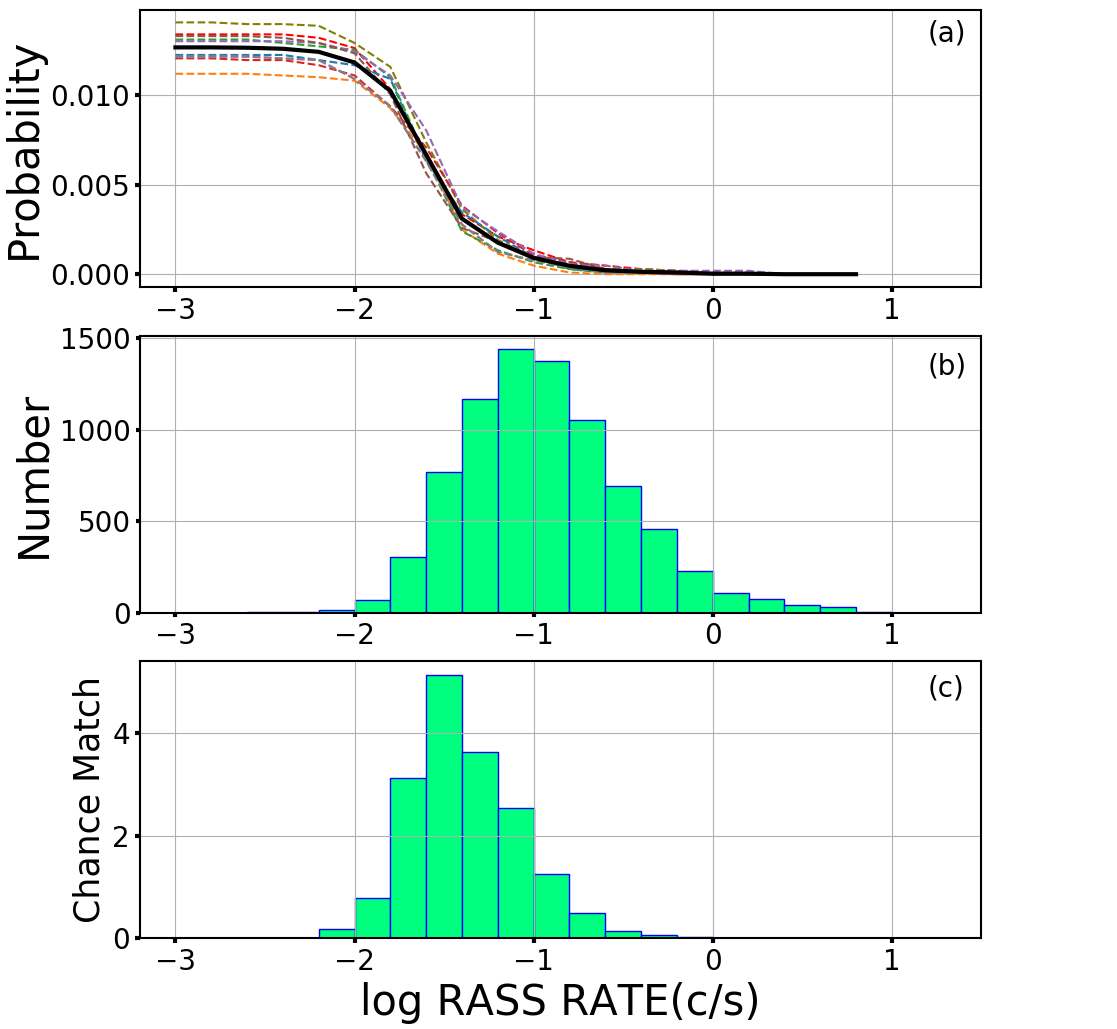}
    \caption{(a) The probability of finding an object brighter than a given count rate within a 120\,arcsec error circle. The coloured dashed line shows the result of each trial and the black solid line shows the averaged result of all the trials.  (b) The RASS2RXS count rate distribution of the RASS2RXS counterparts. (c) Number of potential chance matches as a function of count rate.}
    \label{fig:rass_spurious_rate}
\end{figure}

\begin{figure}
    \centering
    \includegraphics[width=\columnwidth]{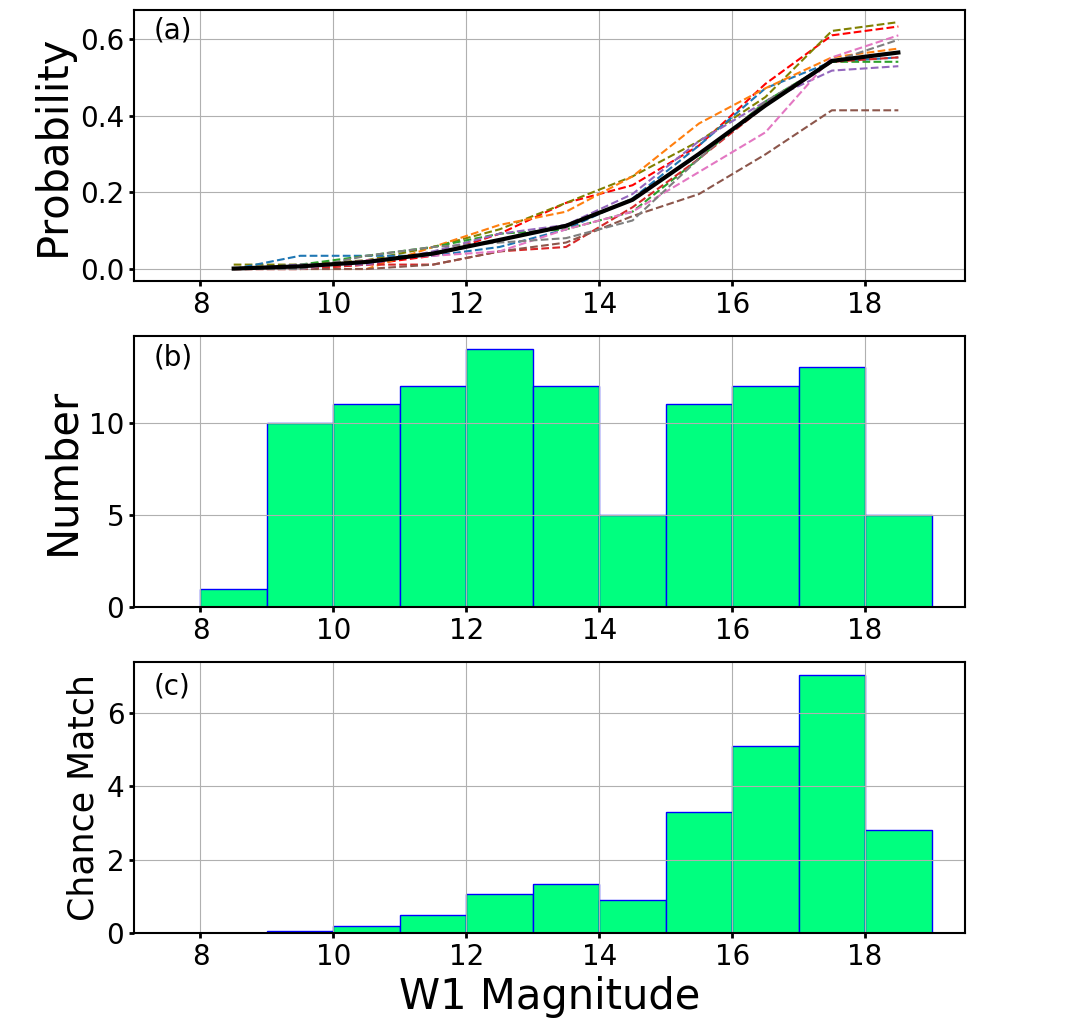}
    \caption{(a) The probability of finding an object brighter than a given W1 magnitude within a 12\,arcsec error circle. The coloured dashed line shows the result of each trial and the black solid line shows the averaged result of all the trials.  (b) The magnitude distribution of the AllWISE counterparts. (c) Number of potential chance matches with in magnitude bin.}
    \label{fig:allwise_spurious_mag}
\end{figure}

\section{Sources with more than one counterpart in SIMBAD or NED}\label{sec:app_mul}
We list the sources, with and without {\xmm} and {\Swift} pointed observations, that have more than one counterparts in SIMBAD or NED in Table~\ref{tab:mul_swiftxmm} and Table~\ref{tab:mul_noxmmswift}, respectively, and give our preference based on the properties of each. 

\begin{table*}
\caption{Sources with {\xmm} or {\Swift} pointed observations that have more than one counterpart in SIMBAD.}
\begin{threeparttable}
    \centering
    \begin{tabular}{cccccc}
    \hline
    No. & XMMSL2 name & Sep.(\,arcsec) & Object name & Object type & Identification\\
    \hline
    1   & XMMSL2 J051935.7-323932 & 0.27 & [SUA2001] J051935.815-323927.81 & Rad & Sy2  \\
        &                        & 0.31 & ESO 362-18                      & Sy2 &      \\
    \hline
    2   & XMMSL2 J060636.5-694932 & 1.38 & XMMSL2 J060636.2-694933          & No* & No*  \\
        &                        & 2.40 & USNO-A2.0 0150-04066298         & *   &      \\
    \hline
    3   & XMMSL2 J104745.6-375930 & 2.89 & 2MASX J10474573-3759323         & G   & AGN  \\
        &                        & 3.13 & XMMSL2 J104745.6-375932         & AGN &      \\
    \hline
    4   & XMMSL2 J124021.7-714852 & 0.80 & \wise\ J124021.21-714857.7        & IrS  & BLL   \\
        &                        & 1.77 & 3FHL J1240.5-7148               & BLL &       \\
    \hline
    5   & XMMSL2 J173616.9-444400 & 0.14 & ...                             & ... & star cluster \\
        &                        & ...  & ...                             & ... &               \\
    \hline
    6   & XMMSL2 J183520.4+611940 & 1.53 & 6C 183446+611710                & QSO & QSO \\
        &                        & 2.20 & [TOS2004] QSO J1835+6119         & ALS &     \\
        &                        & 2.20 & [TOS2004] QSO J1835+6119        & ALS &     \\
    \hline
    7   & XMMSL2 J200112.7+435248 & 1.10 & 2MASS J20011286+4352529         & IrS  & BLL \\
        &                        & 1.18 & ICRF J200112.8+435252           & BLL &    \\
    
    \hline
 \end{tabular}
    Column 3: Separation between the identifier and the X-ray position, in unit of arcsec; column 4: identifier given in SIMBAD. column 5: object type of the identifier given in SIMBAD.  Rad: Radio source. Sy2: Seyfert2 galaxy. No*: Nova. *: Star. G: Galaxy. IrS: Infrared source. BLL: BL Lac-type object. QSO: Quasar. ALS: Absorption line system; column 6: identification adopted in this work.
    \end{threeparttable}
    \label{tab:mul_swiftxmm}
\end{table*}

\begin{table*}
\renewcommand{\arraystretch}{0.9}
\caption{Sources that did not have {\xmm} or {\it Swift} observations and have more than one counterpart in SIMBAD or NED.\\}
\begin{threeparttable}
    \centering
    \begin{tabular}{cccccc}
    \hline
    No. & XMMSL2 name & Sep.(\,arcsec) & Object name & Object type & Identification \\
    \hline
    1   & XMMSL2 J001146.3+505509 & 2.84 &  1RXS J001146.4+505506 & X & *\\
        &                        & 4.52 & TYC 3259-1571-1  & * & \\
    \hline
    2   & XMMSL2 J011527.5-312355 & 0.00 &XMMSL2 J011527.3-312354  & G & G \\
        &                        & 2.34 &2dFGRS TGS448Z231 & X &  \\
    \hline
    3   &  XMMSL2 J034650.3-751500 & 1.92 &WISEA J034650.67-751500.9  & PM* & PM* \\
        &                         & 5.79 & 1RXS J034651.0-751455 & X   & \\
    \hline
    4   &  XMMSL2 J052443.6-613005 & 1.5 &WISEA J052443.74-613006.6 & G & G\\
        &                         & 6.3 & WISEA J052444.31-613000.6& IrS& \\
    \hline
    5   & XMMSL2 J060844.7-601916 & 3.57 & BPS CS 31068-0011  & * & *\\
        &                        & 3.89 & 1RXS J060845.0-601919 & X &  \\
    \hline
    6   & XMMSL2 J063542.5-700542 & 11.08 & 1RXS J063543.6-700551 & X & * \\
        &                        & 11.28 &HD 48136  & * &     \\
    \hline
    7   & XMMSL2 J090822.4-643751$^a$ & 0.00  & {\it Gaia} DR2 5296818376853304448 & BD? & G\\
        &                        & 0.27  & \wise\ J090822.49-643751.7  & G   &    \\
    \hline
    8   & XMMSL2 J092031.0+582243 & 2.9 &WISEA J092030.79+582244.9 & * & * \\
        &                        & 3.0 & WISEA J092031.32+582240.6& IrS &\\
        &                        & 5.1 &2MASS J09203127+5822381 & *   & \\
    \hline
    9  & XMMSL2 J110535.5+095616 & 6.6 &SDSS J110535.45+095622.0 & G & G  \\
        &                        & 9.9 &SDSS J110536.11+095616.5 & G &       \\
        &                        & 10.7& WISEA J110536.12+095618.8  & IrS &      \\
    \hline
    10  & XMMSL2 J112509.4+125432 $^b$ & 1.7  &WISEA J112509.48+125430.3 & * & *  \\
        &                        & 10.1 & SDSS J112510.09+125426.1& G & \\
        &                        & 11.4 & WISEA J112508.82+125436.7& IrS &  \\
    \hline
    11  & XMMSL2 J114359.5-610738 & 6.13 &IGR J11435-6109  & HMXB & HMXB \\
        &                        & 8.33 &UCAC4 145-092986 & *   &    \\
    \hline
    12  & XMMSL2 J123924.8-724407  & 2.13 & {\it Gaia} DR2 5841860932297535488 & *   & *\\
        &                         & 9.07 & TYC 9236-254-1 & *   &  \\
    \hline
    13  & XMMSL2 J134903.0+610730 $^c$ & 2.4 & WISEA J134903.40+610728.6 & * & *\\
        &                        & 10.7 &SDSS J134904.32+610723.8 & G & \\
    \hline
    14  & XMMSL2 J141851.4+002308 & 8.6 &WISEA J141851.71+002315.0 & IrS & * \\
        &                        & 9.3 & 2MASS J14185179+0023148& *   &\\
    \hline
    15  & XMMSL2 J160910.4+680655  & 0.00 & LEDA 2712209 & G & G \\
        &                         & 6.45 & 1RXS J160910.1+680704  & X &\\
    \hline
    16  & XMMSL2 J171238.5+584013 $^d$ & 4.7 & FLS23(R) J171237.8+584014& G & G \\
        &                        & 7.4 &FLS23(R) J171239.3+584011 & G & \\
        &                        & 9.3 &FLS23(R) J171239.4+584017 & G &\\
        &                        & 11.0 &WISEA J171238.37+584024.6  & * & \\
        &                        & 11.9 &FLS23(R) J171238.9+584002 & G & \\
    \hline
    17  & XMMSL2 J172700.3+181422$^e$ & 5.12 & ASASSN-13ag & ? & CV \\
        &                        & 6.20 & GALEX J172700.7+181420 & QSO &\\
    \hline
    18  & XMMSL2 J173449.8+560256 & 1.5 &NVSS J173449+560257 & RadioS & G \\
        &                        & 5.8 & SDSS J173449.60+560251.1& G      &   \\
        &                        & 6.5 &WISEA J173449.69+560249.9 & UvS      &   \\
        &                        & 8.8 &SDSS J173449.81+560247.6 & G      &   \\
    \hline
    19  &  XMMSL2 J180043.5+672026 & 1.81 & [ACM2000] 3-63 B& IrS & *\\
        &                         & 2.64 & 2MASS J18004338+6720240 & *  &  \\
    \hline
    20  & XMMSL2 J184500.0+582554 & 6.4 &WISEA J184500.87+582554.0 & G & G \\
        &                        & 9.6 &WISEA J184501.24+582558.1 & IrS & \\
    \hline
    
    \end{tabular}
     Column 3: Separation between the identifier and the X-ray position, in unit of arcsec; column 4: identifier given in SIMBAD or NED; column 5: Object type of the identifier given in SIMBAD or NED. X: X-ray source. G: galaxy. PM*: High proper-motion star. IrS: Infrared source. BD?: Brown dwarf candidate. UvS: ultraviolet source. \\
    $a$. The nearest counterpart is a candidate brown dwarf at a distance of 97.96 pc and the luminosity in the 
    XMMSL2 detection is $\sim$1.4$\times$ 10$^{31}$ erg\,s$^{-1}$, which is too high for a flaring brown dwarf.\\
    $^{b,c}$ Considering the separation between the counterparts and the X-ray positions and the parallax of the stars (24.88 and 3.8 given in {\it Gaia}), we prefer these two sources as stars. \\
    $^d$ Parallax of the star is 0.297\,mas, corresponding to a distance which is too large for a variable star in this XMMSL2 sample.\\
    $e$ This source is likely to be a CV according to the optical light curve and X-ray detection (\citealp{shappee2013}).
    \end{threeparttable}
    \label{tab:mul_noxmmswift}
\end{table*}

\section{Data release}\label{sec:data_release}

We release the identification of the XMMSL2 variable sources, and the list and description of columns are provided in this section. We provide some columns describing key information about XMMSL detections from the original XMMSL2 catalogue and identifier information from the data bases and optical/IR catalogues. For each source, we also provide a column about the identification resource used. 
\\
{\it Column 1.} {\bf XMMSL2\_unique\_name}: official name for sources detected in the {\it XMM-Newton} slew survey.\\
{\it Columns 2-3.} {\bf XMMSL2\_RA, XMMSL2\_DEC}: XMMSL2 J2000 Right Ascension and Declination, in degrees.\\
{\it Column 4.} {\bf FLUX\_XMMSL2}: the 0.2--2\,keV source flux in the XMMSL2, in unit of erg\,s$^{-1}$\,cm$^{-2}$. Derived from an absorbed power-law spectrum of slope 1.7 and $N_{\rm H}=3\times10^{20}$\,cm$^{-2}$. For sources with more than one XMMSL2 detection, the highest flux is given. \\
{\it Column 5.} {\bf FLUX\_XMMSL2\_ERR}: the error on the XMMSL2 soft band flux, in unit of erg\,s$^{-1}$\,cm$^{-2}$. \\
{\it Column 6.} {\bf EXT\_XMMSL2}: the spatial extension of the source in the XMMSL2 soft band, in unit of pixels. \\
{\it Column 7.} {\bf DET\_ML\_XMMSL2}: the detection likelihoods in the XMMSL2 soft band. \\
{\it Column 8.} {\bf FLUX\_RASS}: the 0.2--2\,keV flux in the RASS at the XMMSL2 source position, in unit of erg\,s$^{-1}$\,cm$^{-2}$. \\
{\it Column 9.} {\bf FLUX\_RASS\_ERR}: the error on the RASS soft flux, in unit of erg\,s$^{-1}$\,cm$^{-2}$. 0 means that the FLUX\_RASS is an upper limit. \\
{\it Column 10.} {\bf FLUX\_RATIO}: flux ratio between FLUX\_XMMSL2 and FLUX\_RASS. \\
{\it Column 11.} {\bf IDENTIFIER}: the catalogue name of the best match. \\
{\it Columns 12-13.} {\bf IDENTIFIER\_RA, IDENTIFIER\_DEC}: RA and Dec of the identifier, in degrees. For sources for which the counterpart remains unclear, {\it Columns 11-13} are empty.\\
{\it Column 14.} {\bf ID\_TYPE}: type of the identifier.\\
{\it Column 15.} {\bf ID\_FLAG}: flag for the type for each source. 1: star, 2: AGN, 3: non-interacting binary, 4: accreting binary, 5: galaxy, 6: unclear, 7: no counterparts.\\
{\it Column 16.} {\bf z}: redshift of the galaxies and AGN when available {in SIMBAD/NED}.\\
{\it Column 17.} {\bf t}: time range (in unit of year) when the fluxes of flaring galaxies declined by a factor of 10.\\
{\it Column 18.} {\bf ID\_RESOURCE}: the astronomical database or catalogue from which the best match has been selected. e.g. SIMBAD, NED, WISEx\_{\it Gaia}x, where x means the number of AllWISE and {\it Gaia}-GR2 associations. For example, WISE1\_{\it Gaia}2 means that the XMMSL2 X-ray source had 1 AllWISE and 2 {\it Gaia}-GR2 associations in the cross-matching. {\bf mul} after SIMBAD or NED means that there is more than one counterpart, and {\bf lc} appended to the resource means that the identification of the source has been updated after checking the X-ray light curves as described in Section~\ref{sec:galaxy} in the paper.


\bsp	
\label{lastpage}
\end{document}